\documentclass[manuscript,screen,review]{acmart}
\AtBeginDocument{%
  }

\usepackage[utf8]{inputenc} %
\usepackage[T1]{fontenc}    %

\usepackage{pifont}
\usepackage{xcolor}         %
\usepackage{booktabs}
\usepackage{hyperref}
\hypersetup{colorlinks,
    linkcolor=red,citecolor=citecolor,urlcolor=blue}
\usepackage{algorithm}
\usepackage{graphicx}
\usepackage{threeparttable}
\usepackage{multirow}
\usepackage{amsmath}
\usepackage{bm}
\usepackage{bbm}
\usepackage{amsfonts}
\usepackage{amsthm}
\usepackage[noend]{algpseudocode}
\usepackage{enumitem} %
\usepackage[subrefformat=parens,labelformat=parens]{subfig}
\usepackage{colortbl}
\usepackage[normalem]{ulem}
\usepackage{soul}
\usepackage{graphicx} %
\usepackage{booktabs}
\usepackage{algpseudocode}
\algtext*{EndIf}

\usepackage{wrapfig}

\usepackage{booktabs}

\usepackage{xspace}

\definecolor{citecolor}{RGB}{34,139,34}
\definecolor{mydarkblue}{rgb}{0,0.08,1}
\definecolor{mydarkgreen}{rgb}{0.02,0.6,0.02}
\definecolor{mydarkred}{rgb}{0.8,0.02,0.02}
\definecolor{mydarkorange}{rgb}{0.40,0.2,0.02}
\definecolor{mypurple}{RGB}{111,0,255}
\definecolor{myred}{rgb}{1.0,0.0,0.0}
\definecolor{mygold}{rgb}{0.75,0.6,0.12}
\definecolor{myblue}{rgb}{0,0.2,0.8}
\definecolor{mydarkgray}{rgb}{0.,0.2,0.2}

\definecolor{lightred}{RGB}{255,235,235}
\definecolor{lightgreen}{RGB}{235,255,235}
\definecolor{lightblue}{RGB}{235,235,255}
\definecolor{lightcyan}{RGB}{235,255,255}
\definecolor{lightmagenta}{RGB}{255,235,255}
\definecolor{lightyellow}{RGB}{255,255,235}

\definecolor{qxkcolor}{RGB}{215,235,255}
\definecolor{softmaxcolor}{RGB}{230,235,255}
\definecolor{probxvcolor}{RGB}{255,255,235}

\definecolor{topkcolor}{RGB}{255,235,235}
\definecolor{zecolor}{RGB}{255,255,235}
\definecolor{dynacolor}{RGB}{235,255,255}

\definecolor{reviewcolor}{RGB}{0,0,200}

\theoremstyle{plain}

\theoremstyle{definition}

\algdef{SE}[DOWHILE]{Do}{doWhile}{\algorithmicdo}[1]{\algorithmicwhile\ #1}%

\newcommand{\squishlist}{
 \begin{list}{$\bullet$}
  { \setlength{\itemsep}{0pt}
     \setlength{\parsep}{3pt}
     \setlength{\topsep}{3pt}
     \setlength{\partopsep}{0pt}
     \setlength{\leftmargin}{1.5em}
     \setlength{\labelwidth}{1em}
     \setlength{\labelsep}{0.5em} } }

\newcommand{\squishend}{
  \end{list}  }

\newcommand{\name}{\texttt{SKYLIGHT}\xspace}

\AtBeginDocument{%
  \fancyhf{} %
  \fancyfoot[C]{\textit{Approved for Public Release; Distribution Unlimited: PA CLEARANCE PENDING - DO NOT DISTRIBUTE}} %
}

\begin{document}

\title{LANTERN: 3D Non-volatile Photonic Crossbar Array Architecture for Energy-Efficient Intelligent Processing}
\title{LANTERN: Large-Area Neuromorphic Three-dimensional AI Engine for Real-time Non-volatile photonic inference}
\title{SKYLIGHT: A \underline{S}ca\underline{l}able \underline{H}undred-Channel 3D Photonic \underline{I}n-Memory \underline{T}ensor Core Architecture for Real-time AI Inference}

\author{Meng Zhang}
\authornote{Both authors contributed equally to this work.}
\affiliation{%
  \institution{Rensselaer Polytechnic Institute}
  \city{Troy}
  \state{NY}
  \country{USA}
}
\email{zhangm19@rpi.edu}

\author{Ziang Yin}
\authornotemark[1]
\affiliation{%
  \institution{Arizona State University}
  \city{Tempe}
  \state{AZ}
  \country{USA}
}
\email{ziangyin@asu.edu}

\author{Nicholas Gangi}
\affiliation{%
  \institution{Rensselaer Polytechnic Institute}
  \city{Troy}
  \state{NY}
  \country{USA}
}
\email{gangin2@rpi.edu}

\author{Alexander Chen}
\affiliation{%
  \institution{Air Force Research Laboratory}
  \city{Rome}
  \state{NY}
  \country{USA}
}
\email{alexander.chen.2@us.af.mil}

\author{Brett Bamfo}
\affiliation{%
  \institution{Rensselaer Polytechnic Institute}
  \city{Troy}
  \state{NY}
  \country{USA}
}
\email{bamfob@rpi.edu}

\author{Tianle Xu}
\affiliation{%
  \institution{Rensselaer Polytechnic Institute}
  \city{Troy}
  \state{NY}
  \country{USA}
}
\email{xut6@rpi.edu}

\author{Jiaqi Gu}
\affiliation{%
  \institution{Arizona State University}
  \city{Tempe}
  \state{AZ}
  \country{USA}
}
\email{jiaqigu@asu.edu}

\author{Zhaoran Rena Huang}
\affiliation{%
  \institution{Rensselaer Polytechnic Institute}
  \city{Troy}
  \state{NY}
  \country{USA}
}
\email{huangz3@rpi.edu}

\renewcommand{\shortauthors}{Zhang et al.}

\begin{abstract}
  
The growing computational demands of artificial intelligence (AI) are challenging conventional electronics, making photonic computing a promising alternative. However, existing photonic architectures face fundamental scalability and reliability barriers. This paper introduces \name, a scalable 3D photonic in-memory tensor core architecture designed for real-time AI inference. By co-designing its topology, wavelength routing, accumulation, and programming in a 3D stack, \name overcomes key limitations. Its innovations include a low-loss 3D Si/SiN crossbar topology, a thermally robust non-micro-ring resonator (MRR)-based wavelength-division multiplexing (WDM) component, a hierarchical signal accumulation using a multi-port photodetector (PD), and optically programmed non-volatile phase-change material (PCM) weights. 
Importantly, \name enables in-situ weight updates that support label-free, layer-local learning (e.g., forward-forward local updates) in addition to inference.
Using SimPhony~\cite{ONN_DAC2025_Gu_SimPhony} for system-level modeling, we show that a single $144\times256$ \name core is feasible within a single reticle and delivers 342.1 TOPS at 23.7 TOPS/W, enabling ResNet-50 inference at 1212 FPS with $\sim$27 mJ per image, and achieves 84.17 FPS/W end-to-end (1.61$\times$ higher than an NVIDIA RTX PRO 6000 Blackwell GPU) under the same workload in real-time measurements.
System-level evaluations on four representative machine learning tasks, including unsupervised local self-learning, demonstrate \name’s robustness to realistic hardware non-idealities (low-bit quantization and signal-proportional analog noise capturing modulation, PCM programming, and readout variations).
With noise-aware training, \name maintains high task accuracy, validating its potential as a comprehensive solution for energy-efficient, large-scale photonic AI accelerators.

\end{abstract}

\maketitle
\thispagestyle{fancy} %

\section{Introduction}
The rapidly growing computational demands of artificial intelligence (AI), particularly for large-scale neural networks, have exposed fundamental limitations of conventional electronic accelerators in terms of energy efficiency, memory bandwidth, and latency, motivating the exploration of alternative computing paradigms that can support high-throughput, massively parallel linear algebra operations.
Photonic computing has emerged as a promising candidate to address these challenges by leveraging the intrinsic properties of light—ultra-high bandwidth, low propagation latency, and parallelism across wavelength, polarization, space, and time. Photonic matrix–vector multiplication (MVM) enables analog computation of dense linear operations at the speed of light, offering the potential for orders-of-magnitude improvements in throughput and energy efficiency compared to digital electronic counterparts. Analog photonic compute approach is particularly attractive for a class of applications such as RF signal processing where high processing throughput is critical with marginal computation error tolerance.  

The paradigm shift of photonic computing is to combine with non-volatile photonic memory to further eliminate repeated data transfers and constant weight holding power consumption, enabling "compute-where-the-data-resides" and significantly reducing system-level overhead.
Phase-change materials (PCMs), as a non-volatile weight storage medium, have emerged as a promising candidate for photonic in-memory computing. Early implementations of PCMs for photonic memory applications faced challenges related to limited endurance, low bit precision, and excess optical loss. Recent advances in materials engineering, however, have demonstrated switching endurance on the order of $10^{6}$ to $10^{8}$ cycles~\cite{sawant2025high}, low loss~\cite{SOREF2024111005} and up to 7-bit precision~\cite{liu2024GSST}. Therefore, PCM cells are now well positioned for integration into large-scale silicon photonic computing networks, offering compact footprint, low insertion loss, multi-level programmability, and long retention for direct encoding of synaptic weights in the optical domain. Recent demonstration in ~\cite{feldmann2021parallel} using PCM-based photonic crossbar arrays validated the feasibility of convolution and matrix operations with high speed and low latency on Si photonic integrated circuit (PIC) chip, establishing a compelling path toward photonic tensor cores (PTCs) for AI inference.

Despite these promising results, existing PCM photonic processors remain limited to small-scale arrays, on the order of $\sim $ 10 channels~\cite{NP_Nature2021_Feldmann}. 
Scaling the architectures to very large-scale photonic integration (VLPI), utilizing crossbars with hundreds of rows and columns, is essential to realizing the inherent advantages of photonics, yet remains exceptionally challenging. These limitations arise from \uline{\textbf{fundamental system-wide bottlenecks}} in topology, accumulation, wavelength routing, and programming methodology.
\ding{202}~\textbf{Loss accumulation from 2D photonic topologies}.~
Most prior crossbar designs rely on planar (2D) photonic layouts, where rows and columns intersect within a single waveguide layer. 
As array sizes grow, the proliferation of waveguide crossings, splitters, and routing detours leads to prohibitive insertion losses and degraded signal integrity. For instance, while a standard foundry-provided silicon waveguide crossing may exhibit a loss as low as 0.13 dB/device, scaling to hundreds of crossbar lines causes the cumulative loss to reach several tens of dB, adding power budget challenge for the light source. 
This accumulated high loss limits the power splitter fan-out depth, making large-scale 2D photonic crossbars impractical even with aggressive amplification.
\ding{203}~\textbf{Accumulation bottlenecks in large-scale photonic MVM}.~
Efficient accumulation of partial sums is a central challenge for scaling photonic computing systems. 
\uline{Coherent accumulation}~\cite{NP_Nature2021_Feldmann} schemes based on cascaded combiners, while compact, suffer from severe phase instability and sensitivity to fabrication and thermal variations, undermining computational reliability. 
\uline{Mode-division multiplexing (MDM) approaches}~\cite{Yang2022MultiDimMicrocombs} are constrained by large multi-mode device footprints (e.g., multi-mode waveguide, mode converters, MDM (de-)mux), high modal crosstalk, and rapidly increasing loss, limiting the number of supported modes~\cite{Su2021MDM, Cristiani2022RoadmapMultimode}.
\uline{Wavelength-division multiplexing (WDM)} provides a more scalable alternative by enabling parallel accumulation across multiple wavelengths. 
However, existing WDM-based photonic crossbars typically rely on microring resonators (MRRs) to couple wavelength channels onto a common bus~\cite{NP_SciRep2017_Tait,Ohno2022MRRCrossbarACSPhotonics,Ning2024HardwareEfficientEPIC,Luan2023HighDensityPhotonicTensorCore}. 
These resonant devices are intrinsically sensitive to temperature ($\sim$70--100 pm/$^\circ$C) and fabrication variations, making them unsuitable for large-scale energy-efficient deployment without complex and power-hungry thermal tuning/locking.
At the other extreme, \uline{purely photocurrent accumulation}~\cite{ONN_ICCAD2024_Gu} where each channel, detected and summed electronically, sacrifices the key advantage of optical parallelism; while the requirement of a large number of photodetectors introduces cumulative electrical noise, and degrades signal-to-noise ratio (SNR).
\ding{204}~\textbf{Limitations of electrically programmed PCM cells}.~
Most existing PCM photonic systems employ electrically controlled programming using integrated microheaters and electrodes. This approach is fundamentally limited by stochastic nucleation and growth dynamics in the crystal and grain-scale variability, which constrain multi-level precision and device-to-device repeatability. Those on-chip interconnected microheaters can also cause thermal crosstalk that limits the density and scalability of the photonic arrays.
Moreover, the large number of electrical routing traces and electrodes to the PCM cells and PDs significantly complicate the electrical interconnect layout for scaling.  %

The scaling limitation to hundreds of channels cannot be achieved through incremental device improvements alone. 
Instead, it requires fundamental innovation that simultaneously addresses: \emph{(1) optical loss and routing congestion issue beyond planar 2D layouts; (2) robust waveguide multiplexing without reliance on thermally sensitive MRRs; (3) 
Statistical manufacture variation resulting in device function non-idealities in individual components, such as deviation from designed power splitting and phase errors, that can accumulate and become detrimental to overall system performance; (4) scalable accumulation that preserves optical parallelism with high signal integrity, and (5) excellent weight programming repeatability and thermal stability of the PCM memory cells}.

To break through these scalability and reliability barriers, we propose \name, a scalable 3D, WDM-enabled, non-volatile photonic tensor core architecture by co-designing topology, wavelength routing, accumulation, and programming in a 3D stack.
By distributing computation across vertically stacked silicon (Si) and silicon nitride (SiN) layers, \name eliminates cascaded crossings within the compute fabric, enabling low-loss scaling to large arrays. 
A non-resonant WDM datapath built from compact dispersion-engineered Mach–Zehnder modulators (MZMs) and Bragg-grating-assisted wavelength-selective couplers (WSCs) provides thermally robust wavelength routing.
Hierarchical partial result accumulation combines multi-port photodetection with photocurrent summation, while heterogeneously integrated III–V SOAs and VCSEL programmed PCM weight banks preserve signal integrity and enable repeatable, low-crosstalk non-volatile weight storage.

Our main contributions are as follows:
\begin{itemize}[noitemsep, topsep=0pt]
    \item \textbf{3D Si/SiN Crossbar Topology} --
    We introduce a 3D Si/SiN photonic crossbar architecture that eliminates in-fabric cascaded crossings, enabling low-loss scaling to large non-volatile arrays (up to $144\times256$).
    \item \textbf{Thermally Robust, Non-Resonant WDM Datapath} -- We replace MRR-based routing with compact ($\sim150\,\mu$m) dispersion-engineered slow-light Si MZMs and Bragg-grating-assisted WSCs, enabling stable WDM over $40^\circ$C--$50^\circ$C without continuous thermal locking.
    \item \textbf{Hierarchical Accumulation at Scale} -- We combine WDM parallelism with multi-port photodetection and photocurrent summation to achieve high-SNR large-scale partial result accumulation.
    \item \textbf{Scalable Non-Volatile PCM Weight Banks and Loss Management:} We develop VCSEL-programmed multi-level PCM weight banks (7 bits, $>10^6$ cycles) and integrate III--V SOAs for power equalization to maintain weight programming repeatability and signal integrity in large fabrics.
     \item \textbf{Advanced Optical Emitter/PCM Cell Packaging Solution} -- Leveraging advanced chip packaging, we introduce a vertical cavity surface emitting laser (VCSEL) array heterogeneously integrated to the Si PIC chip to optically program the weight memory bank. This approach aims to minimize heater-induced thermal crosstalk and enable localized, controlled crystalline nucleation for precise refractive index modulation.
    \item \textbf{System-Level Evaluation} --
    We conduct end-to-end architecture evaluation, including area/power scaling and various machine learning case studies, validating efficiency and compute density under realistic non-idealities.
\end{itemize}

\section{Proposed 3D Photonic In-Memory Tensor Core Architecture \name}

\subsection{Architecture Overview}
\begin{figure}[h]
  \centering
  \includegraphics[width=\linewidth]{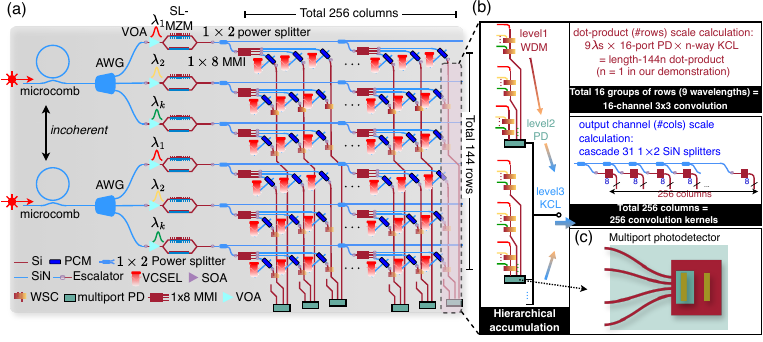}
  \caption{(a) A schematic of \name tensor core architecture; (b) Illustration of 3D fanout network, hierarchical accumulation; and (c) A schematic of a multi-port Ge PD.}
  \label{fig:PTC_Arch}
  \Description{Architecture overview.}
\end{figure}

\noindent\textbf{Convolution-to-MVM mapping.}
\name follows the standard lowering of a convolution to a matrix--vector multiply (e.g., im2col/unfold), as commonly used in crossbar accelerators and prior photonic tensor cores~\cite{feldmann2021parallel, ONN_ICCAD2024_Gu}.
For each output spatial position, the $3\times3\times C_{\mathrm{in}}$ receptive-field patch is flattened into an input vector $\mathbf{x}\in\mathbb{R}^{9C_{\mathrm{in}}}$, and each output-channel kernel is flattened into a weight vector $\mathbf{w}_k\in\mathbb{R}^{9C_{\mathrm{in}}}$; stacking all kernels forms a weight matrix $\mathbf{W}\in\mathbb{R}^{9C_{\mathrm{in}}\times C_{\mathrm{out}}}$.
The PCM crossbar programs $\mathbf{W}$ column-wise (one column per kernel), while $\mathbf{x}$ is broadcast along the rows; the column outputs implement $\mathbf{y}=\mathbf{W}^\top \mathbf{x}$, producing $C_{\mathrm{out}}$ convolution results in parallel.
In \name, one group of nine WDM wavelengths naturally represents the nine spatial taps of a $3\times3$ window, and the 144-row core comprises 16 such groups, enabling a full $3\times3$ convolution over up to $C_{\mathrm{in}}{=}16$ input channels per pass (Fig.~\ref{fig:PTC_Arch}(b)); layers with larger $C_{\mathrm{in}}$ or $C_{\mathrm{out}}$ are handled by tiling/time-multiplexing with accumulation of partial sums.

Fig.~\ref{fig:PTC_Arch} outlines the proposed \name 3D photonic in-memory tensor core. SKYLIGHT employs a scalar dot-product operation, in which the input signal $x$, encoded on the SL-MZM, is multiplied by the weight programmed as intensity in the PCM, forming the fundamental functional block of the 3D crossbar array. The use of scalar (intensity-based) multiplication, rather than phase-based multiplication, relaxes the requirements on optical path-length matching and reduces sensitivity to phase errors and MMI splitter non-idealities.
 \name achieves large-scale convolution/MVM on a compact Si/SiN stack by co-designing \textbf{\uline{four pillars}}: 
\ding{202}~robust WDM datapath, 
\ding{203}~a crossing-free multi-layer crossbar topology, 
\ding{204}~non-volatile PCM weighting with vertical optical programming, and 
\ding{205}~hierarchical accumulation for scalable partial product summation.

\noindent\ding{202}~\textbf{WDM input encoding and robust wavelength routing.}~
Each comb laser provides multiple wavelength channels that are demultiplexed into nine wavelengths (nine rows) to match a $3\times3$ convolutional kernel when mapping CNN inference workloads. 
Each wavelength is power-equalized prior to signal encoding to an ultra-compact slow-light MZM ($\sim150\,\mu\text{m}$ long) that operates at 6- to 8-bit amplitude modulation at $100\,\text{MHz}$--$1\,\text{GHz}$. 
Within the crossbar, wavelength channels are routed and (de-)multiplexed using thermally robust wavelength-selective couplers (WSCs), avoiding sensitive micro-ring resonators (MRRs) and enabling stable multi-wavelength operation under temperature drift.

\noindent\ding{203}~\textbf{Multi-layer Crossing-free Crossbar Topology.}
To distribute the modulated inputs across the array, \name uses a staged crossing-free fanout network: 31 cascaded $1\times2$ SiN splitters provide 32-way broadcast, after which each branch is transferred to the Si layer via a vertical escalator and uniformly distributed by a $1\times8$ Si multi-mode interferometer (MMI). 
This two-stage broadcast realizes fan-out to 256 columns while maintaining a crossing-free compute fabric: row waveguides reside in SiN, and column bus waveguides reside in Si, eliminating cascaded in-plane crossings that otherwise dominate loss and routing congestion in large planar crossbars.

\noindent\ding{204}~\textbf{Non-volatile PCM Weight Bank.}~
Non-volatile weights are stored directly in the photonic crossbar using PCM cells integrated in the SiN layer, which function as absorptive attenuators for in-memory dot-product. 
Weights are programmed optically using a vertically integrated VCSEL array (at wavelength of $1064\,\text{nm}$), where each emitter couples through a vertical coupler to deliver localized programming pulses to the target PCM cell, minimizing dense electrical routing and reducing thermal crosstalk.

\noindent\ding{205}~\textbf{Hierarchical Accumulation.}~
To scale partial-product accumulation to $>100$ rows without fragile coherent combining or parasitics-heavy per-channel electrical readout, \name adopts a hierarchical, multi-level summation strategy (Fig.~\ref{fig:PTC_Arch}(b)) based on multi-port Ge photodetectors. 
At each level, multiple waveguide buses are combined directly at a multi-port PD. 
Recent demonstrations validate the feasibility of multi-port Ge PDs for waveguide-multiplexed MVM~\cite{tang2025waveguide}; here, we build on this approach with AIM-compatible multi-port PDs (leveraging our evanescent PD design experience) to enable low-noise, scalable accumulation across over 100 spatial/spectral channels while maintaining signal integrity at scale.

\subsection{Key Photonic Component Design in \name}

In this section, we introduce the key photonic components, including the comb laser, AWG for wavelength demultiplexing, Bragg grating-assisted WSC, PCM cell integrated with VCSEL array via inversely designed bi-layer grating coupler (BGC), and multi-port Ge photodetectors. Our analysis incorporates both component metrics from existing literatures and original device designs from our group.

\subsubsection{Comb Laser}
Optical frequency combs serve as the backbone for modern high-capacity optical systems, including WDM, by generating a precise set of discrete and equally spaced frequencies. This capability makes them an ideal compact, integrated multi-wavelength source for a crossbar in-memory computing architecture. These wavelength channels ensure the modal orthogonality necessary for signal routing on a shared waveguide bus line. Integrated distributed feedback (DFB) lasers, with emission aligned with the WDM grid frequencies, can also fulfill crossbar functions; however, they require more sophisticated driver controls and occupy a much larger footprint than comb sources. These drawbacks, combined with the challenges of high-density integration on silicon photonic platforms, limit their scalability for high-channel-count crossbar PNN architectures. While the technology readiness level (TRL) of comb lasers is currently lower than that of DFB lasers, this work still focuses on the comb laser due to extensive research activities in this area and the anticipation of its near-future availability in silicon photonic foundries.

The method of generation of comb frequencies is the primary differentiator among the three main types of optical frequency combs: mode-locked lasers, Kerr microcombs, and electro-optic combs \cite{Okawachi_comb_2023}. Mode-locked laser combs generate a broad spectrum through supercontinuum generation, which occurs in highly nonlinear optical waveguides. While traditionally built from bulky solid-state or fiber-based sources, recent advancements have focused on developing compact and scalable PIC. Despite this progress, chip-scale mode-locked lasers remain an emerging technology. Current implementations demonstrate a trade-off between comb spacing (repetition rate) and power. For instance, one device achieves a 3.03 GHz repetition rate with an average power of 12.5 mW \cite{Poelman_mode_locked_2025}, whereas another pushes the repetition rate to 20 GHz but with a lower average power of 1.83 mW \cite{Davenport_mode_locked_2018}. Kerr frequency combs, often called microcombs, leverage the strong nonlinear effects within microresonators. By balancing four-wave mixing with the resonator's group-velocity dispersion, a single-frequency laser pump can be converted into a wide-spanning comb. These microresonators, typically shaped as microrings, confine light for long periods, enabling highly efficient nonlinear interactions. A SiN microring with a 231 GHz free-spectral range (FSR) demonstrated an external conversion efficiency of 31.8\%, with individual comb lines reaching up to 17 dBm \cite{Xue_kerr_2017}. In this device, 5 lines centered around 1550 nm exhibited powers greater than 10 dBm, and nearly 20 lines exceeded 0 dBm. An advanced coupled-ring geometry in SiN with FSRs around 200 GHz achieved an even higher 41\% conversion efficiency, producing 18 distinct lines with powers greater than 1 mW \cite{Kim_kerr_2019}. Electro-optic combs utilize the Pockels effect, where an applied electric field directly modulates a material's refractive index to generate frequency sidebands around a continuous-wave laser. This method offers excellent stability and control. Recent innovations have integrated this technique into compact platforms such as a silicon ring resonator modulator, driven by an electrical signal across a p-n junction, generating a 10-GHz-spaced comb with a maximum power variation of just 0.7 dB between 5 lines and a power reduction of 15 dB relative to the optical input power \cite{Demirtzioglou_ring_2018}. A lithium niobate (LN) microring resonator was fabricated that produced a comb spanning more than 80 nm, with over 900 lines spaced by 10.43 GHz, exhibiting low power variation (<0.1 dB) between adjacent lines and a 40 dB contrast relative to the noise floor near the center wavelength \cite{Zhang_ring_2019}.

The performance of commercially available frequency combs reflects the maturity of these technologies. Standard C-band products offer an average power of 50 mW with selectable channel spacings of 250 MHz, 1 GHz, or 2.5 GHz \cite{menhir_comb_2022}. For O-band applications, commercial systems can deliver a total optical power of 260 mW, providing over 7 mW per channel across 37 channels with a 25 GHz spacing \cite{innolume_comb_2025}. 
Most current optical comb development targets data center applications, where a high channel count is prioritized to achieve aggregate data rates in the tens of Terabits. In contrast, comb lasers for neuromorphic or in-memory computing typically require fewer channels--often 9 or 16--to align with standard $3 \times 3$ or $4 \times 4$ kernel sizes. Furthermore, the output power per comb line must be carefully optimized, as it directly dictates the scalable number of columns in a crossbar architecture, given the constraints of the optical power budget. A summary of the different comb lasers reported in the literature is presented in Table \ref{table:comb_laser}. 

The choice of frequency spacing of a comb source for the cross-bar PNN is primarily determined by the WSC of the column bus waveguide that will be introduced in Section~2.2.4. The grating-based WSC typically supports a channel spacing in the range of $200$--$250\,\mathrm{GHz}$. Therefore, Kerr frequency combs provide the most suitable option for this work as it offers high-power per channel and proper range of channel spacing. To ensure spectral intensity uniformity, variable optical attenuators (VOAs) can be employed to equalize the output power across all channels. Si VOAs are a standard component that is offered at Si photonic foundries. The system-level analysis will be based on a comb laser with 10 dBm per channel \cite{Xue_kerr_2017}. 

\begin{table}[h!]
\centering
\caption{Comparison of different comb lasers.}
\resizebox{0.53\columnwidth}{!}{
\begin{tabular}{|c | c | c | c|} 
 \hline
 Type & Repetition Rate & Power & Reference \\ [0.5ex] 
 \hline
 Mode-locked & 3.03 GHz & 12.5 mW & \cite{Poelman_mode_locked_2025} \\ 
 \hline
 Mode-locked & 20 GHz & 1.83 mW & \cite{Davenport_mode_locked_2018} \\
 \hline
 Kerr & 231 GHz & 10--17 dBm & \cite{Xue_kerr_2017} \\
 \hline
 Kerr & 200 GHz & 1 mW & \cite{Kim_kerr_2019} \\
 \hline
 Electro-optic & 10 GHz & -15 dB to input & \cite{Demirtzioglou_ring_2018} \\
 \hline
 Electro-optic & 10.43 GHz & 40 dB to noise floor & \cite{Zhang_ring_2019} \\ 
 \hline
 Commercial & 2.5 GHz & 50 mW & \cite{menhir_comb_2022} \\
 \hline
 Commercial & 25 GHz & 7 mW & \cite{innolume_comb_2025} \\ 
 \hline
\end{tabular}
\label{table:comb_laser}
}
\end{table}

\subsubsection{Arrayed Waveguide Grating}

Separating closely packed comb laser frequencies requires a scalable approach, while the arrayed waveguide grating (AWG) is well-suited for this.  
An AWG is comprised of large waveguide slabs at the input and output, termed free propagation regions (FPR), connected by \textbf{\textit{N}} waveguides of slightly varying length for separating or combining up to \textbf{\textit{N}} wavelengths.  
In the case of a 9-wavelength demultiplexer, the comb laser output is injected into the first FPR, where it is distributed across the FPR and collected at the 9 arrayed waveguide outputs at approximately equal intensities.  
Light travels along the arrayed waveguides and accumulates phase differences as a result of the intentional path length differences.  
Light from the arrayed waveguides enters the output FPR and creates interference patterns spatially, resulting in peak constructive intensities at each of the 9 FPR output ports, dependent on the wavelength designated for each port. 
Key design parameters of the AWG are the center frequency, frequency spacing, and free spectral range, and important performance metrics of the AWG include loss, crosstalk, and footprint.

Recent literature has produced many high-performance, low-loss AWGs that meet similar requirements to the proposed architecture.  
Silicon and silicon nitride are two optical material platforms of relevance when designing AWGs.  Silicon nitride exhibits low-loss at C-band frequencies at a sizable footprint penalty.  
The existing literature shows partiality towards AWGs with a number of channels equivalent to a power of 2, as is typically most relevant for communication applications.  
However, we anticipate that 8-channel devices in the literature will have very similar metrics to the required 9-frequency AWG.  
Various Si and SiN AWGs are summarized in Table \ref{table:table_awg}.  For this work, we have selected the parameters specified in \cite{shang_low-loss_2017} for system analysis.

\begin{table}[h!]
\centering
\caption{Comparison of Si and SiN AWGs.}
\begin{tabular}{|c | c | c | c | c|} 
 \hline
 Platform & Loss (dB) & Crosstalk (dB) & Footprint & Reference \\ [0.5ex] 
 \hline
 Si & 2.5 $\sim$ 6.045 & -11 & 0.7 mm $\times$ 0.27 mm & \cite{pitris_silicon_2018} \\ 
 \hline
 Si & 2.92 $\sim$ 7.66 & -15.5 $\sim$ -18 & 0.3 mm $\times$ 0.35 mm & \cite{wang_low-loss_2014} \\
 \hline
 SiN & 1.526 $\sim$ 6.13 & -15 $\sim$ -22.5 & 0.3 mm $\times$ 2.3 mm & \cite{fan_compact_2025} \\
 \hline
 SiN & 1.5 & -24 & 0.6 mm $\times$ 1.8 mm & \cite{shang_low-loss_2017} \\
 \hline
\end{tabular}
\label{table:table_awg}
\end{table}

\subsubsection{Slow-light Mach-Zehnder Modulator}

Slow-light Mach-Zehnder modulators (SL-MZMs) are essential for encoding high bitwidth and high-speed electrical signals into the optical domain. While foundry-standard silicon MZMs utilizing the carrier plasma effect typically require a $\pi$ phase-shift (PS) length of several millimeters, SL-MZMs utilize traveling-wave resonance to enhance light-matter interaction. By employing either Bragg grating~\cite{begova_mzm_2023, Han2023BG, Jafari2020BG} or photonic crystal (PhC) phase shifters~\cite{Baba2014PhC, Baba2024PhC, yan2017slow}, SL-MZMs can drastically reduce the device footprint to the 100 $\mu$m to 500 $\mu$m range, representing $5\sim10\times$ size reduction compared to standard foundry component. SL-MZM parameters are adjusted to align with the nine-channel comb frequencies, maximizing the efficiency of the SL modulators. In our prior work, a Bragg grating-based SL-MZM with PS length of $150\,\mu\mathrm{m}$ has demonstrated a 6-bit resolution at up to $3.2\,\text{GHz}$ clock frequency with the lowest achieved normalized mean square error (NMSE) of 0.0018 at a clock frequency of 800 MHz~\cite{begovic2024foundry}.
Compared to Bragg grating-based SL-MZM, photonic crystal (PhC) based SL-MZM might offer slightly higher group indices, advantageous for modulation efficiency and compactness~\cite{begovic2024manufacture}.

In this cross-bar architecture, we adopt a PhC-based SL-MZM design with a PS length of $150\,\mu\text{m}$. Using dispersion engineering~\cite{tamura2015silica, yan2017slow}, the line-shifted PhC SL-MZM will be designed to have a near-constant group index at the band edge of the photonic stop band, allowing thermally stable operation for $\sim 50^\circ\text{C}$. Insertion loss is one of the critical parameters that will affect the overall system performance. In our prior work, we have demonstrated a SL-MZM insertion loss of approximately 5 dB~\cite{begova_mzm_2023}. Through doping engineering and mode matching between the fast-light and slow-light regions, we anticipate an insertion loss can be reduced to $\sim 3\,\text{dB}$.

\subsubsection{Wavelength Selective Coupler}

\begin{figure}[ht]
  \centering
  \subfloat[]{\includegraphics[width=0.65\linewidth]{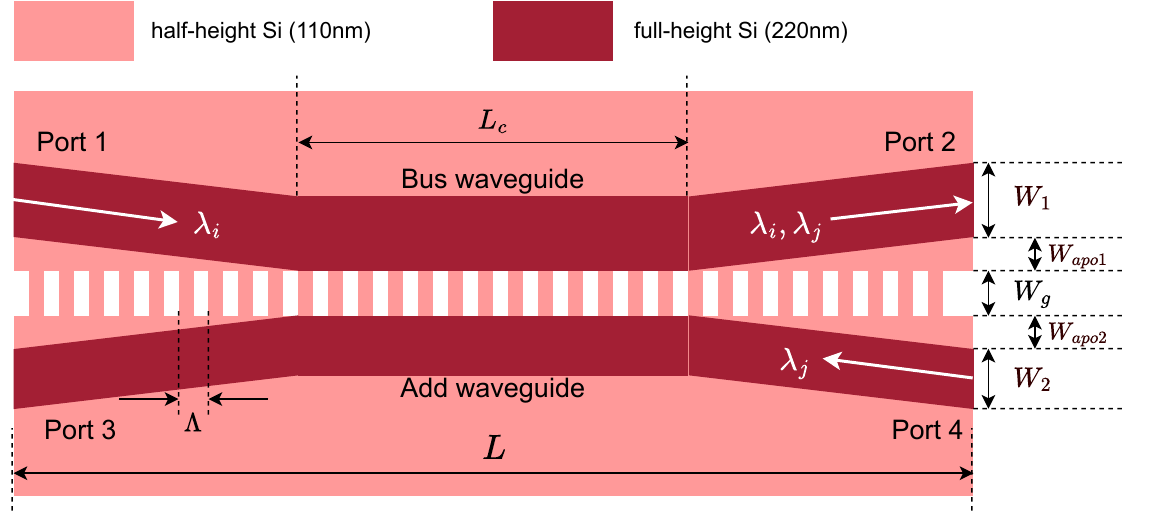}
  \label{fig:WSC_design}
  }
  \subfloat[]{\includegraphics[width=0.35\linewidth]{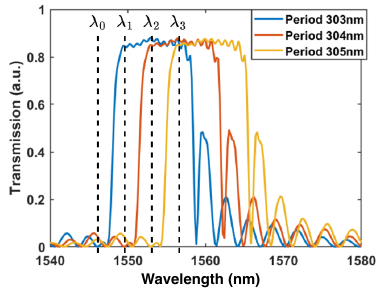}
  \label{fig:WSC_spectrum}
  }
  \caption{(a) Schematic of a Bragg grating-assisted wavelength selective coupler on a SOI substrate utilizing contra-directional coupling to add optical signal from port 4 to bus waveguide. (b) Transmission spectrum of three cascaded WSCs at Port 2 when Port 4 works as the input port. $\lambda_0$ is the original wavelength carried by the bus waveguide, and $\lambda_1 \sim \lambda_3$ are the wavelengths being coupled to the bus waveguide by three WSCs, respectively.
  }
  \label{fig:WSC}
  \Description{WSC design.}
\end{figure}

Micro-ring resonator (MRR) weight banks are commonly used wavelength division multiplexing components for coupling closely spaced wavelengths to a bus waveguide~\cite{feldmann2019all}, but scale poorly due to the need for simultaneous resonance alignment of many MRRs in a crossbar architecture. 
Here, we consider a Bragg grating-assisted WSC that utilizes phase-matching conditions to add specific comb laser frequencies to the vertical bus waveguide in the photonic crossbar. 
Multiple versions of the contra-directional coupler (CDC) have been explored for filter and WDM applications by several groups ~\cite{ma_cdc_2021, minz_cdc_2021, shi_cdc_2013, shi_cdcapo_2013}. 
In this work, we utilize the CDC as a WDM component for adding varying channels of optical signals to the column waveguide for subsequent incoherent summation of CNN matrix multiplication outputs. 
A schematic of the WSC design on a Silicon-on-Insulator (SOI) platform is illustrated in Fig.~\ref{fig:WSC_design} with a full-height Si thickness of 220 nm and half-height Si thickness of 110 nm. The layer thickness choice is consistent with standard silicon photonics foundry offerings. For proof-of-concept validation, a small-scale four-wavelength WSC is designed and implemented on the AIM Photonics platform. The bus waveguide and the add-drop waveguide are designed to be highly asymmetric in the coupling region with a significant width difference, i.e. $W_1 = 600~\text{nm}$, $W_2 = 400~\text{nm}$, to prevent broadband directional coupling between two arms.
The grating in the coupling region has a width of $W_g = 280~\text{nm}$ and a period $\Lambda$ in the range of 300 nm to 310 nm with a 50:50 duty cycle.
The coupler length is selected to be $L=100~\mu\text{m}$, balanced between compactness and grating coupling efficiency $\kappa$.
To suppress the sideband oscillation, we implement a simple linear apodization, noted as $W_{apo1}$ and $W_{apo2}$ in Fig. 2 (a). The central coupling region has a length of $L_c=40~\mu\text{m}$ where the bus waveguide and add waveguide remain parallel, while the waveguides follow a linear detour away from the coupling region in two symmetric input/output sections, with a maximum vertical detour of $W_{apo1}=W_{apo2}=30~\text{nm}$.

Our preliminary design in FDTD simulation demonstrates three cascaded WSCs to add four wavelengths ($\lambda_0 \sim \lambda_3$) in series to the column central bus waveguide.
The optical responses of the three WSCs at Port 2 when input light is injected from Port 4 are shown in Fig.~\ref{fig:WSC_spectrum}, where each WSC injects a distinct wavelength from a contra-directional port (Port 4) with low insertion loss ($\sim 0.5\,\text{dB}$).
The periods of the three cascaded WSC are slightly different so that the coupled wavelengths are intentionally detuned to minimize the crosstalk. 
As a low-Q grating structure, the WSC's band edge drifts only $\sim 0.4\,\text{nm}/10^\circ\text{C}$, as verified in prior measurements~\cite{zhang2025thermally}.

\subsubsection{PCM Cell Engineering }

\begin{figure}[ht]
  \centering
  \includegraphics[width=0.65\linewidth]{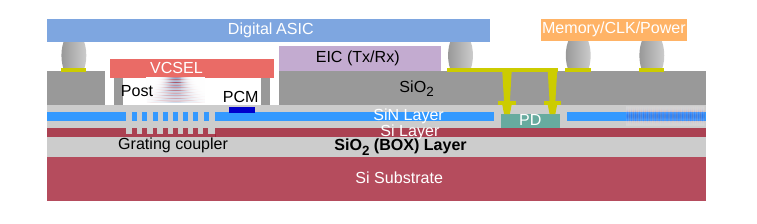}
  \vspace{-10pt}
  \caption{Heterogeneous integration of VCSEL arrays above a SiN/Si photonic computing substrate. 
  Vertical optical coupling is enabled near the PIC, whereas electrical connections are routed through the backend metal stack to the RF front-end EIC (Tx/Rx).}
  \label{fig:VCSEL integration}
  \Description{}
\end{figure}

Optical programming of PCM cells using a VCSEL emitter array is one of the key innovations to enable the scaling of the photonic CNN. %
Li et al. ~\cite{liGratingCouplerDesign2025} reported a coupon-type heterogeneous integration of VCSELs onto a Si PIC using a pick-and-place process, bonded through a submicron-thick benzocyclobutene (BCB) layer. In this approach, the VCSEL is placed in intimate contact with the Si PIC, which may introduce thermal management challenges due to limited heat dissipation. In this work, we advocate a trench/post integration scheme to enable high-precision alignment of the VCSEL to the Si photonic integrated circuit (PIC) while improving thermal dissipation. By varying the post height, the emitted beam diameter at the grating coupler can be adjusted, providing an additional degree of freedom for optimizing coupling efficiency. A schematic of the proposed packaging architecture is illustrated in Fig.~\ref{fig:VCSEL integration}. The VCSEL array chip is bonded to the silicon PIC, with each VCSEL emitter aligned to a corresponding PCM cell. For this packaging scheme, it is more convenient to employ a bottom-emitting VCSEL, in which the distributed Bragg reflector (DBR) stack is designed asymmetrically, and the substrate is transparent at the emission wavelength. Both the cathode and anode electrodes are located on the top side of the VCSEL for wire bonding, thereby eliminating the need for signal routing on the Si PIC. However, this approach requires more complex VCSEL fabrication and would increase the die cost. An alternative packaging strategy involves routing all the ring contacts of the VCSEL to one side of the VCSEL die and interconnected to form a shared common ground (or common anode). Top emitting (surface-emitting) VCSEL die is flip-chip bump-bonded to the Si PIC to form the electrical connection as well as mechanical support. The complementary electrodes are accessed from the backside of the VCSEL and wire-bonded to the PCB. In this configuration, minimal signal routing is required on the Si PIC. Overall, heterogeneous integration of a VCSEL die to a Si PIC, while maintaining precise alignment and manageable thermal and power dissipation, is nontrivial. As this paper primarily focuses on crossbar architecture computing analysis, the detailed discussion of VCSEL packaging strategies for photonic computing will be presented in a separate work.

\begin{figure}[ht]
  \centering
  \includegraphics[width=0.6\linewidth]{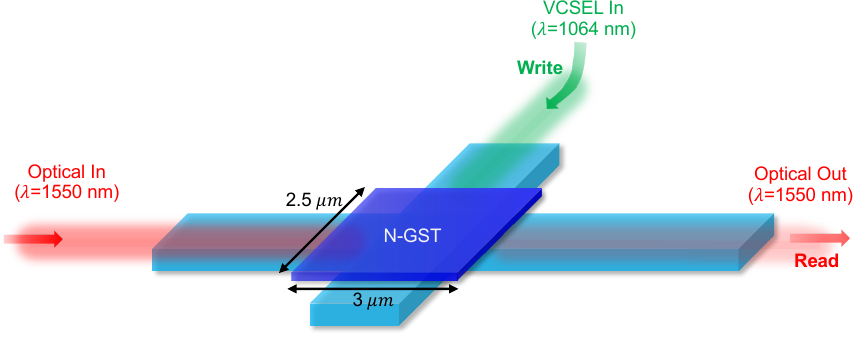}
  \vspace{-10pt}
  \caption{Schematic of the vertically programmed integrated phase-change photonic memory cell. A silicon waveguide carries the signal light at $\lambda = 1550\ \mathrm{nm}$ from the optical input to the optical output, while a vertically incident VCSEL programming beam at $\lambda = 1064\ \mathrm{nm}$ is coupled from above to switch the phase-change material. The N-GST memory element is positioned at the waveguide intersection with dimensions of $2.5\ \mu\mathrm{m} \times 3\ \mu\mathrm{m}$, enabling localized optical modulation through vertical optical programming and in-plane signal transmission.} 
  \label{fig: PCM Crossbar}
  \Description{}
\end{figure}

The optical emitter wavelength for PCM cell programming can span a broad range, as the absorbed optical energy is ultimately converted into heat to trigger crystalline nucleation. The emission intensity and pulse duration of each VCSEL are electrically controlled, enabling precise optical programming of the PCM memory weights.  To avoid stochastic nucleation associated with direct surface illumination, the VCSEL emission is not incident directly on the PCM patch but coupled through a grating coupler into a SiN waveguide. The guided light is then routed to the PCM region positioned at the intersection with the optical signal waveguide, as illustrated in Fig.~\ref{fig: PCM Crossbar}.

GaAs-based VCSEL arrays with emission at 850~nm and 940~nm are rather mature technology. At these wavelengths, silicon exhibits strong absorption (e.g., $\alpha \approx 1000\,\mathrm{cm^{-1}}$ at 940~nm, corresponding to an absorption length of $\sim 10\,\mu\mathrm{m}$), which limits the grating coupler implementation to the SiN layer, while restricted from Si layer. Quantum-dot embedded GaAs VCSELs can be employed to induce material strain, thus shifting the emission wavelength beyond the silicon absorption edge toward O-band operation \cite{ref-VCSEL-403,ref-VCSEL-404,ref-VCSEL-405}. However, O-band VCSELs generally exhibit lower internal quantum efficiency and wall-plug efficiency compared to GaAs-based devices. InGaAs/GaAs VCSELs operating at 1064~nm represent a promising alternative. At this wavelength, silicon absorption is significantly reduced ($\alpha \sim 10\,\mathrm{cm^{-1}}$, corresponding to an absorption length of approximately 1~mm), as it is very close to the Si energy band edge.  Since the VCSEL signals only needs to propagate a short distance ($\sim 80\,\mu\mathrm{m}$, corresponding to the Si/SiN escalator lateral length) in a Si waveguide before being fully transferred to the SiN waveguide, the absorption in Si is deemed acceptable provided the coupling efficiency can be significantly improved using a bilayer SiN/Si grating coupler design. 

To utilize a VCSEL for PCM programming via a grating coupler, the device must exhibit high brightness and stable lasing characteristics with excellent beam quality. PCM cells typically respond on microsecond time scales (from a few to tens of microseconds), so a modulation speed in the range of $10$--$100\,\mathrm{kHz}$ is sufficient for programming. High-power VCSEL operation at $1064\,\mathrm{nm}$ with modulation speed at tens of GHz has been reported~\cite{ref-VCSEL-401,ref-VCSEL-402,ref-VCSEL-409}. For efficient grating-coupler coupling, single-mode operation, i.e., fundamental transverse mode with suppression of higher-order modes, is strongly preferred. Several transverse-mode control techniques have been explored, such as antiresonant reflecting optical waveguide (ARROW) structures \cite{ref-VCSEL-410, ref-VCSEL-411}, impurity-induced disorder \cite{ref-VCSEL-408},  surface-relief engineering combined with multi-junction designs \cite{ref-VCSEL-407}, antiresonant oxide islands \cite{ref-VCSEL-406}, etc. More recently, an oxide-confined VCSEL structure incorporating double oxide islands was reported to sustain single-transverse-mode output power up to $36.2\,\mathrm{mW}$ at a wavelength of $\lambda = 1064\,\mathrm{nm}$, with a power conversion efficiency (PCE) of $54.8\%$ \cite{ref-VCSE-412}. 

Large-scale VCSEL arrays comprising up to several thousand emitters at emission wavelengths of 850 nm or 940 nm are widely available from commercial vendors. Array VCSEL emitters at 1064 nm, though less common, are also commercially available, offering continuous-wave (CW) output powers ranging from several watts to 30 W (e.g., Princeton Optronics, PQCW-CS6-30-W1064). In such arrays, individual emitters typically deliver optical power on the order of ~5 mW. In this work, low power VCSEL emitters can also be employed, with multiple emitters integrated to program one PCM cell. This approach can not only enhance operational reliability and beam quality of the VCSEL emitters but may also potentially enable improved bit precision in weight encoding combined with multi-segment programming schemes. Moreover, PCM programming requires only pulsed optical operation, while a VCSEL can provide much higher peak power in pulsed operation than in its CW mode. However, most commercial 1064 nm VCSEL arrays are optimized for specialized sensing, free-space optical communication, and various industrial applications. Implementing a 1064 nm VCSEL array specifically tailored for heterogeneous integration with PCM-based weight encoding on a silicon PIC will require additional research and development.

\begin{figure}[ht]
  \centering
  \includegraphics[width=1\linewidth]{figs/VCSEL_inverse_design_v3.png}
  \vspace{-10pt}
  \caption{(a) Schematic of the vertical coupling configuration. The simulation includes a thin SiO$_2$ layer between the 220 nm Si and SiN layers. (b) Inverse-designed SiN etch pattern. (c) Inverse-designed Si etch pattern. (d) Electric-field intensity distribution in the x-y plane in the middle of the SiN layer. (e) Electric-field intensity distribution in the x-y plane in the middle of the Si layer. (f) Electric-field intensity in the x–z plane at $y = 0\ \mu\text{m}$.}
  \label{fig:inverse design}
  \Description{}
\end{figure}

To simplify fabrication and assembly, vertical coupling is required because the VCSEL emitters operate under normal incidence. Conventional uniform Bragg gratings require oblique incidence to satisfy the phase-matching condition, while under normal incidence, the coupling efficiency is nearly zero. \cite{yangHighperformanceGratingCouplers2023}. In this work, we explore an inverse design method to overcome this constraint by providing additional degrees of freedom to tailor the optical field scattering profile and enable symmetry breaking. We explored a fabrication-aware grating coupler compatible with the AIM process design kit (PDK), in which a minimum feature size of 100~nm is enforced to meet the DUV photolithography resolution. 
For a feasibility study, we explored an inverse-designed bilayer grating coupler (BGC) for VCSEL light coupling at $\lambda = 1064\,\mathrm{nm}$. The BGC consists of a fully etched SiN layer and a partially etched silicon layer with an etch depth of 160~nm. The thickness of both layers is fixed at 220~nm. The overall vertical coupling configuration is illustrated in Fig.~\ref{fig:inverse design}(a). The design footprint is set to \(8~\mu\text{m} \times 8~\mu\text{m}\), corresponding to the VCSEL beam waist. Adjoint optimization is carried out in Tidy3D for 30 iterations, yielding a simulated fundamental-mode coupling efficiency of $72.0\%$ (insertion loss of 1.43 dB ) under normal incidence with the fabrication penalty by Tidy3D. 
To ensure optimization stability, the projection factor \(\beta\) is kept constant at 5 throughout the entire process. The optimized etch patterns of the Si and SiN layers are presented in Fig.~\ref{fig:inverse design}(b) and (c), respectively. 
Fig.~\ref{fig:inverse design}(d) provides the corresponding side-view intensity distribution of the device, while Fig.~\ref{fig:inverse design}(e) and (f) show the electric-field intensity distributions in the SiN and Si layers.
For system power/loss link budget analysis, the grating loss is set at 2.7 dB. In future studies, we will consider incorporating 
distributed Bragg reflector (DBR) beneath the substrate \cite{huangCompactInverseDesigned2025} and meta-surface \cite{yoonInverseDesignSibased2023} to the grating coupler to improve grating coupling efficiency.

Phase-change materials provide a versatile platform for photonic in-memory computing, but their performance strongly depends on material choice. Ge$_2$Sb$_2$Te$_5$ (GST) offers large refractive index contrast and fast switching speed, enabling compact, high-modulation-efficiency photonic devices \cite{Guo2019, Bedeschi2009}; however, its extinction coefficient increases significantly from $\sim$0.08 in the amorphous phase to $\sim$1.23 in the crystalline phase at $\lambda = 1550\,\mathrm{nm}$, resulting in high optical absorption and insertion loss that limits large-scale scalability \cite{Aryana2023}. Emerging materials such as Ge$_2$Sb$_2$Se$_4$Te$_1$ (GSST) and Sb$_2$Se$_3$ exhibit significantly lower optical absorption due to their reduced extinction coefficients and larger band gaps, enabling improved transparency in both amorphous and crystalline phases while maintaining sufficient refractive index contrast for efficient optical modulation. This reduction in absorption arises from the substitution of tellurium with selenium, which increases the band gap and suppresses free-carrier absorption. However, this increased structural and amorphous-phase stability results in slower crystallization kinetics and growth-dominated phase transitions, leading to relatively slower switching speed \cite{Zhang2019, Delaney2020}.

\begin{table}[htbp]
\centering
\caption{Optical properties of N-GST in amorphous and crystalline phases, where $L$ ($\mu$m) corresponds to the length where 86\% of the optical power is absorbed}
\label{tab:NGST_properties}

\begin{tabular}{cccccc}
\toprule
Phase & $\lambda$ (nm) & $n$ \cite{Xia2024} & $k$ \cite{Xia2024} & $\alpha$ (dB/$\mu$m) & $L$ ($\mu$m) \\
\midrule

A-N-GST & 1064 & 3.80 & 0.20 & 1.1 & 2.0 \\
        & 1550 & 3.58 & 0.06 & 0.27 & 7.10 \\

C-N-GST & 1064 & 5.10 & 1.70 & 61.05 & 0.032 \\
        & 1550 & 5.67 & 0.75 & 14.90 & 0.134 \\

\bottomrule
\end{tabular}

\end{table}

To balance multilevel stability, optical loss, and device scalability, a nitrogen-doped GST (N-GST) was selected as the PCM for this work. Nitrogen doping induces grain refinement and progressive crystallization, improving intermediate-state stability and multilevel programmability, and reducing optical absorption compared to GST. In this work, we consider a 40~nm-thick N-GST film as the PCM memory element. The complex refractive index at $\lambda = 1550\,\mathrm{nm}$ of amorphous N-GST is $n_a = 3.58 + 0.06i$, corresponding to an insertion loss of approximately $0.27~\mathrm{dB/\mu m}$, while in the crystalline state $n_c = 5.67 + 0.75i$, corresponding to an insertion loss of approximately $14.9~\mathrm{dB/\mu m}$. The small extinction coefficient in the amorphous phase ensures near-transparent transmission, whereas the strong optical contrast between phases enables efficient modulation, where the extinction ratio is calculated to be approximately $36.58~\mathrm{dB}$.
In N-GST devices demonstrated previously, crystallization is electrically induced using 14.5--18.4~mW, corresponding to switching energies of approximately 3--3.63~$\mu$J and crystallization times of $\sim$20~$\mu$s. Amorphization requires approximately 42~mW, with a switching energy of $\sim$2.1~$\mu$J and a melt-quench time of $\sim$9.6~$\mu$s~\cite{Xia2024}, highlighting the relatively high power consumption of electrical programming.

\begin{figure}[ht]
  \centering
  \includegraphics[width=0.75\linewidth]{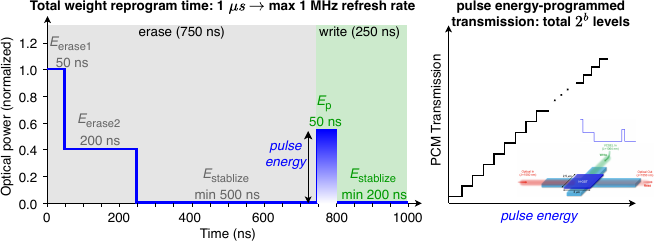}
  \vspace{-10pt}
  \caption{
    Multilevel erase and write pulse sequence used for phase-change memory programming.
    Optical pulse profile showing a two-step erase sequence consisting of an initial erase pulse ($E_{\mathrm{erase1}}$, 50~ns) and a secondary erase pulse ($E_{\mathrm{erase2}}$, 200~ns), followed by a stabilization period ($E_{\mathrm{stabilize}}$, minimum 500~ns).
    After erase/reset, the write pulse consists of a 50 ns pulse ($E_p$) with variable pulse energy, followed by a final stabilization period ($E_{\mathrm{stabilize}}$, minimum 200~ns), allowing the material transmission to reach a steady state.
    The total PCM weight update latency is 1 $\mu s$, leading to a maximum 1 MHz refresh rate.
    Multiple programming pulse energies (i.e., pulse height) enable deterministic access to distinct intermediate transmission levels, enabling $2^b$ multilevel operation.
    }
  \label{fig: Multilevel programming and erase pulse sequence}
  \Description{}
\end{figure}

In this work, we aim to utilize optical programming at $\lambda = 1064\,\mathrm{nm}$ via a VCSEL array on a Si photonic CNN chip \cite{Nevzorov2023}. The size of the PCM patch is $2.5~\mu\mathrm{m} \times 3~\mu\mathrm{m}$, with the geometry engineered to optimize optical absorption and mode overlap with the underlying waveguide. Since the first step of the amorphization process requires the highest optical energy, it is used as the benchmark for the required power delivered through the grating coupler. The grating coupler loss was simulated to be 2.7~dB. Therefore, to achieve the required 6~mW of peak power at the PCM, the VCSEL must provide approximately 12~mW of peak optical power to compensate for the coupling loss. For crystallization, a 50~ns optical pulse with a peak power of 2.7~mW is applied, yielding an energy of 135~pJ. Erase (amorphization) is achieved using a two-step pulse sequence consisting of an initial 50~ns pulse followed by a 200~ns pulse, each at an average power of 2.75~mW. These correspond to energies of 138~pJ and 55~pJ, respectively, for a total erase energy of approximately 69~pJ. This pulse scheme enables controlled melt-quench dynamics and reliable phase reset \cite{Li2019}. Overall, the total time required for one cycle is 50~ns for programming, followed by a minimum of 200~ns for stabilization, 250~ns for erase, and finally a minimum of 500~ns for stabilization of the amorphous state. This results in a theoretical total cycle time of approximately 1~$\mu$s (i.e., 1~MHz). By increasing the stabilization times, the operating frequency of the PCM can be reduced to the 10--100~kHz regime. 

Optical programming was selected primarily because it enables dramatically lower programming energy in the phase-change material, reducing the required energy from the nanojoule range for electrical switching to the picojoule level. This substantial reduction improves overall device efficiency and minimizes thermal load, which is critical for dense photonic integration. Additionally, optical excitation enables ultrafast switching, with phase transitions occurring on sub‐nanosecond timescales, allowing for significantly faster operation than electrical programming \cite{Sun2022PlasmonicPCM}. To implement this efficiently, a grating coupler is used to deliver the optical pulse directly into the waveguide, where the light remains confined and interacts strongly with the PCM through near‐field coupling. In contrast, free‐space programming suffers from incomplete absorption because a portion of the optical pulse passes through the thin PCM film, resulting in significantly higher programming energies. Waveguide delivery, therefore, enables much lower switching energies and faster, more controlled phase transitions. Furthermore, grating coupler integration eliminates bulky free‐space optical components, providing a compact, robust, and scalable solution for photonic memory arrays \cite{Zhou2022PCMPhotonicMemory}. Importantly, phase-change materials such as Ge$_2$Sb$_2$Te$_5$ exhibit crystallization temperatures on the order of $\sim$150$^\circ$C, ensuring that normal photonic circuit operating temperatures remain well below the threshold required to induce unintended phase transitions. As a result, the PCM state remains thermally stable during standard device operation, and phase changes occur only when intentionally triggered by localized optical programming pulses \cite{Ran2025GeRichGST}.

Some silicon photonic foundries, such as AIM Photonics, offer customization options that allow top trenches to be etched down to either the SiN or the Si layer by selectively removing the SiO$_2$ cladding. This capability provides process flexibility for introducing non-standard materials, such as various phase change materials, into a CMOS fabrication platform. 
The top trench opening can also facilitate more efficient VCSEL light coupling. However, mechanical integrity must be carefully evaluated, and the maximum allowable top-trench opening size should be calibrated to ensure structural stability.

\subsubsection{Integrated Semiconductor Optical Amplifier on Si Platform}

Recent progress in heterogeneous integration has enabled the bonding of III-V semiconductor optical amplifiers (SOAs) onto silicon photonic platforms. Micro-transfer printing ($\mu$TP) and direct heterogeneous bonding are two commonly adopted integration approaches. Using $\mu$TP, recent work has demonstrated $8.1\,\mathrm{dB/mm}$ small-signal gain ~\cite{ref-SOA-101,ref-SOA-102}. Through direct heterogeneous integration, Tang \textit{et al.} reported a small-signal gain of $21\,\mathrm{dB}$ and a saturation output power of $13\,\mathrm{dBm}$~\cite{ref-SOA-103}. On-chip C-band SOAs have begun transitioning from research prototypes to commercial products offered by silicon photonic foundries (such as Dream Photonics), as SOAs are increasingly regarded as critical components for AI data centers and co-packaged optics (CPO) applications.

One challenge of SOAs is their amplified spontaneous emission (ASE), which can substantially degrade the signal-to-noise ratio (SNR) of a system. In cascaded architectures, each SOA contributes additional noise, leading to cumulative noise buildup. Without effective suppression, the noise power may eventually exceed the signal power. Considerable attention has therefore been devoted to reducing the SOA noise figure~\cite{ref-SOA-104}. Slab-coupled SOAs typically exhibit a noise figure of approximately $5\,\mathrm{dB}$~\cite{ref-SOA-105}, while optimized designs with a trade-off in gain for improved noise performance have demonstrated noise figures as low as $3.7\,\mathrm{dB}$~\cite{ref-SOA-106}.
In this work, for system-level analysis, we adopt representative SOA parameters corresponding to an active length of $3\,\mathrm{mm}$, a C-band gain of $7$--$9\,\mathrm{dB/mm}$, and an insertion loss of $1\,\mathrm{dB/facet}$ for a heterogeneously bonded device.

\subsubsection{Multi-Port Photodetector}

Recently, a multi-port Ge photodetector (PD) was developed at Advanced Micro Foundry (AMF) that can be used to linearly accumulate waveguide inputs up to 16 ports~\cite{tang2025waveguide}. A schematic of the multi-port PD is shown in Fig.~\ref{fig:PTC_Arch}(c). For the photonic crossbar architecture, each port performs incoherent summation of optical power across nine wavelength channels from a single comb laser (Level 1). The outputs from multiple comb lasers are then further combined at the multi-port PDs (Level 2, Level 3, and beyond), enabling fast and efficient multi-level accumulation while significantly simplifying the overall architecture and signal routing.

The dark current of a PD scales approximately with its photosensitive area. 
As the number of ports increases, the total Ge active area and consequently the total dark current will also increase.
The reported 16-port PD has a Ge active area of $35~\mu\mathrm{m}\times13~\mu\mathrm{m}$ and exhibits a dark current ($I_d$) of 43~nA at a reverse bias of $-3$~V. 
Taking the waveguide coupling region into consideration results in an estimated total area of $\sim4000~\mu\mathrm{m}^2$.
It is worth noting that dark current is not equivalent to noise spectral density, as the DC component can be easily removed at the circuit level using a DC current filter. In photonic computing systems, quantization noise, thermal noise, shot noise, and nonlinear distortion all contribute to the degradation of the effective number of bits (ENOB). In the shot-noise-limited regime, the RMS noise current can be expressed as
\[
i_n = \sqrt{2 q (I_{ph} + I_d) B},
\]
which yields a minimum required optical power of approximately $3.17\,\mu\mathrm{W}$ for the multi-port Ge PD, corresponding to a PD sensitivity of $-25.0\,\mathrm{dBm}$. The 3 dB bandwidth for the 16-port PD is reported to be 11.8 GHz, sufficient for high speed data processing, and is expected to maintain a 6.1 GHz bandwidth when the PD is further scaled up to 250 ports~\cite{tang2025waveguide}. A Ge PD responsivity of $0.82\,\mathrm{A/W}$ is assumed in the photocurrent (I$_{ph}$) calculation. The required signal-to-noise ratio (SNR) increases considerably with ADC resolution. For a quantization-noise-limited ADC, the SNR follows
\[
\mathrm{SNR}_{dB} \approx 6.02N + 1.76,
\]
where $N$ is the bit resolution. In this work, an 8-bit ADC resolution is assumed, and a PD sensitivity of $-25.0\,\mathrm{dBm}$ (approximately $3.17\,\mu\mathrm{W}$) is adopted for system-level analysis.

\section{Evaluation Results}
In this section, we evaluate the proposed 3D WDM–PCM photonic tensor core from two complementary angles: (i) Architecture-level feasibility, focusing on whether the 3D Si/SiN compute fabric (crossbar + routing + peripherals) can be realized within a single reticle and how the layout scales with array dimension. (ii) End-to-end inference robustness under hardware-realistic low-bit quantization and relative analog noise, reflecting non-idealities in input modulation, PCM weight programming, and the opto-electronic readout chain. The goal is not to re-optimize floating-point baselines, but to validate that the proposed operating point (low-bit + noise) preserves usable task accuracy.

\subsection{Architecture-Level Performance}
\begin{table}[]
\caption{Component parameters used in \name system-level evaluation.}
\label{tab:devicetab}
\resizebox{0.8\columnwidth}{!}{
\begin{tabular}{|c|c|c|}
\hline
Device                                            & Parameter                & Value                                                                                                            \\ \hline
\multirow{2}{*}{Comb Laser~\cite{Xue_kerr_2017}}                       & Channel Power            & 10 -- 17 dBm                                                                                                           \\
                      & Repetition Rate          & 231 GHz                                                                                                          \\ \hline
\multirow{3}{*}{AWG~\cite{shang_low-loss_2017}}                              & Area                     & 0.6 mm $\times$ 1.8 mm                                                                                                  \\
                      & Loss                     & 1.5 dB                                                                                                           \\
                      & Crosstalk                & -24 dB                                                                                                           \\ \hline
\multirow{3}{*}{VOA~\cite{timurdogan2018aim}}                              & Area                     & 116 $\mu$m $\times$ 20 $\mu$m                                                                                                  \\
                      & Loss                     & 0.18 dB                                                                                                           \\
                      & Power                &  70 mW @ 6 dB attenuation                                                                                                           \\ \hline
\multirow{4}{*}{SL-MZM~\cite{begovic2024foundry}}                            & Area        & 250 $\mu$m $\times$ 25 $\mu$m                                                                                                       \\
                      & Loss                     & 3 dB (foreseeable)                                                                                                              \\
                      & Extinction ratio         & 1.17 dB                                                                                                           \\
                      & Energy per switch        & 131.6 fJ @4-bits, 119.8 fJ @6-bits, 117.1 fJ @8-bits                                                                \\ \hline
\multirow{3}{*}{WSC (\emph{proposed})}                              & Area       & 100 $\mu$m $\times$ 10 $\mu$m                                                                                                       \\
                      & Loss                     & $\sim$0.25dB                                                                                                     \\
                      & Crosstalk                & -20 dB                                                                                                  \\ \hline
\multirow{5}{*}{PCM Unit~\cite{Li2019}}                         & Energy Consumption           & 66 -- 134 pJ/Programming, 680 pJ/Erase                                                                       \\
                      & Programming Time        & 50 ns (pulse) + 200 ns (stabilization)           
                           \\
                      & Erase Time        & 250 ns (pulse) + 500 ns (stabilization)           
                           \\                                               
                      & Programming Resolution        & 5-bit (GST), 7-bit (N-GST)                                                                                                  \\
                      & Standard Deviation       & 0.005 (GST), 0.01 (N-GST) (Programming)                                                                                                \\ \hline
\multirow{3}{*}{SOA~\cite{VanGasse:19}}                              & Area                     & 1.2 mm long                                                                                                         \\
                      & Gain                     & 24.7 dB                                                                                                             \\
                      & Drive power           & 410 mW                                                                                                             \\ \hline
\multirow{5}{*}{Multi-port PD~\cite{tang2025waveguide}}                    & Area          & 40 $\mu$m $\times$ 100 $\mu$m                                                                                          \\
                      & Responsivity             & 0.82 A/W                                                                                                          \\
                      & Dark current             & 43 nA                                                                                                            \\
                      & Bandwidth                & 11.8 GHz @ -3V bias                                                                                                \\
                      & Sensitivity              &  3.17 $\mu$W (-25 dBm)
                           \\ \hline
\multirow{2}{*}{Inverse Designed Grating Coupler (\emph{proposed})} & Area        & 8 $\mu$m $\times$ 8 $\mu$m                                                                                                          \\
                      & Loss                     & 1.43 dB                                                                                                           \\ \hline
\multirow{2}{*}{1$\times$8 MMI~\cite{zhang2024tempo}}                          & Area       & 27.8 $\mu$m $\times$ 11.3 $\mu$m                                                                                                    \\
                      & Loss                     & 0.14 dB                                                                                                           \\ \hline
\multirow{2}{*}{1$\times$2 Power Splitter~\cite{zhu2021compact}}        & Area           & 80 $\mu$m $\times$ 10 $\mu$m                                                                                                    \\
                      & Loss                     & 0.02 dB                                                                                               \\ \hline
\multirow{2}{*}{Foundry waveguide crossing~\cite{NP_OFC2020_Rakowski}}        & Area           & 8 $\mu$m $\times$ 8 $\mu$m                                                                                                    \\
                      & Loss                     & 0.23 dB                                                                                               \\ \hline
\end{tabular}
}
\vspace{-5pt}
\end{table}

Based on the component summarized in Table~\ref{tab:devicetab}, we leverage the open-source photonic AI architecture modeling tool \textsc{SimPhony}~\cite{ONN_DAC2025_Gu_SimPhony} to conduct a comprehensive system-level performance evaluation on \name.

\subsubsection{Area Estimation}

\begin{figure}[ht]
  \centering
  \includegraphics[width=0.75\linewidth]{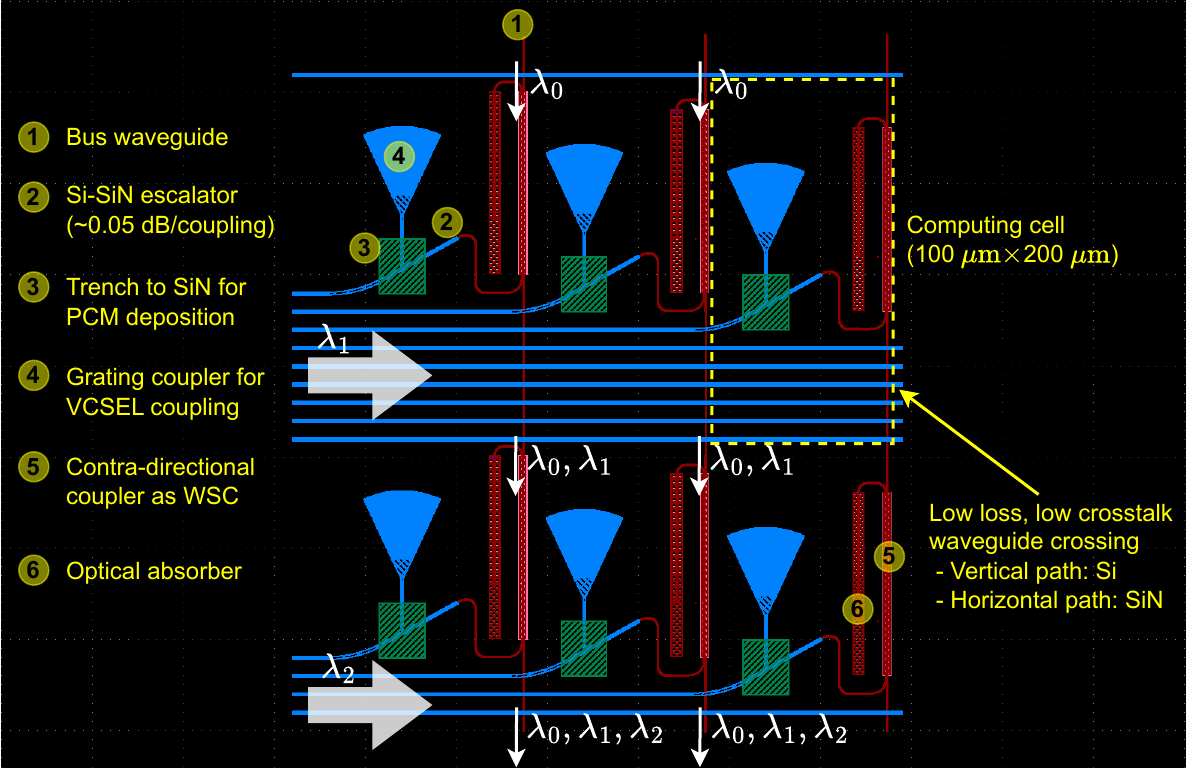}
  \vspace{-10pt}
  \caption{Layout of a 2$\times$3 computing array for illustration. Each computing unit, including PCM cells, occupies a chip area of 100~$\mu\text{m}\times$ 200~$\mu\text{m}$. Red paths represent the Si layer while blue paths represent the SiN layer.}
  \label{fig:Computing_unit}
  \Description{3-by-2 computing unit example.}
\end{figure}

\begin{figure}
    \centering
    \includegraphics[width=0.75\linewidth]{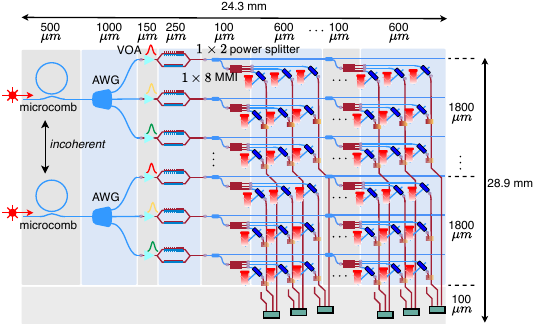}
    \caption{Detailed area estimation of \name layout.}
    \label{fig:AreaOverlay}
\end{figure}

To evaluate the area estimation for our proposed architecture, we must first ensure the $144 \times 256$ photonic tensor core array can be implemented in a single PIC die.
The maximum reticle (die) size for a PIC chip is $26 \times 33\,\text{mm}^2$ without lithography stitching.
While multi-die integration could enable even larger scales, this work targets single-die realization. 
The layout of a $3 \times 2$ crossbar computing unit is illustrated in Fig.~\ref{fig:Computing_unit}, including the horizontal SiN routing waveguides from MMIs for power distribution.
Each computing unit cell occupies an area of $100\,\mu\text{m} \times 200\,\mu\text{m}$ and by replacing the area-consuming optical absorber with a compact custom design, this area is expected to be scaled down to $\sim 75\,\mu\text{m} \times 200\,\mu\text{m}$.
Fig.~\ref{fig:AreaOverlay} illustrates the detailed area estimation including every photonic component used in \name architecture.
The horizontal dimension of 24.3 mm is divided between the fan-out components and the computing unites. 
The input routing section on the left consists of 500 $\mu$m micro combs, 1000 $\mu$m AWGs, 150 $\mu$m VOAs, and 250 $\mu$m SL-MZMs.
The remaining width of 256-column crossbar array can be divided into 32 groups, each of which includes a 1$\times$2 power splitter, a 1$\times$8 MMI and 8 columns of computing units, summed up to 700 $\mu$m.
The vertical direction, spanning a total of 28.9 mm, is primarily determined by the row-wise scaling of the computing array, with an additional 100 $\mu$m at the bottom distributed for multi-port PDs. 
Hence, the 144 $\times$ 256 computing array yields a total crossbar area of $24.3 \times 28.9\,\text{mm}^2$. 
The remaining $\sim 155\,\text{mm}^2$ of die area will be dedicated to the routing waveguides and alignment structures.

\subsubsection{Power Budget}

The power consumption of a photonic tensor core is largely dictated by the optical link budget (which determines the required laser power) and the mixed-signal peripheral overheads (DAC/ADC, modulator drivers, and weight programming). 
We use \textsc{SimPhony}~\cite{ONN_DAC2025_Gu_SimPhony} to model the end-to-end power of a 144$\times$256 \name core. 
The total power includes: (i) the inference comb laser, (ii) input
DACs and SL-MZMs for activation modulation, (iii) VOAs for comb-line equalization, (iv) multi-port PD + TIA and output ADCs for readout/digitization, and (v) DAC-weight + VCSEL energy for PCM programming/erase when weights are reconfigured.

\noindent\textbf{Critical Path Insertion Loss}.~
We first compute the worst-case (critical-path) optical insertion loss $IL$ from the comb laser output (after VOA equalization) to the PD input, including the unavoidable fanout splitting loss across the core width. 
For \name, the critical path corresponds to the farthest (last) column, which traverses the maximum number of cascaded splitters in the row-wise distribution network. 
Using Table~\ref{tab:devicetab} component losses, we estimate:
\begin{equation}
\small
\begin{aligned}
IL \;=\;&
IL_{AWG}
+ 5 \cdot IL_{Escalator}
+ IL_{MMI}
+ IL_{SL-MZM} \\
&+ IL_{1 \times 2~splitter} \cdot \left(\frac{W}{8}-1\right)
+ 8 \cdot IL_{WSC}
+ IL_{PCM}
+ IL_{VOA}
+ 10\log_{10}(W),
\end{aligned}
\label{eq:ILcrit}
\end{equation}
where $W$ is the core width (number of columns).
The first four terms correspond to AWG, five Si/SiN escalators, the $1\times 8$ MMI, and the SL-MZM insertion loss. 
The $IL_{1 \times 2~splitter}\cdot(\frac{W}{8}-1)$ term captures the \emph{excess} loss of cascaded $1\times 2$ power splitters used to expand each of the 8 coarse branches to $W/8$ columns (critical path at the far end passes the most splitters). 
The $8 \cdot IL_{WSC}$ term accounts for the 8 WSCs along the 9-wavelength hierarchical
accumulation route, followed by the PCM unit loss and VOA loss. 
Finally, $10\log_{10}(W)$ is the \emph{ideal} fanout loss from splitting power across $W$ columns. 
For $W{=}256$, Eq.~(\ref{eq:ILcrit}) yields $IL\approx 30$\,dB.

\noindent\textbf{Inference Laser Power}.~
Given the critical-path insertion loss, we compute the required electrical laser power from the PD sensitivity and the input modulation precision. 
Let $S$ denote the PD sensitivity in dBm (Table~\ref{tab:devicetab} reports $S{=}{-}25$\,dBm), and $ER$ denote the SL-MZM extinction ratio in dB (Table~\ref{tab:devicetab} reports $ER{=}1.17$\,dB). 
The minimum optical power at the PD required for reliable detection is $10^{S/10}$ (in mW). Propagating this requirement backward through the optical path gives the required launched optical power $10^{(S+IL)/10}$. 
To preserve $b_{out}$-bit output resolution under intensity modulation, the full-scale optical swing must scale with the number of quantization levels, hence the $2^{b_{out}}$ factor. 
Moreover, a finite extinction ratio reduces the usable modulation depth: if the ``off'' leakage is $10^{-ER/10}$ of the ``on'' level, only a fraction $(1-10^{-ER/10})$ contributes to signal swing; therefore we compensate by $\frac{1}{1-10^{-ER/10}}$.
Finally, we convert optical power to electrical power using the laser wall-plug efficiency $\eta_{WPE}$ (we assume $\eta_{WPE}{=}20\%$ for the comb laser). 
This yields:
\begin{equation}
\label{eq:LaserPower}
   \small
    P_{laser} = \frac{10^{(S + IL) / 10} \cdot 2^{b_{out}}}{\eta_{WPE}} \cdot \frac{1}{1 - 0.1^{ER / 10}},
\end{equation}

\noindent\textbf{DAC/ADC Power Models}.~
We model the power of input DACs and output ADCs using a standard resolution--rate scaling law: converter power grows approximately linearly with sampling rate $f$ and exponentially with resolution due to capacitor/quantization
noise scaling. 
We use:
\begin{equation}
    \small
    P_{DAC}(b_{in},f)=P_{0,DAC}\frac{2^{b_{in}}}{b_{in}+1}f, 
\end{equation}

\begin{equation}
    \small
    P_{ADC}(b_{out},f)=P_{0,ADC}\frac{2^{b_{out}}}{b_{out}+1}f,
\end{equation}
where $P_{0,DAC}$ and $P_{0,ADC}$ are technology-dependent coefficients calibrated to representative high-speed mixed-signal designs, and the $\frac{2^b}{b+1}$ term captures the dominant exponential dependence on bit resolution (with a mild normalization for small $b$). 
In our evaluation, $b_{in}$ and $b_{out}$ follow the hardware-aware precision
setting (INT6 inputs and INT8 outputs in Sec.~\ref{sec:NoiseModel}), and $f$ is the core operating rate used in the system-level model.

\noindent\textbf{VCSEL Power for PCM Programming.}
We use the reported per-event optical PCM programming/erase energy, $E_{\mathrm{PCM,opt}}$ (Table~\ref{tab:devicetab}), and convert it to \emph{electrical} VCSEL energy by accounting for the vertical grating-coupler loss ($L_{GC}=1.43$~dB, Table~\ref{tab:devicetab}) and the VCSEL electrical-to-optical efficiency $\eta_{\mathrm{VCSEL}}$ (54.8\%):
\begin{equation}
\small
E_{\mathrm{VCSEL,elec}} = E_{\mathrm{PCM,opt}} \cdot 10^{L_{GC}/10} \big/ \eta_{\mathrm{VCSEL}}.
\end{equation}
Because PCM weights are non-volatile, this energy is incurred only during weight reprogramming and introduces no static hold power during inference (in contrast to volatile weighting units).

\begin{figure}
    \centering
    \includegraphics[width=0.6\linewidth]{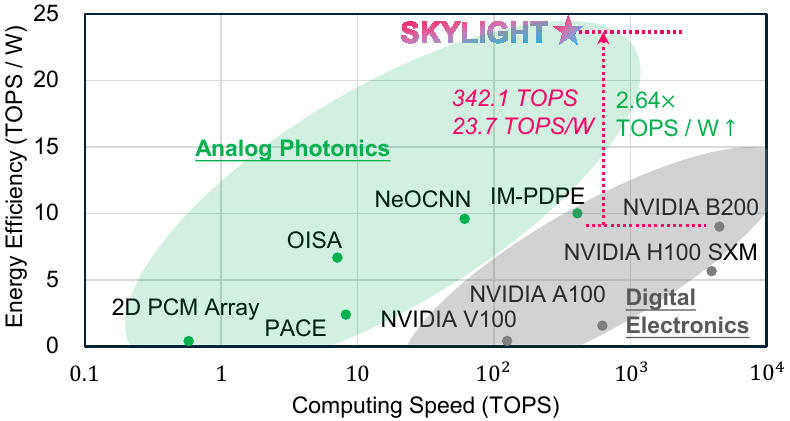}
    \caption{Our proposed \name advances the Pareto frontier in energy efficiency versus computing throughput, outperforming prior photonic accelerators~\cite{Feldmann2021,Zhou2023,Hua2025,10.1145/3650200.3656609,a557a39c698546f1902a12c137b1913f} and state-of-the-art GPUs~\cite{nvidiaV100Datasheet2020,nvidiaA100Datasheet2022,nvidiaH100Datasheet2023,nvidiaHGXB200PCFSummary2025}.}
    \label{fig:ParetoFront}
\end{figure}

\noindent\textbf{System-level Performance Comparison to Other Photonic/Electronic Accelerators}.~
Using the above link- and power-budget models, Fig.~\ref{fig:ParetoFront} places \name at 342.1 TOPS and 23.7 TOPS/W, advancing the throughput-efficiency Pareto frontier relative to prior photonic accelerators and state-of-the-art GPUs.

\subsection{Ablation Study on Proposed Techniques}
\begin{figure}
    \centering
    \includegraphics[width=0.95\linewidth]{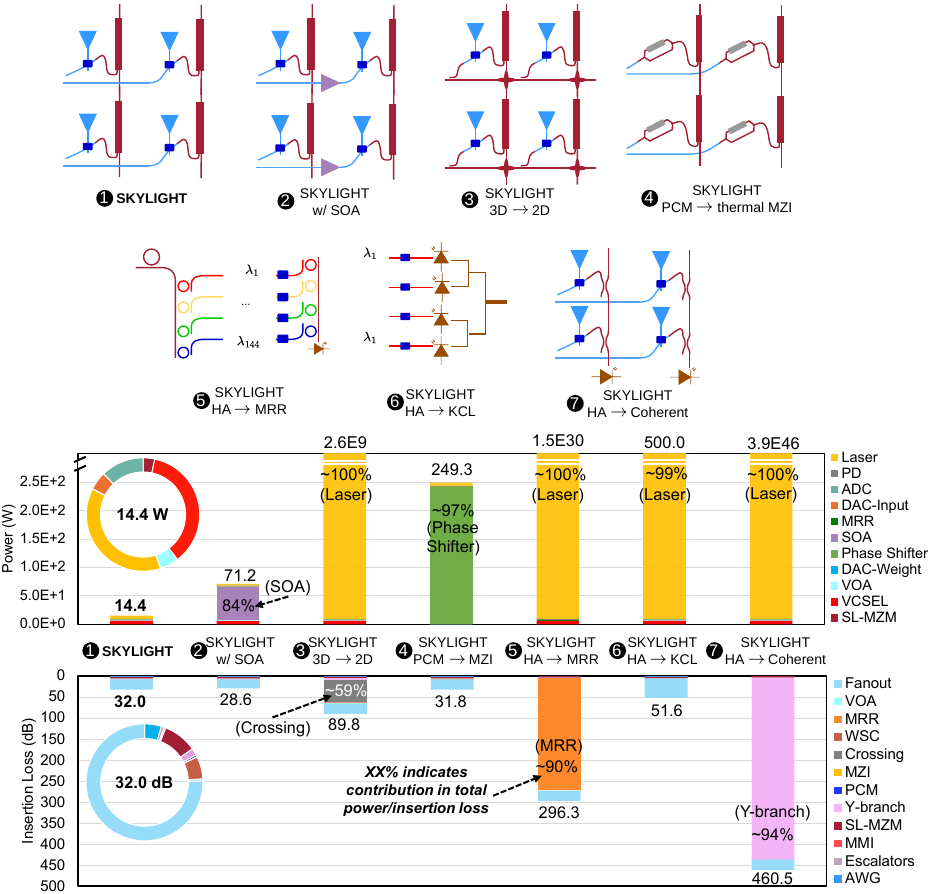}
    \caption{Compare different \name variants to validate the importance of our proposed techniques to scalability, i.e., 3D crossbar topology, non-volatile PCM weighting, and hierarchical accumulation (HA). 
    In the top graph comparing power consumption, the percentage indicated within each color-coded bar represents the power contribution of the corresponding component. For example, in configuration \ding{203}, the SOA (shown in purple) accounts for 84\% of the total power consumption. }
    \label{fig:ArchVariant}
\end{figure}

We conduct a thorough ablation study to quantify how each proposed technique affects (i) end-to-end optical link insertion loss (IL) and (ii) system-level power. 
All variants are evaluated for the same 144$\times$256 tensor core under the component parameters in Table~\ref{tab:devicetab}, and the results are summarized in Fig.~\ref{fig:ArchVariant} (top: power breakdown; bottom: IL breakdown).

\subsubsection{Impact of On-chip Amplification using SOAs (\ding{203})}
Fanout-induced attenuation is a primary scalability limiter in large photonic crossbars: as the number of columns increases, the per-row optical power must be split across more destinations, and the required laser power grows rapidly to maintain the target bit resolution at PD. 
To examine whether on-chip amplification improves scalability, we evaluate a variant that inserts SOAs after 128 columns to restore the optical power budget for the remaining fanout. 
In this configuration, the comb laser effectively supports only a 128-column fanout prior to amplification.
As shown in Fig.~\ref{fig:ArchVariant} (\ding{203}), SOA insertion reduces the end-to-end insertion loss from 32~dB (baseline \name) to 28.6~dB, alleviating laser power requirements. 
However, this benefit comes at a substantial system-level energy cost: total power increases from 14.4~W to 70.6~W, with SOA drive power accounting for approximately 83\% of the total budget.
This study highlights a key design trade-off. 
While optical amplification can relax laser power constraints, it shifts the scaling burden to electrical drive energy and amplifier noise, which, unfortunately, will be inevitable for coherent tensor cores to scale up.
In contrast, \name avoids reliance on amplification by leveraging multiple incoherent comb sources to support 256-column fanout. 
Our incoherent PTC design with only 9 wavelengths per comb source preserves scalability without incurring the dominant energy and noise penalties associated with SOA-based scaling.

\subsubsection{3D vs. 2D Crossbar Topology (\ding{204})}
We compare our crossing-free low-loss 3D Si/SiN topology against a 2D planar topology~\cite{Dong2023PhotonicTensorCoreContinuousTime, NP_Nature2021_Feldmann}, using a serially cascaded Y-branch for column fanout. 
In such a 2D layout, the dot-product/accumulation path inevitably incurs a \emph{linear} number of cascaded waveguide crossings and Y-branches as the core width grows. 
For a 144$\times$256-scale core, the 2D topology introduces approximately one crossing and one y-branch per column along the critical path, plus additional crossings in the vertical accumulation portion (in our case, core width + 8 crossings).
Given the foundry-level crossing loss (0.23 dB per crossing, Table~\ref{tab:devicetab}), crossings alone contribute \textasciitilde53 dB loss at this scale.
As a result, the planar 2D variant (\ding{204} in Fig.~\ref{fig:ArchVariant}) exhibits an end-to-end insertion loss of 89.8~dB, which drives the required optical launch power to an unrealistic $2.6\times10^{9}$~W and renders large-core operation infeasible.
In contrast, the proposed 3D topology separates the row and column routing onto SiN and Si layers, respectively, and uses low-loss inter-layer transitions to avoid the $\mathcal{O}(\mathrm{core\ size})$ cascaded crossing penalty on the critical path. 
\textbf{The shift from crossing-limited planar scaling to crossing-free 3D topology} is a necessary condition for realizing hundreds-scale photonic tensor cores within practical power budgets.

\subsubsection{PCM vs. Thermo-optic MZI as Weight Unit (\ding{205})}
Non-volatile weighting is a central requirement for scaling photonic in-memory tensor cores while maintaining the energy advantage. 
To emphasize its significance, we replace each PCM weight with a thermo-optic MZI weight unit, whose state must be maintained by continuous heater bias.
As shown in Fig.~\ref{fig:ArchVariant} (\ding{205}), the optical insertion loss is essentially unchanged.
The dominant difference is \emph{system power}: the MZI variant incurs a large static ``weight-holding'' cost that scales with the number of weights, increasing total power to 248.9~W, with thermo-optic phase shifters contributing $\sim$97\% of the budget. 
In addition, thermal phase tuning limits weight update bandwidth (tens of $\mu$s timescale).
In contrast, non-volatile PCM weights incur energy primarily during reprogramming and require near-zero static power during inference. 
This ablation demonstrates that non-volatility is critical for energy efficiency; otherwise, static weight maintenance dominates the power budget and negates the efficiency gains of large-scale photonic cores.

\subsubsection{Hierarchical Accumulation (HA) vs. Existing Accumulation Methods (\ding{206}/\ding{207}/\ding{208})}

Scaling partial product accumulation is a primary architectural bottleneck in large photonic tensor cores: purely WDM-based summation pushes wavelength count (and routing complexity) to grow rapidly, whereas purely electrical fan-in pushes detector count and delay overhead to grow with array size. 
\name{} addresses this by introducing \emph{hierarchical accumulation} (HA), a bounded-degree reduction network that combines (i) a small, fixed WDM degree per group (e.g., 9 wavelengths), (ii) multi-port photodetectors for optical aggregation, and (iii) current-domain summation (KCL) \emph{before} digitization. 
This design keeps the wavelength usage bounded while enabling optical-domain partial product summation without requiring a dedicated PD at every dot-product site and thus avoiding quadratically many PDs with accumulated PD noises.

We compare our hierarchical accumulation strategy with existing approaches for partial product accumulation.

\noindent\textbf{MRR-based Multi-wavelength Accumulation \ding{206}}.~
A direct WDM accumulation approach~\cite{NP_SciRep2017_Tait,Ohno2022MRRCrossbarACSPhotonics,Ning2024HardwareEfficientEPIC,Luan2023HighDensityPhotonicTensorCore,Miscuglio2020PhotonicTensorCores} would require a wavelength count proportional to the core height (144 wavelengths in our case) and a large number of microrings to couple wavelengths onto an accumulation bus waveguide (=2$\times$144 rings).
This approach is fundamentally limited by active tuning/locking across a large ring bank, while cumulative resonant insertion loss compounds along the accumulation path. 
Consistent with this scaling behavior, the MRR-based baseline (\ding{206}) becomes infeasible at this scale, exhibiting 296.3~dB end-to-end insertion loss and driving the required power to $1.5\times10^{30}$~W.

\noindent\textbf{Pure KCL-based Photocurrent Summation \ding{207}}.~
In a purely electrical accumulation baseline, each dot-product output must be detected locally at a PD before any summation can occur in the current domain~\cite{ONN_ICCAD2024_Gu}. 
This eliminates optical-domain partial-sum accumulation and, consequently, voids the key scaling benefit of accumulation \emph{before} detection: the optical distribution network must still deliver sufficient power to \emph{every} PD to satisfy sensitivity and resolution constraints. 
As the array scales, this requirement drives a substantially tighter link budget and rapidly increases the required laser power.
In Fig.~\ref{fig:ArchVariant} (variant \ding{207}), IL increases to 51.6 dB, and system power rises to 500 W, dominated (\textasciitilde99\%) by the laser.
These results indicate that PD-per-site detection shifts the scaling bottleneck from accumulation to optical power delivery, making large-core operation laser-limited.

\noindent\textbf{Coherent Light Combining \ding{208}}.~
A coherent-combining baseline forms partial sums by interfering phase-aligned optical fields through a combiner/coupler chain~\cite{NP_Nature2021_Feldmann,Totovic2022WDMUniversalLinearOptics,Dong2023PhotonicTensorCoreContinuousTime}. 
First, the combiner network introduces a deep cascade of combiners/couplers on the critical path, compounding insertion loss as the fan-in grows.
Second, coherent addition requires tight phase alignment across many paths; fabrication and thermal/polarization variations introduce phase errors that translate directly into imperfect interference, adding combining loss on top of the passive network loss (and, in the worst case, approaching a $\sim$3~dB penalty per combining stage).
Figure~\ref{fig:ArchVariant} (\ding{208}) shows that coherent combining becomes infeasible for large cores.
These results highlight that \textbf{coherent summation is fundamentally brittle at scale, simultaneously limited by phase stability and by the compounded loss} of deep combiner networks, which is resolved by \name’s incoherent, hierarchical accumulation strategy.

\subsubsection{Benefit of Large Core Size in Inference Energy and Throughput}
\begin{figure}
    \centering
    \includegraphics[width=0.7\linewidth]{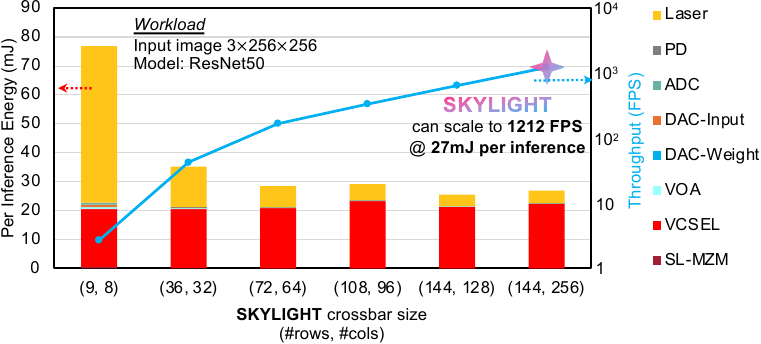}
    \caption{Energy and throughput scaling with increasing \name core size. 
    Energy consumption is for a single ImageNet image inference through ResNet50, simulated by \textsc{SimPhony}~\cite{ONN_DAC2025_Gu_SimPhony}. 
    The large core size enabled by our \name design can support 1212 FPS at 27 mJ per image inference in a single core.
    }
    \label{fig:CoreSizeSweep}
\end{figure}

We next quantify the system-level value of \name’s key capability: scaling a \emph{single} photonic in-memory tensor core to hundred-scale channels. 
We sweep core size and evaluate end-to-end inference throughput and energy for a representative workload (ResNet-50 with $3\times256\times256$ input), as shown in Fig.~\ref{fig:CoreSizeSweep}.

Scaling the core size directly translates into real-time performance. 
As the core grows, more computations are executed at one shot within light propagation delay, sharply reducing time-multiplexing and driving throughput upward. 
At the full $144\times256$ core enabled by \name, throughput reaches 1212~frames-per-second (FPS). 
At the same time, energy per inference drops because larger cores complete the same inference task in shorter time frames: many overheads contribute approximately constant \emph{power} during operation (e.g., comb generation and mixed-signal I/O), so higher throughput directly reduces runtime and thus energy (power$\times$time). 
A small $9\times8$ core consumes $\sim$80~mJ per inference, whereas the $144\times256$ core reaches $\sim$27~mJ per inference.
From an end-to-end system efficiency perspective, \name (SKYLight) achieves 84.17~FPS/W, compared to 52.27~FPS/W measured for an NVIDIA RTX PRO 6000 Blackwell GPU under the same workload in real-time measurement, making \name 1.61$\times$ more efficient.

These results underscore a central message: photonic computing’s system-level advantages emerge at large tensor core scales. 
By sustaining single-core scaling to hundreds of channels, \name converts optical parallelism into real-time inference capability (kilo-FPS) at low per-inference energy, enabling high-performance, energy-efficient perception and decision making for latency-critical and endurance-constrained deployments. 

Beyond single-core scaling, \name naturally extends to \uline{multi-core systems} to increase aggregate throughput and support larger models. 
\textbf{Multiple \name tensor cores can be potentially integrated as chiplets and composed via electronic or photonic die-to-die interconnects} for fanout and reduction operations that support high bandwidth-density and energy efficiency at scale. 
Such heterogeneous multi-chiplet integration can scale \name from a single hundred-channel core to a tiled photonic compute fabric with thousands-scale aggregate channels, supporting larger networks and higher end-to-end throughput without sacrificing the architectural benefits established by \name at the core level.

\subsection{Accuracy and Robustness Evaluation on Various Real-time Supervised/Unsupervised ML Tasks}
\begin{figure}
    \centering
    \includegraphics[width=0.95\linewidth]{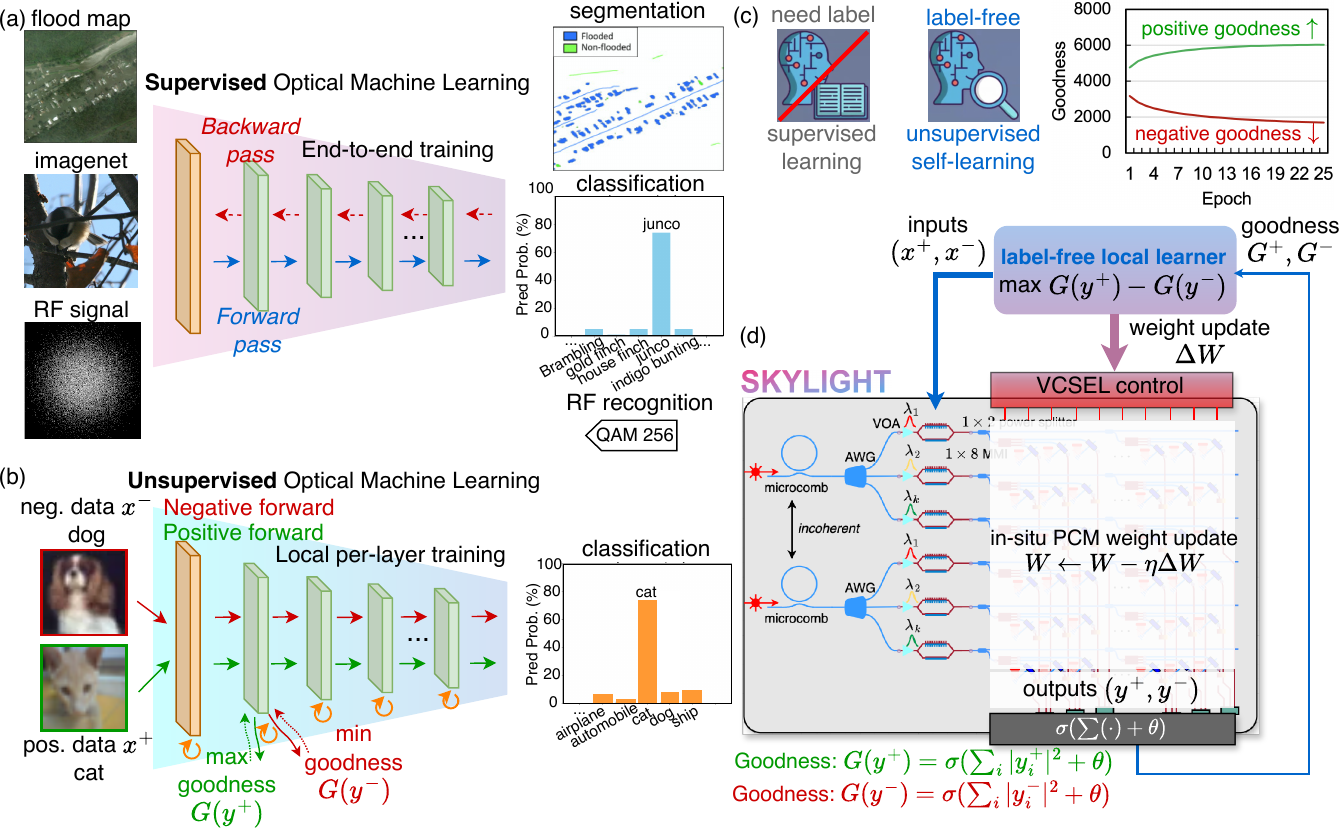}
    \vspace{-5pt}
    \caption{Evaluate our \name on (a) supervised and (b) unsupervised trained ONNs on various ONN models and vision/RF tasks.
    Our \name can support unsupervised self-learning using a label-free forward-forward PCM weight update algorithm via local layer-wise positive goodness maximization and negative goodness minimization.~\cite{hinton2022forwardforwardalgorithmpreliminaryinvestigations,Chen2025}.
    (c) Unsupervised learning curve of \name on CIFAR-10.
    (d) Illustration of local unsupervised learning architecture.}
    \label{fig:Training}
    \vspace{-10pt}
\end{figure}

To evaluate \name beyond hardware performance metrics, we study four representative workloads that reflect diverse application domains in edge autonomy, RF sensing, large-scale vision, and adaptive unsupervised learning. 

We first clarify the non-ideal hardware effects considered in our accuracy evaluation, i.e., quantization and noise.

\noindent\textbf{Hardware Non-ideality Modeling: Quantization and Noises}.~
\label{sec:NoiseModel}
We employ a noise-aware quantization workflow in which both training and inference use
low-bit quantization and signal-proportional (relative) noise injection.
For activations/inputs, we use a relative noise standard deviation
\(\sigma_{\mathrm{in}}=0.0031\), matching the experimentally measured noise level of our SL-MZM device~\cite{Zhang:24}.
For other analog non-idealities, including photodetector/readout fluctuations and potential
PCM weight variability, we assume a relative noise level of \(0.01\).

Accommodating device characteristic, we set \textbf{SL-MZM inputs} with INT6, \textbf{PCM weights} with INT7, and \textbf{PD+ADC output} with INT8.
Let \(Q_b(\cdot)\) denote uniform \(b\)-bit quantization. We inject relative Gaussian noise as
\begin{equation}
    x_{\mathrm{eq}} \;=\; Q_b(x)\;+\;\delta x, \quad \delta x\sim \mathcal{N}\!\left(0,\;(\sigma\,|Q_b(x)|)^2\right),
\end{equation}
and similarly for weights,
\begin{equation}
    w_{\mathrm{eq}} \;=\; Q_b(w)\;+\; \delta w, \quad \delta w \sim \mathcal{N}\!\left(0,\;(\sigma_w\,|Q_b(w)|)^2\right).
\end{equation}
This model captures the scaled-with-signal behavior common to analog intensity fluctuations
and device-level variability.

\begin{table}[]
\caption{Accuracy evaluation of \name across diverse learning tasks (image/RF time series classification and segmentation) and model scales under quantization and hardware noise injection. 
We report performance for noise-free training (baseline), noise-injected inference, and noise-aware training. 
Quantization/noise settings: input INT6 with Gaussian noise (std = 0.0031), weights INT7 with Gaussian noise (std = 0.01), and output INT8 with Gaussian noise (std = 0.01).}
\label{tab:compresults}
\resizebox{\textwidth}{!}{
\begin{tabular}{|c|c|c|c||c|c|c|}
\hline
\multirow{2}{*}{Tasks}                                                            & \multirow{2}{*}{Model}                                                         & \multirow{2}{*}{Method}                                                                  & \multirow{2}{*}{Eval. Metric}                              & \multicolumn{3}{c|}{Performance}                                                              \\
                                                                                  &                                                                                &                                                                                          &                                                            & \begin{tabular}[c]{@{}c@{}}Noise-free\\ (Baseline)\end{tabular} & \begin{tabular}[c]{@{}c@{}}Noise-injected\\ (Baseline)\end{tabular} & \begin{tabular}[c]{@{}c@{}}Noise-injected\\ (Noise-aware Train)\end{tabular} \\ \hline
RF class.                                                                         & Dual-channel Tiny-ResNet                                                       & Supervised                                                                               & Accuracy                                                   & 0.883                                                           & 0.841        & 0.873       \\ \hline
\begin{tabular}[c]{@{}c@{}}Large-scale vision class.\\ (ImageNet-1k)\end{tabular} & Spark + ResNet50                                                               & \begin{tabular}[c]{@{}c@{}}Self-supervised pretrain\\  + supervised finetune\end{tabular} & \begin{tabular}[c]{@{}c@{}}Accuracy\\ (Top 1)\end{tabular} & 0.791                                                           & 0.738        & 0.752       \\ \hline
\begin{tabular}[c]{@{}c@{}}Small-scale vision class.\\ (CIFAR-10)\end{tabular}    & \begin{tabular}[c]{@{}c@{}}3 layer CNN\\ C96k3 - C384k3 - C1536k3\end{tabular} & \begin{tabular}[c]{@{}c@{}}Unsupervised\\ (SCFF)\end{tabular}                            & Accuracy                                                   & 0.792                                                           & 0.763        & 0.773       \\ \hline
Flood mapping Segm.                                                               & HRNet+OCR                                                                      & Supervised                                                                               & Mean IoU                                                   & 0.515                                                           & 0.428        & 0.493       \\ \hline
\end{tabular}
}
\end{table}

Under the above quantization bitwidth and relative-noise settings, we evaluate end-to-end task accuracy of \name and quantify the benefit of noise-aware training in robustness boost across representative supervised and unsupervised workloads. 
Figure~\ref{fig:Training} illustrates the evaluated tasks and training flows and emphasizes the unsupervised local learning capability of \name with layer-wise goodness optimization. 
Table~\ref{tab:compresults} shows the results across different tasks.

\noindent\textbf{Real-time Edge Radio-frequency (RF) Signal Classification.}~
We evaluate 8-class RF signal classification on CSPB-ML-2018R2~\cite{9883741}, representative of real-time spectrum monitoring and RF modulation recognition in contested environments (BPSK, QPSK, 256QAM, FM, etc). 
Such workloads prioritize high-throughput inference with tolerance to moderate analog noise. 
We deploy a lightweight dual-channel residual architecture (DualBranchTinyResNet) operating on two-channel STFT spectrogram inputs.
Two shallow convolutional stems extract frequency-time features before fusion into residual blocks and a final classifier, preserving ResNet-style inductive bias while maintaining channel widths compatible with PCM array constraints. 
Training is fully supervised using cross-entropy loss.
Direct noise injection without robustness training causes a 3.7\% relative drop from the baseline. 
Noise-aware training successfully recovers the accuracy to 0.873, close to the noise-free reference.
Importantly, RF spectrum analytics often tolerates moderate analog inaccuracy in exchange for high-throughput, low-latency processing; the achieved accuracy demonstrates that photonic in-memory execution can satisfy such operational requirements without sacrificing energy efficiency or core scalability. 

\noindent\textbf{Large-scale Vision Inference (ImageNet-1K~\cite{NN_imagenet2009}) with Self-Supervised Pretraining}.~ 
We evaluate ImageNet-1K classification as a reference workload for high-parameter, memory- and IO-intensive inference tasks. 
This task stresses both sustained throughput and robustness in deep residual architectures. 
We deploy a ResNet50~\cite{NN_CVPR2016_He} backbone, consisting of stacked bottleneck residual convolution blocks, representative of deployment-grade vision pipelines.
To strengthen representation quality and downstream adaptability, we adopt SparK~\cite{tian2023designing} masked self-supervised pretraining prior to supervised fine-tuning. 
SparK is a masked image modeling framework tailored for CNNs that \emph{learns transferable representations} via sparse/hierarchical masking and reconstruction objectives. 
This pretraining strategy \uline{enhances generalization and reduces the amount of task-specific retraining required for downstream adaptation, making it well-aligned with hardware-constrained or on-device fine-tuning scenarios}.
The SparK-pretrained ResNet50, when mapped to \name, can achieve over 1000 FPS inference throughout while sustaining 0.752 Top-1 accuracy after noise-aware training.

\noindent\textbf{In-situ Unsupervised (Label-free) Machine Vision Learning (CIFAR-10~\cite{NN_cifar2009})}.~
We evaluate label-free learning as a powerful augmented feature for \name on latency-sensitive, connectivity-constrained deployment scenarios where ground-truth labels and cloud retraining are unavailable. 
Such settings arise in autonomous sensing platforms that must adapt to new environments, targets, or visual domains without centralized supervision.
We deploy a compact 3-layer CNN with channel configuration C96k3 – C384k3 – C1536k3, where C\#k3 denotes a $3\times3$ convolution with \# output channels (followed by standard nonlinearity/normalization). 
This shallow, high-channel regime maps efficiently onto photonic matrix-vector primitives and stresses local feature extraction under constrained depth.
As shown in Fig.~\ref{fig:Training}(b-d), training is performed using self-contrastive forward-forward (SCFF)~\cite{Chen2025}, a forward-only, \uline{layer-wise local learning algorithm that eliminates the use of labeled data or global backpropagation}. 
SCFF trains each layer in \name independently using a self-contrastive “goodness” objective that separates positive and negative input examples, enabling representation learning without labels or global error signals. 
This learning rule aligns naturally with photonic in-memory execution on edge, as it avoids high-bandwidth weight-gradient communication and large activation storage.
The unsupervised model achieves 0.773 accuracy with noise-aware training (vs.\ 0.783 noise-free reference), demonstrating that local, forward-only adaptation remains robust under realistic non-idealities in \name.
For \uline{edge AI environments where data distributions shift and supervisory signals are scarce, such local unsupervised learning enables continual refinement without costly backpropagation, centralized retraining, or persistent connectivity}.

\noindent\textbf{Segmentation of Remote-Sensing Flood Mapping from Pre-/post-event Satellite Imagery~\cite{9883741}}.~
We evaluate a high-resolution, decision-relevant remote-sensing task that is directly aligned with situational awareness: mapping flooded regions from paired pre- and post-event satellite images. 
This workload requires dense, pixel-level segmentation and change attribution (e.g., identifying inundated areas while preserving critical structure such as roads and buildings), thereby stressing multi-stage perception pipelines, spatial reasoning, and post-processing under large image resolutions.
We adopt an HRNet~\cite{9052469} backbone with Optical Character Recognition (OCR)~\cite{10.1007/978-3-030-58539-6_11} context refinement for dense prediction on the SpaceNet-8 flood dataset~\cite{9883741}. 
HRNet maintains high-resolution representations throughout the network via parallel multi-resolution streams with repeated fusion, which is essential for boundary-sensitive outputs. 
OCR further augments pixel features with object-/region-level context, improving segmentation consistency by coupling per-pixel predictions to scene-level structure. 
For flood mapping, \name with noise-aware training can realize mIoU of 0.493 under hardware noises.

\section{Conclusion}

In this work, we introduce \name{}, a scalable 3D photonic in-memory tensor core architecture that breaks the long-standing loss and power barriers that limit existing photonic processors to small arrays. 
\name is built on four tightly integrated innovations: (i) a thermally robust, non-resonant WDM datapath that eliminates microring sensitivity and locking overhead, (ii) a crossing-free 3D Si/SiN crossbar fabric that suppresses loss accumulation and routing congestion at scale, (iii) heterogeneously-integrated optically-programmed PCM weight banks that provide low-power in-memory computation, and (iv) hierarchical accumulation that composes spatial and spectral parallelism to support hundred-channel summation. 
Together, these core contributions realize large-scale convolution/MVM with superior throughput and energy efficiency suited for real-time AI inference.
\name establishes a cross-layer scaling paradigm for PTC design: stability, efficiency, and programmability are ensured at the physical layer, while loss- and SNR-scalable computation are enabled by the 3D crossbar topology and hierarchical accumulation. 
By aligning device physics with system structure, \name enables robust, hundred-channel photonic inference fabrics delivering over 342 TOPS computing capability per tensor core at 14.4 W with unsupervised local learning ability, fundamentally beyond prior 2D planar, resonance-locked, and non-hierarchical approaches.

\begin{acks}
The authors would like to thank the SMART fellowship program, the Naval Engineering Education Consortium at NSWC Crane Division, the New York State Focus Center for their partial support, and Rensselaer Polytechnic Institute  for their internal support.  
\end{acks}

\bibliographystyle{ACM-Reference-Format}
\bibliography{./ref/Top, ./ref/Top_sim, ./ref/NN, ./ref/NP, ./ref/PD, ./ref/rena, ./ref/inverse_design_gc, ./ref/PCM_Material}


\begin{thebibliography}{103}


\ifx \showCODEN    \undefined \def \showCODEN     #1{\unskip}     \fi
\ifx \showISBNx    \undefined \def \showISBNx     #1{\unskip}     \fi
\ifx \showISBNxiii \undefined \def \showISBNxiii  #1{\unskip}     \fi
\ifx \showISSN     \undefined \def \showISSN      #1{\unskip}     \fi
\ifx \showLCCN     \undefined \def \showLCCN      #1{\unskip}     \fi
\ifx \shownote     \undefined \def \shownote      #1{#1}          \fi
\ifx \showarticletitle \undefined \def \showarticletitle #1{#1}   \fi
\ifx \showURL      \undefined \def \showURL       {\relax}        \fi
\providecommand\bibfield[2]{#2}
\providecommand\bibinfo[2]{#2}
\providecommand\natexlab[1]{#1}
\providecommand\showeprint[2][]{arXiv:#2}

\bibitem[Anderson et~al\mbox{.}(2023)]%
        {begova_mzm_2023}
\bibfield{author}{\bibinfo{person}{Stephen~R. Anderson}, \bibinfo{person}{Amir Begović}, \bibinfo{person}{Hao Jiang}, {and} \bibinfo{person}{Z.~Rena Huang}.} \bibinfo{year}{2023}\natexlab{}.
\newblock \showarticletitle{Compact Slow-Light Integrated Silicon Electro-Optic Modulators With Low Driving Voltage}.
\newblock \bibinfo{journal}{\emph{IEEE Photonics Technology Letters}} \bibinfo{volume}{35}, \bibinfo{number}{13} (\bibinfo{year}{2023}), \bibinfo{pages}{697--700}.
\newblock
\href{https://doi.org/10.1109/LPT.2023.3273178}{doi:\nolinkurl{10.1109/LPT.2023.3273178}}


\bibitem[Aryana et~al\mbox{.}(2023)]%
        {Aryana2023}
\bibfield{author}{\bibinfo{person}{K. Aryana}, \bibinfo{person}{H.~J. Kim}, \bibinfo{person}{Md.~R. Islam}, \bibinfo{person}{N. Hong}, \bibinfo{person}{C.-C. Popescu}, \bibinfo{person}{S. Makarem}, \bibinfo{person}{T. Gu}, \bibinfo{person}{J. Hu}, {and} \bibinfo{person}{P.~E. Hopkins}.} \bibinfo{year}{2023}\natexlab{}.
\newblock \showarticletitle{Optical and Thermal Properties of {Ge$_2$Sb$_2$Te$_5$}, {Sb$_2$Se$_3$}, and {Sb$_2$S$_3$} for Reconfigurable Photonic Devices [Invited]}.
\newblock \bibinfo{journal}{\emph{Optical Materials Express}} \bibinfo{volume}{13}, \bibinfo{number}{11} (\bibinfo{year}{2023}), \bibinfo{pages}{3277--3291}.
\newblock
\href{https://doi.org/10.1364/OME.503178}{doi:\nolinkurl{10.1364/OME.503178}}


\bibitem[Bedeschi et~al\mbox{.}(2009)]%
        {Bedeschi2009}
\bibfield{author}{\bibinfo{person}{F. Bedeschi}, \bibinfo{person}{R. Fackenthal}, \bibinfo{person}{C. Resta}, \bibinfo{person}{E.~M. Donze}, \bibinfo{person}{M. Jagasivamani}, \bibinfo{person}{E.~C. Buda}, \bibinfo{person}{F. Pellizzer}, \bibinfo{person}{D.~W. Chow}, \bibinfo{person}{A. Cabrini}, \bibinfo{person}{G.~M.~A. Calvi}, \bibinfo{person}{R. Faravelli}, \bibinfo{person}{A. Fantini}, \bibinfo{person}{G. Torelli}, \bibinfo{person}{D. Mills}, \bibinfo{person}{R. Gastaldi}, {and} \bibinfo{person}{G. Casagrande}.} \bibinfo{year}{2009}\natexlab{}.
\newblock \showarticletitle{A Bipolar-Selected Phase Change Memory Featuring Multi-Level Cell Storage}.
\newblock \bibinfo{journal}{\emph{IEEE Journal of Solid-State Circuits}} \bibinfo{volume}{44}, \bibinfo{number}{1} (\bibinfo{year}{2009}), \bibinfo{pages}{217--227}.
\newblock
\href{https://doi.org/10.1109/JSSC.2008.2006439}{doi:\nolinkurl{10.1109/JSSC.2008.2006439}}


\bibitem[Begovi{\'c} et~al\mbox{.}(2024a)]%
        {begovic2024manufacture}
\bibfield{author}{\bibinfo{person}{Amir Begovi{\'c}}, \bibinfo{person}{Amar Maksumi{\'c}}, \bibinfo{person}{Alexander Chen}, \bibinfo{person}{Nicholas~M Fahrenkopf}, \bibinfo{person}{Christopher Baiocco}, {and} \bibinfo{person}{Z~Rena Huang}.} \bibinfo{year}{2024}\natexlab{a}.
\newblock \showarticletitle{Manufacture reliability assessment of Si photonic foundry fabricated slow-light photonic crystal waveguides}.
\newblock \bibinfo{journal}{\emph{Applied Optics}} \bibinfo{volume}{63}, \bibinfo{number}{12} (\bibinfo{year}{2024}), \bibinfo{pages}{3359--3365}.
\newblock


\bibitem[Begovi{\'c} et~al\mbox{.}(2024b)]%
        {begovic2024foundry}
\bibfield{author}{\bibinfo{person}{Amir Begovi{\'c}}, \bibinfo{person}{Meng Zhang}, \bibinfo{person}{Dennis Yin}, \bibinfo{person}{Nicholas Gangi}, \bibinfo{person}{Jiaqi Gu}, {and} \bibinfo{person}{Z Rena~Huang}.} \bibinfo{year}{2024}\natexlab{b}.
\newblock \showarticletitle{Foundry fabricated compact slow-light Mach-Zehnder modulator and photodetector for on-chip analog photonic computing}.
\newblock \bibinfo{journal}{\emph{Optics Express}} \bibinfo{volume}{32}, \bibinfo{number}{23} (\bibinfo{year}{2024}), \bibinfo{pages}{42016--42030}.
\newblock


\bibitem[Chen and Yu(2004)]%
        {ref-VCSEL-411}
\bibfield{author}{\bibinfo{person}{N.S. Chen} {and} \bibinfo{person}{S.F. Yu}.} \bibinfo{year}{2004}\natexlab{}.
\newblock \showarticletitle{Transient response of ARROW VCSELs under external optical feedback}.
\newblock \bibinfo{journal}{\emph{IEEE Photonics Technology Letters}} \bibinfo{volume}{16}, \bibinfo{number}{7} (\bibinfo{year}{2004}), \bibinfo{pages}{1610--1612}.
\newblock
\href{https://doi.org/10.1109/LPT.2004.827966}{doi:\nolinkurl{10.1109/LPT.2004.827966}}


\bibitem[Chen et~al\mbox{.}(2025)]%
        {Chen2025}
\bibfield{author}{\bibinfo{person}{Xing Chen}, \bibinfo{person}{Dongshu Liu}, \bibinfo{person}{J{\'e}r{\'e}mie Laydevant}, {and} \bibinfo{person}{Julie Grollier}.} \bibinfo{year}{2025}\natexlab{}.
\newblock \showarticletitle{Self-Contrastive Forward-Forward algorithm}.
\newblock \bibinfo{journal}{\emph{Nature Communications}} \bibinfo{volume}{16}, \bibinfo{number}{1} (\bibinfo{date}{01 Jul} \bibinfo{year}{2025}), \bibinfo{pages}{5978}.
\newblock
\showISSN{2041-1723}
\href{https://doi.org/10.1038/s41467-025-61037-0}{doi:\nolinkurl{10.1038/s41467-025-61037-0}}


\bibitem[Cristiani et~al\mbox{.}(2022)]%
        {Cristiani2022RoadmapMultimode}
\bibfield{author}{\bibinfo{person}{Ilaria Cristiani} {et~al\mbox{.}}} \bibinfo{year}{2022}\natexlab{}.
\newblock \showarticletitle{Roadmap on multimode photonics}.
\newblock \bibinfo{journal}{\emph{Journal of Optics}} \bibinfo{volume}{24}, \bibinfo{number}{8} (\bibinfo{year}{2022}), \bibinfo{pages}{083001}.
\newblock
\href{https://doi.org/10.1088/2040-8986/ac7a48}{doi:\nolinkurl{10.1088/2040-8986/ac7a48}}


\bibitem[Davenport et~al\mbox{.}(2018)]%
        {Davenport_mode_locked_2018}
\bibfield{author}{\bibinfo{person}{Michael~L. Davenport}, \bibinfo{person}{Songtao Liu}, {and} \bibinfo{person}{John~E. Bowers}.} \bibinfo{year}{2018}\natexlab{}.
\newblock \showarticletitle{Integrated heterogeneous silicon/{III}\&{V} mode-locked lasers}.
\newblock \bibinfo{journal}{\emph{Photon. Res.}} \bibinfo{volume}{6}, \bibinfo{number}{5} (\bibinfo{date}{May} \bibinfo{year}{2018}), \bibinfo{pages}{468--478}.
\newblock
\href{https://doi.org/10.1364/PRJ.6.000468}{doi:\nolinkurl{10.1364/PRJ.6.000468}}


\bibitem[Delaney et~al\mbox{.}(2020)]%
        {Delaney2020}
\bibfield{author}{\bibinfo{person}{Mark Delaney}, \bibinfo{person}{Ioannis Zeimpekis}, \bibinfo{person}{Daniel Lawson}, \bibinfo{person}{David~W. Hewak}, {and} \bibinfo{person}{Otto~L. Muskens}.} \bibinfo{year}{2020}\natexlab{}.
\newblock \showarticletitle{A New Family of Ultralow Loss Reversible Phase‐Change Materials for Photonic Integrated Circuits: Sb$_2$S$_3$ and Sb$_2$Se$_3$}.
\newblock \bibinfo{journal}{\emph{Advanced Functional Materials}} \bibinfo{volume}{30}, \bibinfo{number}{36} (\bibinfo{year}{2020}), \bibinfo{pages}{2002447}.
\newblock
\href{https://doi.org/10.1002/adfm.202002447}{doi:\nolinkurl{10.1002/adfm.202002447}}


\bibitem[Demirtzioglou et~al\mbox{.}(2018)]%
        {Demirtzioglou_ring_2018}
\bibfield{author}{\bibinfo{person}{Iosif Demirtzioglou}, \bibinfo{person}{Cosimo Lacava}, \bibinfo{person}{Kyle R.~H. Bottrill}, \bibinfo{person}{David J.Thomson}, \bibinfo{person}{Graham~T. Reed}, \bibinfo{person}{David~J. Richardson}, {and} \bibinfo{person}{Periklis Petropoulos}.} \bibinfo{year}{2018}\natexlab{}.
\newblock \showarticletitle{Frequency comb generation in a silicon ring resonator modulator}.
\newblock \bibinfo{journal}{\emph{Opt. Express}} \bibinfo{volume}{26}, \bibinfo{number}{2} (\bibinfo{date}{Jan} \bibinfo{year}{2018}), \bibinfo{pages}{790--796}.
\newblock
\href{https://doi.org/10.1364/OE.26.000790}{doi:\nolinkurl{10.1364/OE.26.000790}}


\bibitem[Deng et~al\mbox{.}(2009)]%
        {NN_imagenet2009}
\bibfield{author}{\bibinfo{person}{Jia Deng}, \bibinfo{person}{Wei Dong}, \bibinfo{person}{Richard Socher}, \bibinfo{person}{Li-Jia Li}, \bibinfo{person}{Kai Li}, {and} \bibinfo{person}{Li Fei-Fei}.} \bibinfo{year}{2009}\natexlab{}.
\newblock \showarticletitle{ImageNet: A large-scale hierarchical image database}. In \bibinfo{booktitle}{\emph{Proc.~CVPR}}. \bibinfo{pages}{248--255}.
\newblock


\bibitem[Dong et~al\mbox{.}(2023)]%
        {Dong2023PhotonicTensorCoreContinuousTime}
\bibfield{author}{\bibinfo{person}{Bowei Dong}, \bibinfo{person}{Samarth Aggarwal}, \bibinfo{person}{Wen Zhou}, \bibinfo{person}{Utku~Emre Ali}, \bibinfo{person}{Nikolaos Farmakidis}, \bibinfo{person}{June~Sang Lee}, \bibinfo{person}{Yuhan He}, \bibinfo{person}{Xuan Li}, \bibinfo{person}{Dim-Lee Kwong}, \bibinfo{person}{C.~D. Wright}, \bibinfo{person}{Wolfram H.~P. Pernice}, {and} \bibinfo{person}{H. Bhaskaran}.} \bibinfo{year}{2023}\natexlab{}.
\newblock \showarticletitle{Higher-dimensional processing using a photonic tensor core with continuous-time data}.
\newblock \bibinfo{journal}{\emph{Nature Photonics}}  \bibinfo{volume}{17} (\bibinfo{year}{2023}), \bibinfo{pages}{1080--1088}.
\newblock
\href{https://doi.org/10.1038/s41566-023-01313-x}{doi:\nolinkurl{10.1038/s41566-023-01313-x}}


\bibitem[Fan et~al\mbox{.}(2025)]%
        {fan_compact_2025}
\bibfield{author}{\bibinfo{person}{Zhuping Fan}, \bibinfo{person}{Xiongshuo Yan}, \bibinfo{person}{Xuan Li}, \bibinfo{person}{Xiao Ma}, \bibinfo{person}{Xuyang Wang}, {and} \bibinfo{person}{Jun Zou}.} \bibinfo{year}{2025}\natexlab{}.
\newblock \showarticletitle{Compact {Arrayed} {Waveguide} {Gratings} {Fabricated} on 800-nm-{Thick} {Si}$_{\textrm{3}}$ {N}$_{\textrm{4}}$ {Photonic} {Integration} {Platform}}.
\newblock \bibinfo{journal}{\emph{Journal of Lightwave Technology}} \bibinfo{volume}{43}, \bibinfo{number}{11} (\bibinfo{date}{June} \bibinfo{year}{2025}), \bibinfo{pages}{5366--5373}.
\newblock
\showISSN{0733-8724, 1558-2213}
\href{https://doi.org/10.1109/JLT.2025.3547494}{doi:\nolinkurl{10.1109/JLT.2025.3547494}}


\bibitem[Feldmann et~al\mbox{.}(2021a)]%
        {NP_Nature2021_Feldmann}
\bibfield{author}{\bibinfo{person}{Johannes Feldmann}, \bibinfo{person}{Nathan Youngblood}, \bibinfo{person}{Maxim Karpov}, \bibinfo{person}{Helge Gehring}, \bibinfo{person}{Xuan Li}, \bibinfo{person}{Maik Stappers}, \bibinfo{person}{Manuel~Le Gallo}, \bibinfo{person}{Xin Fu}, \bibinfo{person}{Anton Lukashchuk}, \bibinfo{person}{Arslan Raja}, \bibinfo{person}{Junqiu Liu}, \bibinfo{person}{David Wright}, \bibinfo{person}{Abu Sebastian}, \bibinfo{person}{Tobias Kippenberg}, \bibinfo{person}{Wolfram Pernice}, {and} \bibinfo{person}{Harish Bhaskaran}.} \bibinfo{year}{2021}\natexlab{a}.
\newblock \showarticletitle{Parallel convolutional processing using an integrated photonic tensor core}.
\newblock \bibinfo{journal}{\emph{Nature}} (\bibinfo{year}{2021}).
\newblock


\bibitem[Feldmann et~al\mbox{.}(2021b)]%
        {feldmann2021parallel}
\bibfield{author}{\bibinfo{person}{Johannes Feldmann}, \bibinfo{person}{Nathan Youngblood}, \bibinfo{person}{Maxim Karpov}, \bibinfo{person}{Helge Gehring}, \bibinfo{person}{Xuan Li}, \bibinfo{person}{Maik Stappers}, \bibinfo{person}{Manuel Le~Gallo}, \bibinfo{person}{Xin Fu}, \bibinfo{person}{Anton Lukashchuk}, \bibinfo{person}{Arslan~S Raja}, {et~al\mbox{.}}} \bibinfo{year}{2021}\natexlab{b}.
\newblock \showarticletitle{Parallel convolutional processing using an integrated photonic tensor core}.
\newblock \bibinfo{journal}{\emph{Nature}} \bibinfo{volume}{589}, \bibinfo{number}{7840} (\bibinfo{year}{2021}), \bibinfo{pages}{52--58}.
\newblock


\bibitem[Feldmann et~al\mbox{.}(2021c)]%
        {Feldmann2021}
\bibfield{author}{\bibinfo{person}{J. Feldmann}, \bibinfo{person}{N. Youngblood}, \bibinfo{person}{M. Karpov}, \bibinfo{person}{H. Gehring}, \bibinfo{person}{X. Li}, \bibinfo{person}{M. Stappers}, \bibinfo{person}{M. Le~Gallo}, \bibinfo{person}{X. Fu}, \bibinfo{person}{A. Lukashchuk}, \bibinfo{person}{A.~S. Raja}, \bibinfo{person}{J. Liu}, \bibinfo{person}{C.~D. Wright}, \bibinfo{person}{A. Sebastian}, \bibinfo{person}{T.~J. Kippenberg}, \bibinfo{person}{W.~H.~P. Pernice}, {and} \bibinfo{person}{H. Bhaskaran}.} \bibinfo{year}{2021}\natexlab{c}.
\newblock \showarticletitle{Parallel convolutional processing using an integrated photonic tensor core}.
\newblock \bibinfo{journal}{\emph{Nature}} \bibinfo{volume}{589}, \bibinfo{number}{7840} (\bibinfo{date}{01 Jan} \bibinfo{year}{2021}), \bibinfo{pages}{52--58}.
\newblock
\showISSN{1476-4687}
\href{https://doi.org/10.1038/s41586-020-03070-1}{doi:\nolinkurl{10.1038/s41586-020-03070-1}}


\bibitem[Feldmann et~al\mbox{.}(2019)]%
        {feldmann2019all}
\bibfield{author}{\bibinfo{person}{Johannes Feldmann}, \bibinfo{person}{Nathan Youngblood}, \bibinfo{person}{C~David Wright}, \bibinfo{person}{Harish Bhaskaran}, {and} \bibinfo{person}{Wolfram~HP Pernice}.} \bibinfo{year}{2019}\natexlab{}.
\newblock \showarticletitle{All-optical spiking neurosynaptic networks with self-learning capabilities}.
\newblock \bibinfo{journal}{\emph{Nature}} \bibinfo{volume}{569}, \bibinfo{number}{7755} (\bibinfo{year}{2019}), \bibinfo{pages}{208--214}.
\newblock


\bibitem[Gasse et~al\mbox{.}(2019)]%
        {VanGasse:19}
\bibfield{author}{\bibinfo{person}{Kasper~Van Gasse}, \bibinfo{person}{Ruijun Wang}, {and} \bibinfo{person}{Gunther Roelkens}.} \bibinfo{year}{2019}\natexlab{}.
\newblock \showarticletitle{27 dB gain III--V-on-silicon semiconductor optical amplifier with \&gt; 17 dBm output power}.
\newblock \bibinfo{journal}{\emph{Opt. Express}} \bibinfo{volume}{27}, \bibinfo{number}{1} (\bibinfo{date}{Jan} \bibinfo{year}{2019}), \bibinfo{pages}{293--302}.
\newblock
\href{https://doi.org/10.1364/OE.27.000293}{doi:\nolinkurl{10.1364/OE.27.000293}}


\bibitem[Guo et~al\mbox{.}(2019)]%
        {Guo2019}
\bibfield{author}{\bibinfo{person}{P. Guo}, \bibinfo{person}{A.~M. Sarangan}, {and} \bibinfo{person}{I. Agha}.} \bibinfo{year}{2019}\natexlab{}.
\newblock \showarticletitle{A Review of Germanium-Antimony-Telluride Phase Change Materials for Non-Volatile Memories and Optical Modulators}.
\newblock \bibinfo{journal}{\emph{Applied Sciences}} \bibinfo{volume}{9}, \bibinfo{number}{3} (\bibinfo{year}{2019}), \bibinfo{pages}{530}.
\newblock
\href{https://doi.org/10.3390/app9030530}{doi:\nolinkurl{10.3390/app9030530}}


\bibitem[Han et~al\mbox{.}(2023)]%
        {Han2023BG}
\bibfield{author}{\bibinfo{person}{Changhao Han}, \bibinfo{person}{Zhao Zheng}, \bibinfo{person}{Haowen Shu}, \bibinfo{person}{Ming Jin}, \bibinfo{person}{Jun Qin}, \bibinfo{person}{Ruixuan Chen}, \bibinfo{person}{Yuansheng Tao}, \bibinfo{person}{Bitao Shen}, \bibinfo{person}{Bowen Bai}, \bibinfo{person}{Fenghe Yang}, \bibinfo{person}{Yimeng Wang}, \bibinfo{person}{Haoyu Wang}, \bibinfo{person}{Feifan Wang}, \bibinfo{person}{Zixuan Zhang}, \bibinfo{person}{Shaohua Yu}, \bibinfo{person}{Chao Peng}, {and} \bibinfo{person}{Xingjun Wang}.} \bibinfo{year}{2023}\natexlab{}.
\newblock \showarticletitle{Slow-light silicon modulator with 110-GHz bandwidth}.
\newblock \bibinfo{journal}{\emph{Science Advances}} \bibinfo{volume}{9}, \bibinfo{number}{42} (\bibinfo{year}{2023}), \bibinfo{pages}{eadi5339}.
\newblock
\href{https://doi.org/10.1126/sciadv.adi5339}{doi:\nolinkurl{10.1126/sciadv.adi5339}}


\bibitem[Hasegawa et~al\mbox{.}(2010)]%
        {ref-SOA-106}
\bibfield{author}{\bibinfo{person}{H. Hasegawa}, \bibinfo{person}{M. Funabashi}, \bibinfo{person}{Yokouchi N.}, \bibinfo{person}{K. Kiyota}, {and} \bibinfo{person}{K. Maruyama}.} \bibinfo{year}{2010}\natexlab{}.
\newblock \showarticletitle{Design and fabrication of semiconductor optical amplifier with low noise figure}.
\newblock \bibinfo{journal}{\emph{In IEEE OECC Technical Digest 204–205}} (\bibinfo{year}{2010}).
\newblock


\bibitem[He et~al\mbox{.}(2016)]%
        {NN_CVPR2016_He}
\bibfield{author}{\bibinfo{person}{Kaiming He}, \bibinfo{person}{Xiangyu Zhang}, \bibinfo{person}{Shaoqing Ren}, {and} \bibinfo{person}{Jian Sun}.} \bibinfo{year}{2016}\natexlab{}.
\newblock \showarticletitle{Deep Residual Learning for Image Recognition}. In \bibinfo{booktitle}{\emph{Proc.~CVPR}}. \bibinfo{pages}{770--778}.
\newblock


\bibitem[Hinton(2022)]%
        {hinton2022forwardforwardalgorithmpreliminaryinvestigations}
\bibfield{author}{\bibinfo{person}{Geoffrey Hinton}.} \bibinfo{year}{2022}\natexlab{}.
\newblock \bibinfo{title}{The Forward-Forward Algorithm: Some Preliminary Investigations}.
\newblock
\showeprint[arxiv]{2212.13345}~[cs.LG]
\urldef\tempurl%
\url{https://arxiv.org/abs/2212.13345}
\showURL{%
\tempurl}


\bibitem[Hua et~al\mbox{.}(2025)]%
        {Hua2025}
\bibfield{author}{\bibinfo{person}{Shiyue Hua}, \bibinfo{person}{Erwan Divita}, \bibinfo{person}{Shanshan Yu}, \bibinfo{person}{Bo Peng}, \bibinfo{person}{Charles Roques-Carmes}, \bibinfo{person}{Zhan Su}, \bibinfo{person}{Zhang Chen}, \bibinfo{person}{Yanfei Bai}, \bibinfo{person}{Jinghui Zou}, \bibinfo{person}{Yunpeng Zhu}, \bibinfo{person}{Yelong Xu}, \bibinfo{person}{Cheng-kuan Lu}, \bibinfo{person}{Yuemiao Di}, \bibinfo{person}{Hui Chen}, \bibinfo{person}{Lushan Jiang}, \bibinfo{person}{Lijie Wang}, \bibinfo{person}{Longwu Ou}, \bibinfo{person}{Chaohong Zhang}, \bibinfo{person}{Junjie Chen}, \bibinfo{person}{Wen Zhang}, \bibinfo{person}{Hongyan Zhu}, \bibinfo{person}{Weijun Kuang}, \bibinfo{person}{Long Wang}, \bibinfo{person}{Huaiyu Meng}, \bibinfo{person}{Maurice Steinman}, {and} \bibinfo{person}{Yichen Shen}.} \bibinfo{year}{2025}\natexlab{}.
\newblock \showarticletitle{An integrated large-scale photonic accelerator with ultralow latency}.
\newblock \bibinfo{journal}{\emph{Nature}} \bibinfo{volume}{640}, \bibinfo{number}{8058} (\bibinfo{date}{01 Apr} \bibinfo{year}{2025}), \bibinfo{pages}{361--367}.
\newblock
\showISSN{1476-4687}
\href{https://doi.org/10.1038/s41586-025-08786-6}{doi:\nolinkurl{10.1038/s41586-025-08786-6}}


\bibitem[Huang and Barz(2025)]%
        {huangCompactInverseDesigned2025}
\bibfield{author}{\bibinfo{person}{Shiang-Yu Huang} {and} \bibinfo{person}{Stefanie Barz}.} \bibinfo{year}{2025}\natexlab{}.
\newblock \showarticletitle{Compact Inverse Designed Vertical Coupler with Bottom Reflector for Sub-Decibel Fiber-to-Chip Coupling on Silicon on Insulator Platform}.
\newblock \bibinfo{journal}{\emph{Scientific Reports}} \bibinfo{volume}{15}, \bibinfo{number}{1} (\bibinfo{date}{Jan.} \bibinfo{year}{2025}), \bibinfo{pages}{2925}.
\newblock
\showISSN{2045-2322}
\href{https://doi.org/10.1038/s41598-025-86161-1}{doi:\nolinkurl{10.1038/s41598-025-86161-1}}


\bibitem[Hänsch et~al\mbox{.}(2022)]%
        {9883741}
\bibfield{author}{\bibinfo{person}{Ronny Hänsch}, \bibinfo{person}{Jacob Arndt}, \bibinfo{person}{Matthew Gibb}, \bibinfo{person}{Arnold Boedihardjo}, \bibinfo{person}{Tyler Pedelose}, {and} \bibinfo{person}{Todd~M. Bacastow}.} \bibinfo{year}{2022}\natexlab{}.
\newblock \showarticletitle{The SpaceNet 8 Challenge - From Foundation Mapping to Flood Detection}. In \bibinfo{booktitle}{\emph{IGARSS 2022 - 2022 IEEE International Geoscience and Remote Sensing Symposium}}. \bibinfo{pages}{5073--5076}.
\newblock
\href{https://doi.org/10.1109/IGARSS46834.2022.9883741}{doi:\nolinkurl{10.1109/IGARSS46834.2022.9883741}}


\bibitem[Innolume(2025)]%
        {innolume_comb_2025}
\bibfield{author}{\bibinfo{person}{Innolume}.} \bibinfo{year}{2025}\natexlab{}.
\newblock \bibinfo{title}{Comb-lasers}.
\newblock
\urldef\tempurl%
\url{https://www.innolume.com/innoproducts/comb-laser/}
\showURL{%
\tempurl}
\newblock
\shownote{[Online; accessed 22-January-2026]}.


\bibitem[Jafari et~al\mbox{.}(2020)]%
        {Jafari2020BG}
\bibfield{author}{\bibinfo{person}{O. Jafari}, \bibinfo{person}{W. Shi}, {and} \bibinfo{person}{S. Larochelle}.} \bibinfo{year}{2020}\natexlab{}.
\newblock \showarticletitle{Mach-Zehnder Silicon Photonic Modulator Assisted by Phase-Shifted Bragg Gratings}.
\newblock \bibinfo{journal}{\emph{IEEE Photonics Technology Letters}} \bibinfo{volume}{32}, \bibinfo{number}{8} (\bibinfo{year}{2020}), \bibinfo{pages}{445--448}.
\newblock
\href{https://doi.org/10.1109/LPT.2020.2978793}{doi:\nolinkurl{10.1109/LPT.2020.2978793}}


\bibitem[Juodawlkis et~al\mbox{.}(2011)]%
        {ref-SOA-105}
\bibfield{author}{\bibinfo{person}{Paul~W. Juodawlkis}, \bibinfo{person}{Jason~J. Plant}, \bibinfo{person}{William Loh}, \bibinfo{person}{Leo~J. Missaggia}, \bibinfo{person}{Frederick~J. O’Donnell}, \bibinfo{person}{Douglas~C. Oakley}, \bibinfo{person}{Antonio Napoleone}, \bibinfo{person}{Jonathan Klamkin}, \bibinfo{person}{Juliet~T. Gopinath}, \bibinfo{person}{Daniel~J. Ripin}, \bibinfo{person}{Sangyoun Gee}, \bibinfo{person}{Peter~J. Delfyett}, {and} \bibinfo{person}{Joseph~P. Donnelly}.} \bibinfo{year}{2011}\natexlab{}.
\newblock \showarticletitle{High-power, low-noise 1.5-$\mu$m slab-coupled optical waveguide (SCOW) emitters: physics, devices, and applications.}
\newblock \bibinfo{journal}{\emph{IEEE J. Sel. Top. Quant. Electron. 17, 1698–1714}} (\bibinfo{year}{2011}).
\newblock


\bibitem[Kawahara et~al\mbox{.}(2024)]%
        {Baba2024PhC}
\bibfield{author}{\bibinfo{person}{Keisuke Kawahara}, \bibinfo{person}{Tai Tsuchizawa}, \bibinfo{person}{Noritsugu Yamamoto}, \bibinfo{person}{Yuriko Maegami}, \bibinfo{person}{Koji Yamada}, \bibinfo{person}{Shinsuke Hara}, {and} \bibinfo{person}{Toshihiko Baba}.} \bibinfo{year}{2024}\natexlab{}.
\newblock \showarticletitle{High-speed, low-voltage, low-bit-energy silicon photonic crystal slow-light modulator with impedance-engineered distributed electrodes}.
\newblock \bibinfo{journal}{\emph{Optica}} \bibinfo{volume}{11}, \bibinfo{number}{9} (\bibinfo{date}{Sep} \bibinfo{year}{2024}), \bibinfo{pages}{1212--1219}.
\newblock
\href{https://doi.org/10.1364/OPTICA.531372}{doi:\nolinkurl{10.1364/OPTICA.531372}}


\bibitem[Kim et~al\mbox{.}(2019)]%
        {Kim_kerr_2019}
\bibfield{author}{\bibinfo{person}{Bok~Young Kim}, \bibinfo{person}{Yoshitomo Okawachi}, \bibinfo{person}{Jae~K. Jang}, \bibinfo{person}{Mengjie Yu}, \bibinfo{person}{Xingchen Ji}, \bibinfo{person}{Yun Zhao}, \bibinfo{person}{Chaitanya Joshi}, \bibinfo{person}{Michal Lipson}, {and} \bibinfo{person}{Alexander~L. Gaeta}.} \bibinfo{year}{2019}\natexlab{}.
\newblock \showarticletitle{Turn-key, high-efficiency {Kerr} comb source}.
\newblock \bibinfo{journal}{\emph{Opt. Lett.}} \bibinfo{volume}{44}, \bibinfo{number}{18} (\bibinfo{date}{Sep} \bibinfo{year}{2019}), \bibinfo{pages}{4475--4478}.
\newblock
\href{https://doi.org/10.1364/OL.44.004475}{doi:\nolinkurl{10.1364/OL.44.004475}}


\bibitem[Krizhevsky et~al\mbox{.}(2009)]%
        {NN_cifar2009}
\bibfield{author}{\bibinfo{person}{Alex Krizhevsky}, \bibinfo{person}{Geoffrey Hinton}, {et~al\mbox{.}}} \bibinfo{year}{2009}\natexlab{}.
\newblock \showarticletitle{Learning Multiple Layers of Features from Tiny Images}.
\newblock  (\bibinfo{year}{2009}).
\newblock


\bibitem[Larsson et~al\mbox{.}(2018)]%
        {ref-VCSEL-409}
\bibfield{author}{\bibinfo{person}{A. Larsson}, \bibinfo{person}{E. Simpanen}, \bibinfo{person}{J.S. Gustavsson}, \bibinfo{person}{E. Haglund}, \bibinfo{person}{E.P. Haglund}, \bibinfo{person}{T. Lengyel}, \bibinfo{person}{P.A. Andrekson}, \bibinfo{person}{W.V. Sorin}, \bibinfo{person}{S. Mathai}, \bibinfo{person}{M. Tan}, {and} \bibinfo{person}{S.R. Bickham}.} \bibinfo{year}{2018}\natexlab{}.
\newblock \showarticletitle{1060 nm VCSELs for long-reach optical interconnects}.
\newblock \bibinfo{journal}{\emph{Optical Fiber Technology}}  \bibinfo{volume}{44} (\bibinfo{year}{2018}), \bibinfo{pages}{36--42}.
\newblock
\showISSN{1068-5200}
\href{https://doi.org/10.1016/j.yofte.2018.01.001}{doi:\nolinkurl{10.1016/j.yofte.2018.01.001}}
\newblock
\shownote{Special Issue on Data Center Communications}.


\bibitem[Ledentsov et~al\mbox{.}(2025)]%
        {ref-VCSEL-405}
\bibfield{author}{\bibinfo{person}{N.N. Ledentsov}, \bibinfo{person}{Jr. Ledentsov, N.}, \bibinfo{person}{V.A. Shchukin}, \bibinfo{person}{A.N. Ledentsov}, \bibinfo{person}{O.Y. Makarov}, \bibinfo{person}{I.E. Titkov}, \bibinfo{person}{M. Lindemann}, \bibinfo{person}{T. de Adelsburg~Ettmayer}, \bibinfo{person}{N.C. Gerhardt}, \bibinfo{person}{M.R. Hofmann}, {et~al\mbox{.}}} \bibinfo{year}{2025}\natexlab{}.
\newblock \showarticletitle{VCSELs: Influence of Design on Performance and Data Transmission over Multi-Mode and Single-Mode Fibers}.
\newblock \bibinfo{journal}{\emph{Photonics, 12, 1037.}} (\bibinfo{year}{2025}).
\newblock


\bibitem[Ledentsov et~al\mbox{.}(1994)]%
        {ref-VCSEL-404}
\bibfield{author}{\bibinfo{person}{N.N. Ledentsov}, \bibinfo{person}{V.M. Ustinov}, \bibinfo{person}{A.Y. Egorov}, \bibinfo{person}{A.E. Zhukov}, \bibinfo{person}{M.V. Maximov}, \bibinfo{person}{I.G. Tabatadze}, {and} \bibinfo{person}{P.S. Kop’ev}.} \bibinfo{year}{1994}\natexlab{}.
\newblock \showarticletitle{Optical properties of heterostructures with InGaAs-GaAs quantum clusters}.
\newblock \bibinfo{journal}{\emph{Semiconductors, 28, 832–834}} (\bibinfo{year}{1994}).
\newblock


\bibitem[Li and Roelkens(2025)]%
        {liGratingCouplerDesign2025}
\bibfield{author}{\bibinfo{person}{He Li} {and} \bibinfo{person}{G{\"u}nther Roelkens}.} \bibinfo{year}{2025}\natexlab{}.
\newblock \showarticletitle{Grating Coupler Design for {{VCSELs}} Heterogeneously Integrated on Silicon Nitride}. In \bibinfo{booktitle}{\emph{2025 {{IEEE Silicon Photonics Conference}} ({{SiPhotonics}})}}. \bibinfo{publisher}{IEEE}, \bibinfo{address}{London, United Kingdom}, \bibinfo{pages}{1--2}.
\newblock
\showISBNx{979-8-3315-0618-6}
\href{https://doi.org/10.1109/SiPhotonics64386.2025.10984830}{doi:\nolinkurl{10.1109/SiPhotonics64386.2025.10984830}}


\bibitem[Li et~al\mbox{.}(2025)]%
        {ref-VCSEL-402}
\bibfield{author}{\bibinfo{person}{Shengnan Li}, \bibinfo{person}{Jingjing Dai}, \bibinfo{person}{Wei Li}, \bibinfo{person}{Chong Li}, {et~al\mbox{.}}} \bibinfo{year}{2025}\natexlab{}.
\newblock \showarticletitle{Modeling of high-peak-power vertically integrated VCSEL composite cavity with cascaded energy conversion}.
\newblock \bibinfo{journal}{\emph{Opt. Express 33, 16704-16716}} (\bibinfo{year}{2025}).
\newblock


\bibitem[Li et~al\mbox{.}(2024)]%
        {10.1145/3650200.3656609}
\bibfield{author}{\bibinfo{person}{Xianbin Li}, \bibinfo{person}{Yinyi Liu}, \bibinfo{person}{Fan Jiang}, \bibinfo{person}{Chengeng Li}, \bibinfo{person}{Yuxiang Fu}, \bibinfo{person}{Wei Zhang}, {and} \bibinfo{person}{Jiang Xu}.} \bibinfo{year}{2024}\natexlab{}.
\newblock \showarticletitle{NEOCNN: NTT-Enabled Optical Convolution Neural Network Accelerator}. In \bibinfo{booktitle}{\emph{Proceedings of the 38th ACM International Conference on Supercomputing}} (Kyoto, Japan) \emph{(\bibinfo{series}{ICS '24})}. \bibinfo{publisher}{Association for Computing Machinery}, \bibinfo{address}{New York, NY, USA}, \bibinfo{pages}{352–362}.
\newblock
\showISBNx{9798400706103}
\href{https://doi.org/10.1145/3650200.3656609}{doi:\nolinkurl{10.1145/3650200.3656609}}


\bibitem[Li et~al\mbox{.}(2019)]%
        {Li2019}
\bibfield{author}{\bibinfo{person}{X. Li}, \bibinfo{person}{N. Youngblood}, \bibinfo{person}{C. R{\'i}os}, \bibinfo{person}{Z. Cheng}, \bibinfo{person}{C.~D. Wright}, \bibinfo{person}{W.~H.~P. Pernice}, {and} \bibinfo{person}{H. Bhaskaran}.} \bibinfo{year}{2019}\natexlab{}.
\newblock \showarticletitle{Fast and Reliable Storage Using a 5 Bit, Nonvolatile Photonic Memory Cell}.
\newblock \bibinfo{journal}{\emph{Optica}} \bibinfo{volume}{6}, \bibinfo{number}{1} (\bibinfo{year}{2019}), \bibinfo{pages}{1--8}.
\newblock
\href{https://doi.org/10.1364/OPTICA.6.000001}{doi:\nolinkurl{10.1364/OPTICA.6.000001}}


\bibitem[Liu et~al\mbox{.}(2024)]%
        {liu2024GSST}
\bibfield{author}{\bibinfo{person}{Shichang Liu}, \bibinfo{person}{Xing Yang}, \bibinfo{person}{Liangjun Lu}, \bibinfo{person}{Yu Li}, \bibinfo{person}{Jianping Chen}, {and} \bibinfo{person}{Linjie Zhou}.} \bibinfo{year}{2024}\natexlab{}.
\newblock \showarticletitle{{Non-volatile silicon-GSST multilevel optical switch}}. In \bibinfo{booktitle}{\emph{AOPC 2024: Optical Devices and Integration}}, \bibfield{editor}{\bibinfo{person}{Linjie Zhou}} (Ed.), Vol.~\bibinfo{volume}{13499}. International Society for Optics and Photonics, \bibinfo{publisher}{SPIE}, \bibinfo{pages}{134990F}.
\newblock
\href{https://doi.org/10.1117/12.3046167}{doi:\nolinkurl{10.1117/12.3046167}}


\bibitem[Lott et~al\mbox{.}(2005)]%
        {ref-VCSEL-403}
\bibfield{author}{\bibinfo{person}{J.A. Lott}, \bibinfo{person}{A.R. Kovsh}, \bibinfo{person}{N.N. Ledentsov}, {and} \bibinfo{person}{D. Bimberg}.} \bibinfo{year}{2005}\natexlab{}.
\newblock \showarticletitle{GaAs-Based InAs/InGaAs Quantum Dot Vertical Cavity and Vertical External Cavity Surface Emitting Lasers Emitting Near 1300 nm}.
\newblock \bibinfo{journal}{\emph{In Proceedings of the 2005 Pacific Rim Conference on Lasers \& Electro-Optics, Tokyo, Japan; pp. 160–161}} (\bibinfo{year}{2005}).
\newblock


\bibitem[Luan et~al\mbox{.}(2023)]%
        {Luan2023HighDensityPhotonicTensorCore}
\bibfield{author}{\bibinfo{person}{Enxiao Luan}, \bibinfo{person}{Shangxuan Yu}, \bibinfo{person}{Mahsa Salmani}, \bibinfo{person}{Mohammadreza Sanadgol~Nezami}, \bibinfo{person}{Bhavin~J. Shastri}, \bibinfo{person}{Lukas Chrostowski}, {and} \bibinfo{person}{Armaghan Eshaghi}.} \bibinfo{year}{2023}\natexlab{}.
\newblock \showarticletitle{Towards a high-density photonic tensor core enabled by intensity-modulated microrings and photonic wire bonding}.
\newblock \bibinfo{journal}{\emph{Scientific Reports}}  \bibinfo{volume}{13} (\bibinfo{year}{2023}), \bibinfo{pages}{1260}.
\newblock
\href{https://doi.org/10.1038/s41598-023-28264-5}{doi:\nolinkurl{10.1038/s41598-023-28264-5}}


\bibitem[Ma et~al\mbox{.}(2021)]%
        {ma_cdc_2021}
\bibfield{author}{\bibinfo{person}{Yuxin Ma}, \bibinfo{person}{Yong Zhao}, \bibinfo{person}{Yuechun Shi}, \bibinfo{person}{Lijun Hao}, \bibinfo{person}{Zhenxing Sun}, \bibinfo{person}{Ziming Hong}, {and} \bibinfo{person}{Xiangfei Chen}.} \bibinfo{year}{2021}\natexlab{}.
\newblock \showarticletitle{Silicon Add-Drop Multiplexer Based on $\pi$ Phase-Shifted Antisymmetric Bragg Grating}.
\newblock \bibinfo{journal}{\emph{IEEE Journal of Quantum Electronics}} \bibinfo{volume}{57}, \bibinfo{number}{4} (\bibinfo{year}{2021}), \bibinfo{pages}{1--8}.
\newblock
\href{https://doi.org/10.1109/JQE.2021.3078060}{doi:\nolinkurl{10.1109/JQE.2021.3078060}}


\bibitem[Minz and Sonkar(2021)]%
        {minz_cdc_2021}
\bibfield{author}{\bibinfo{person}{Manoranjan Minz} {and} \bibinfo{person}{Ramesh~Kumar Sonkar}.} \bibinfo{year}{2021}\natexlab{}.
\newblock \showarticletitle{Design of a hybrid mode and wavelength division (de) multiplexer based on contra-directional grating assisted couplers on the SOI platform}.
\newblock \bibinfo{journal}{\emph{Applied Optics}} \bibinfo{volume}{60}, \bibinfo{number}{9} (\bibinfo{year}{2021}), \bibinfo{pages}{2640--2646}.
\newblock


\bibitem[Miscuglio and Sorger(2020)]%
        {Miscuglio2020PhotonicTensorCores}
\bibfield{author}{\bibinfo{person}{Mario Miscuglio} {and} \bibinfo{person}{Volker~J. Sorger}.} \bibinfo{year}{2020}\natexlab{}.
\newblock \showarticletitle{Photonic tensor cores for machine learning}.
\newblock \bibinfo{journal}{\emph{Applied Physics Reviews}} \bibinfo{volume}{7}, \bibinfo{number}{3} (\bibinfo{year}{2020}), \bibinfo{pages}{031404}.
\newblock
\href{https://doi.org/10.1063/5.0001942}{doi:\nolinkurl{10.1063/5.0001942}}


\bibitem[Morsali et~al\mbox{.}(2024)]%
        {a557a39c698546f1902a12c137b1913f}
\bibfield{author}{\bibinfo{person}{Mehrdad Morsali}, \bibinfo{person}{Sepehr Tabrizchi}, \bibinfo{person}{Deniz Najafi}, \bibinfo{person}{Mohsen Imani}, \bibinfo{person}{Mahdi Nikdast}, \bibinfo{person}{Arman Roohi}, {and} \bibinfo{person}{Shaahin Angizi}.} \bibinfo{year}{2024}\natexlab{}.
\newblock \showarticletitle{OISA: Architecting an Optical In-Sensor Accelerator for Efficient Visual Computing}. In \bibinfo{booktitle}{\emph{2024 Design, Automation and Test in Europe Conference and Exhibition, DATE 2024 - Proceedings}} \emph{(\bibinfo{series}{Proceedings -Design, Automation and Test in Europe, DATE})}. \bibinfo{publisher}{Institute of Electrical and Electronics Engineers Inc.}, \bibinfo{address}{United States}.
\newblock
\newblock
\shownote{Publisher Copyright: {\textcopyright} 2024 EDAA.; 2024 Design, Automation and Test in Europe Conference and Exhibition, DATE 2024 ; Conference date: 25-03-2024 Through 27-03-2024}.


\bibitem[Nevzorov et~al\mbox{.}(2023)]%
        {Nevzorov2023}
\bibfield{author}{\bibinfo{person}{A.~A. Nevzorov}, \bibinfo{person}{V.~A. Mikhalevsky}, \bibinfo{person}{A.~V. Kiselev}, \bibinfo{person}{A.~A. Burtsev}, \bibinfo{person}{N.~N. Eliseev}, \bibinfo{person}{V.~V. Ionin}, {and} \bibinfo{person}{A.~A. Lotin}.} \bibinfo{year}{2023}\natexlab{}.
\newblock \showarticletitle{Controlling Optical Properties of GST Thin Films by Ultrashort Laser Pulses Series Impact}.
\newblock \bibinfo{journal}{\emph{Optical Materials}}  \bibinfo{volume}{141} (\bibinfo{year}{2023}), \bibinfo{pages}{113925}.
\newblock
\href{https://doi.org/10.1016/j.optmat.2023.113925}{doi:\nolinkurl{10.1016/j.optmat.2023.113925}}


\bibitem[Ning et~al\mbox{.}(2024)]%
        {Ning2024HardwareEfficientEPIC}
\bibfield{author}{\bibinfo{person}{Shupeng Ning}, \bibinfo{person}{Jiaqi Gu}, \bibinfo{person}{Chenghao Feng}, \bibinfo{person}{Rongxing Tang}, \bibinfo{person}{Hanqing Zhu}, \bibinfo{person}{David~Z. Pan}, {and} \bibinfo{person}{Ray~T. Chen}.} \bibinfo{year}{2024}\natexlab{}.
\newblock \showarticletitle{A Hardware-Efficient Silicon Electronic-Photonic Chip for Optical Structured Neural Networks}. In \bibinfo{booktitle}{\emph{SPIE Photonics West}}.
\newblock


\bibitem[{NVIDIA}(2020)]%
        {nvidiaV100Datasheet2020}
\bibfield{author}{\bibinfo{person}{{NVIDIA}}.} \bibinfo{year}{2020}\natexlab{}.
\newblock \bibinfo{title}{{NVIDIA V100 Tensor Core GPU Datasheet}}.
\newblock
\urldef\tempurl%
\url{https://images.nvidia.com/content/technologies/volta/pdf/volta-v100-datasheet-update-us-1165301-r5.pdf}
\showURL{%
\tempurl}
\newblock
\shownote{US-1165301-R5, Jan 2020}.


\bibitem[{NVIDIA}(2022)]%
        {nvidiaA100Datasheet2022}
\bibfield{author}{\bibinfo{person}{{NVIDIA}}.} \bibinfo{year}{2022}\natexlab{}.
\newblock \bibinfo{title}{{NVIDIA A100 Tensor Core GPU Datasheet}}.
\newblock
\urldef\tempurl%
\url{https://www.nvidia.com/content/dam/en-zz/Solutions/Data-Center/a100/pdf/nvidia-a100-datasheet-nvidia-us-2188504-web.pdf}
\showURL{%
\tempurl}
\newblock
\shownote{2188504, May 2022}.


\bibitem[{NVIDIA}(2023)]%
        {nvidiaH100Datasheet2023}
\bibfield{author}{\bibinfo{person}{{NVIDIA}}.} \bibinfo{year}{2023}\natexlab{}.
\newblock \bibinfo{title}{{NVIDIA H100 Tensor Core GPU Datasheet}}.
\newblock
\urldef\tempurl%
\url{https://www.cisco.com/c/dam/en/us/products/collateral/servers-unified-computing/ucs-c-series-rack-servers/nvidia-h100-80-gpu.pdf}
\showURL{%
\tempurl}
\newblock
\shownote{2569583, Feb 2023}.


\bibitem[{NVIDIA}(2025)]%
        {nvidiaHGXB200PCFSummary2025}
\bibfield{author}{\bibinfo{person}{{NVIDIA}}.} \bibinfo{year}{2025}\natexlab{}.
\newblock \bibinfo{title}{{PCF Summary for NVIDIA HGX B200 (Datasheet)}}.
\newblock
\urldef\tempurl%
\url{https://images.nvidia.com/aem-dam/Solutions/documents/HGX-B200-PCF-Summary.pdf}
\showURL{%
\tempurl}
\newblock
\shownote{4069550, Jul 2025}.


\bibitem[Ohno et~al\mbox{.}(2022)]%
        {Ohno2022MRRCrossbarACSPhotonics}
\bibfield{author}{\bibinfo{person}{Shuhei Ohno}, \bibinfo{person}{Rui Tang}, \bibinfo{person}{Kasidit Toprasertpong}, \bibinfo{person}{Shinichi Takagi}, {and} \bibinfo{person}{Mitsuru Takenaka}.} \bibinfo{year}{2022}\natexlab{}.
\newblock \showarticletitle{Si Microring Resonator Crossbar Array for On-Chip Inference and Training of the Optical Neural Network}.
\newblock \bibinfo{journal}{\emph{ACS Photonics}} \bibinfo{volume}{9}, \bibinfo{number}{8} (\bibinfo{year}{2022}), \bibinfo{pages}{2614--2622}.
\newblock
\href{https://doi.org/10.1021/acsphotonics.1c01777}{doi:\nolinkurl{10.1021/acsphotonics.1c01777}}


\bibitem[Okawachi et~al\mbox{.}(2023)]%
        {Okawachi_comb_2023}
\bibfield{author}{\bibinfo{person}{Yoshitomo Okawachi}, \bibinfo{person}{Bok~Young Kim}, \bibinfo{person}{Michal Lipson}, {and} \bibinfo{person}{Alexander~L. Gaeta}.} \bibinfo{year}{2023}\natexlab{}.
\newblock \showarticletitle{Chip-scale frequency combs for data communications in computing systems}.
\newblock \bibinfo{journal}{\emph{Optica}} \bibinfo{volume}{10}, \bibinfo{number}{8} (\bibinfo{date}{Aug} \bibinfo{year}{2023}), \bibinfo{pages}{977--995}.
\newblock
\href{https://doi.org/10.1364/OPTICA.460175}{doi:\nolinkurl{10.1364/OPTICA.460175}}


\bibitem[Photonics(2022)]%
        {menhir_comb_2022}
\bibfield{author}{\bibinfo{person}{Menhir Photonics}.} \bibinfo{year}{2022}\natexlab{}.
\newblock \bibinfo{title}{Menhir-1550 {Series}}.
\newblock
\urldef\tempurl%
\url{https://menhir-photonics.com/menhir-1550/}
\showURL{%
\tempurl}
\newblock
\shownote{[Online; accessed 22-January-2026]}.


\bibitem[Pitris et~al\mbox{.}(2018)]%
        {pitris_silicon_2018}
\bibfield{author}{\bibinfo{person}{Stelios Pitris}, \bibinfo{person}{George Dabos}, \bibinfo{person}{Charoula Mitsolidou}, \bibinfo{person}{Theoni Alexoudi}, \bibinfo{person}{Peter De~Heyn}, \bibinfo{person}{Joris Van~Campenhout}, \bibinfo{person}{Ronald Broeke}, \bibinfo{person}{George~T. Kanellos}, {and} \bibinfo{person}{Nikos Pleros}.} \bibinfo{year}{2018}\natexlab{}.
\newblock \showarticletitle{Silicon photonic 8 × 8 cyclic {Arrayed} {Waveguide} {Grating} {Router} for {O}-band on-chip communication}.
\newblock \bibinfo{journal}{\emph{Optics Express}} \bibinfo{volume}{26}, \bibinfo{number}{5} (\bibinfo{date}{March} \bibinfo{year}{2018}), \bibinfo{pages}{6276}.
\newblock
\showISSN{1094-4087}
\href{https://doi.org/10.1364/OE.26.006276}{doi:\nolinkurl{10.1364/OE.26.006276}}


\bibitem[Poelman et~al\mbox{.}(2025)]%
        {Poelman_mode_locked_2025}
\bibfield{author}{\bibinfo{person}{Stijn Poelman}, \bibinfo{person}{Tom Reep}, \bibinfo{person}{Maximilien Billet}, {and} \bibinfo{person}{Bart Kuyken}.} \bibinfo{year}{2025}\natexlab{}.
\newblock \showarticletitle{High-power heterogeneously integrated mode-locked laser enabled by a booster amplifier}.
\newblock \bibinfo{journal}{\emph{Opt. Express}} \bibinfo{volume}{33}, \bibinfo{number}{26} (\bibinfo{date}{Dec} \bibinfo{year}{2025}), \bibinfo{pages}{54747--54756}.
\newblock
\href{https://doi.org/10.1364/OE.581535}{doi:\nolinkurl{10.1364/OE.581535}}


\bibitem[Qin et~al\mbox{.}(2025)]%
        {ref-SOA-101}
\bibfield{author}{\bibinfo{person}{Senbiao Qin}, \bibinfo{person}{Emadreza Soltanian}, \bibinfo{person}{Joan Ramirez}, \bibinfo{person}{Delphine Neel}, \bibinfo{person}{Nicolas Vaissiere}, {et~al\mbox{.}}} \bibinfo{year}{2025}\natexlab{}.
\newblock \showarticletitle{Microtransfer printing of InP SOAs on advanced silicon photonics platform for C-band preamplified receivers}.
\newblock \bibinfo{journal}{\emph{Proceedings Volume 13371, Silicon Photonics XX; 133710K}} (\bibinfo{year}{2025}).
\newblock


\bibitem[Rakowski et~al\mbox{.}(2020)]%
        {NP_OFC2020_Rakowski}
\bibfield{author}{\bibinfo{person}{Michal Rakowski}, \bibinfo{person}{Colleen Meagher}, \bibinfo{person}{Karen Nummy}, \bibinfo{person}{Abdelsalam Aboketaf}, \bibinfo{person}{Javier Ayala}, \bibinfo{person}{Yusheng Bian}, \bibinfo{person}{Brendan Harris}, \bibinfo{person}{Kate Mclean}, \bibinfo{person}{Kevin McStay}, \bibinfo{person}{Asli Sahin}, \bibinfo{person}{Louis Medina}, \bibinfo{person}{Bo Peng}, \bibinfo{person}{Zoey Sowinski}, \bibinfo{person}{Andy Stricker}, \bibinfo{person}{Thomas Houghton}, \bibinfo{person}{Crystal Hedges}, \bibinfo{person}{Ken Giewont}, \bibinfo{person}{Ajey Jacob}, \bibinfo{person}{Ted Letavic}, \bibinfo{person}{Dave Riggs}, \bibinfo{person}{Anthony Yu}, {and} \bibinfo{person}{John Pellerin}.} \bibinfo{year}{2020}\natexlab{}.
\newblock \showarticletitle{45nm CMOS — Silicon Photonics Monolithic Technology (45CLO) for Next-Generation, Low Power and High Speed Optical Interconnects}. In \bibinfo{booktitle}{\emph{2020 Optical Fiber Communications Conference and Exhibition (OFC)}}. \bibinfo{pages}{1--3}.
\newblock


\bibitem[Ran et~al\mbox{.}(2025)]%
        {Ran2025GeRichGST}
\bibfield{author}{\bibinfo{person}{Sheng Ran}, \bibinfo{person}{Elisa Petroni}, \bibinfo{person}{Laurent Laurin}, \bibinfo{person}{Matteo Baldo}, \bibinfo{person}{Andrea Serafini}, \bibinfo{person}{Minh-Anh Luong}, \bibinfo{person}{Alessandro Motta}, \bibinfo{person}{Andrea Redaelli}, {and} \bibinfo{person}{Alain Claverie}.} \bibinfo{year}{2025}\natexlab{}.
\newblock \showarticletitle{Phase Transitions and Chemical Segregation in Ge-Rich GST Based Phase Change Memory Cells}.
\newblock \bibinfo{journal}{\emph{Scientific Reports}} \bibinfo{volume}{15}, \bibinfo{number}{1} (\bibinfo{year}{2025}), \bibinfo{pages}{11357}.
\newblock
\href{https://doi.org/10.1038/s41598-025-95227-z}{doi:\nolinkurl{10.1038/s41598-025-95227-z}}


\bibitem[Roelkens et~al\mbox{.}(2024)]%
        {ref-SOA-102}
\bibfield{author}{\bibinfo{person}{Gunther Roelkens}, \bibinfo{person}{Jing Zhang}, \bibinfo{person}{Laurens Bogaert}, \bibinfo{person}{Emadreza Soltanian}, \bibinfo{person}{Maximilien Billet}, \bibinfo{person}{Ali Uzun}, \bibinfo{person}{Biwei Pan}, \bibinfo{person}{Yang Liu}, \bibinfo{person}{Evangelia Delli}, \bibinfo{person}{Dongbo Wang}, \bibinfo{person}{Valeria~Bonito Oliva}, \bibinfo{person}{Lam~Thi Ngoc~Tran}, \bibinfo{person}{Xin Guo}, \bibinfo{person}{He Li}, \bibinfo{person}{Senbiao Qin}, \bibinfo{person}{Konstantinos Akritidis}, \bibinfo{person}{Ye Chen}, \bibinfo{person}{Yu Xue}, \bibinfo{person}{Margot Niels}, \bibinfo{person}{Dennis Maes}, \bibinfo{person}{Max Kiewiet}, \bibinfo{person}{Tom Reep}, \bibinfo{person}{Tom Vanackere}, \bibinfo{person}{Tom Vandekerckhove}, \bibinfo{person}{Isaac~Luntadila Lufungula}, \bibinfo{person}{Jasper De~Witte}, \bibinfo{person}{Luis Reis}, \bibinfo{person}{Stijn Poelman}, \bibinfo{person}{Ying Tan}, \bibinfo{person}{Hong Deng}, \bibinfo{person}{Wim Bogaerts},
  \bibinfo{person}{Geert Morthier}, \bibinfo{person}{Dries Van~Thourhout}, {and} \bibinfo{person}{Bart Kuyken}.} \bibinfo{year}{2024}\natexlab{}.
\newblock \showarticletitle{Present and future of micro-transfer printing for heterogeneous photonic integrated circuits}.
\newblock \bibinfo{journal}{\emph{APL Photonics}} \bibinfo{volume}{9}, \bibinfo{number}{1} (\bibinfo{date}{01} \bibinfo{year}{2024}), \bibinfo{pages}{010901}.
\newblock


\bibitem[Sawant et~al\mbox{.}(2025)]%
        {sawant2025high}
\bibfield{author}{\bibinfo{person}{Rajath Sawant}, \bibinfo{person}{Anthony Albanese}, \bibinfo{person}{Arnaud Rogemont}, \bibinfo{person}{Gregorio Gonzalez-Cortes}, \bibinfo{person}{Yoann Br{\^u}l{\'e}}, \bibinfo{person}{Lara Karam}, \bibinfo{person}{Jean-baptiste Jager}, \bibinfo{person}{Stephane Malhouitre}, \bibinfo{person}{Benoit Charbonnier}, \bibinfo{person}{Aur{\'e}lien Coillet}, \bibinfo{person}{Pierre No{\'e}}, {and} \bibinfo{person}{Benoit Cluzel}.} \bibinfo{year}{2025}\natexlab{}.
\newblock \showarticletitle{High-Endurance and Energy-Efficient All-Optical Programming of Scalable Silicon Waveguides with Integrated Phase-Change Material Patches}.
\newblock \bibinfo{journal}{\emph{Advanced Optical Materials}} \bibinfo{volume}{13}, \bibinfo{number}{25} (\bibinfo{year}{2025}), \bibinfo{pages}{e00775}.
\newblock


\bibitem[Shang et~al\mbox{.}(2017)]%
        {shang_low-loss_2017}
\bibfield{author}{\bibinfo{person}{Kuanping Shang}, \bibinfo{person}{Shibnath Pathak}, \bibinfo{person}{Chuan Qin}, {and} \bibinfo{person}{S.~J.~Ben Yoo}.} \bibinfo{year}{2017}\natexlab{}.
\newblock \showarticletitle{Low-{Loss} {Compact} {Silicon} {Nitride} {Arrayed} {Waveguide} {Gratings} for {Photonic} {Integrated} {Circuits}}.
\newblock \bibinfo{journal}{\emph{IEEE Photonics Journal}} \bibinfo{volume}{9}, \bibinfo{number}{5} (\bibinfo{date}{Oct.} \bibinfo{year}{2017}), \bibinfo{pages}{1--5}.
\newblock
\showISSN{1943-0655}
\href{https://doi.org/10.1109/JPHOT.2017.2751003}{doi:\nolinkurl{10.1109/JPHOT.2017.2751003}}


\bibitem[Shi et~al\mbox{.}(2013a)]%
        {shi_cdc_2013}
\bibfield{author}{\bibinfo{person}{Wei Shi}, \bibinfo{person}{Xu Wang}, \bibinfo{person}{Charlie Lin}, \bibinfo{person}{Han Yun}, \bibinfo{person}{Yang Liu}, \bibinfo{person}{Tom Baehr-Jones}, \bibinfo{person}{Michael Hochberg}, \bibinfo{person}{Nicolas~AF Jaeger}, {and} \bibinfo{person}{Lukas Chrostowski}.} \bibinfo{year}{2013}\natexlab{a}.
\newblock \showarticletitle{Silicon photonic grating-assisted, contra-directional couplers}.
\newblock \bibinfo{journal}{\emph{Optics express}} \bibinfo{volume}{21}, \bibinfo{number}{3} (\bibinfo{year}{2013}), \bibinfo{pages}{3633--3650}.
\newblock


\bibitem[Shi et~al\mbox{.}(2013b)]%
        {shi_cdcapo_2013}
\bibfield{author}{\bibinfo{person}{Wei Shi}, \bibinfo{person}{Han Yun}, \bibinfo{person}{Charlie Lin}, \bibinfo{person}{Jonas Flueckiger}, \bibinfo{person}{Nicolas~AF Jaeger}, {and} \bibinfo{person}{Lukas Chrostowski}.} \bibinfo{year}{2013}\natexlab{b}.
\newblock \showarticletitle{Coupler-apodized Bragg-grating add--drop filter}.
\newblock \bibinfo{journal}{\emph{Optics letters}} \bibinfo{volume}{38}, \bibinfo{number}{16} (\bibinfo{year}{2013}), \bibinfo{pages}{3068--3070}.
\newblock


\bibitem[Soref et~al\mbox{.}(2024)]%
        {SOREF2024111005}
\bibfield{author}{\bibinfo{person}{Richard Soref}, \bibinfo{person}{Francesco {De Leonardis}}, \bibinfo{person}{Martino {De Carlo}}, {and} \bibinfo{person}{Vittorio~M.N. Passaro}.} \bibinfo{year}{2024}\natexlab{}.
\newblock \showarticletitle{Compact non-volatile multilevel Sb2Se3 electro-optical switching in the mid-infrared group-IV-photonics platform}.
\newblock \bibinfo{journal}{\emph{Optics \& Laser Technology}}  \bibinfo{volume}{176} (\bibinfo{year}{2024}), \bibinfo{pages}{111005}.
\newblock
\showISSN{0030-3992}
\href{https://doi.org/10.1016/j.optlastec.2024.111005}{doi:\nolinkurl{10.1016/j.optlastec.2024.111005}}


\bibitem[Su et~al\mbox{.}(2021b)]%
        {ref-VCSEL-408}
\bibfield{author}{\bibinfo{person}{Patrick Su}, \bibinfo{person}{Kevin~P. Pikul}, \bibinfo{person}{Mark~D. Kraman}, {and} \bibinfo{person}{John~M. Dallesasse}.} \bibinfo{year}{2021}\natexlab{b}.
\newblock \showarticletitle{High-power single-mode vertical-cavity surface-emitting lasers using strain-controlled disorder-defined apertures}.
\newblock \bibinfo{journal}{\emph{Applied Physics Letters}} \bibinfo{volume}{119}, \bibinfo{number}{24} (\bibinfo{date}{12} \bibinfo{year}{2021}), \bibinfo{pages}{241101}.
\newblock
\showISSN{0003-6951}
\href{https://doi.org/10.1063/5.0068713}{doi:\nolinkurl{10.1063/5.0068713}}


\bibitem[Su et~al\mbox{.}(2021a)]%
        {Su2021MDM}
\bibfield{author}{\bibinfo{person}{Yikai Su}, \bibinfo{person}{Yu He}, \bibinfo{person}{Haoshuo Chen}, \bibinfo{person}{Xiaoying Li}, {and} \bibinfo{person}{Guifang Li}.} \bibinfo{year}{2021}\natexlab{a}.
\newblock \showarticletitle{Perspective on mode-division multiplexing}.
\newblock \bibinfo{journal}{\emph{Applied Physics Letters}} \bibinfo{volume}{118}, \bibinfo{number}{20} (\bibinfo{year}{2021}), \bibinfo{pages}{200502}.
\newblock
\href{https://doi.org/10.1063/5.0046071}{doi:\nolinkurl{10.1063/5.0046071}}


\bibitem[Sun et~al\mbox{.}(2022)]%
        {Sun2022PlasmonicPCM}
\bibfield{author}{\bibinfo{person}{Wei Sun}, \bibinfo{person}{Yu Lu}, \bibinfo{person}{Liang Miao}, {and} \bibinfo{person}{Yiming Zhang}.} \bibinfo{year}{2022}\natexlab{}.
\newblock \showarticletitle{All-Optical Phase-Change Memory with Improved Performance by Plasmonic Effect}.
\newblock \bibinfo{journal}{\emph{Photonics}} \bibinfo{volume}{9}, \bibinfo{number}{3} (\bibinfo{year}{2022}), \bibinfo{pages}{132}.
\newblock
\href{https://doi.org/10.3390/photonics9030132}{doi:\nolinkurl{10.3390/photonics9030132}}


\bibitem[Tait et~al\mbox{.}(2017)]%
        {NP_SciRep2017_Tait}
\bibfield{author}{\bibinfo{person}{Alexander~N. Tait}, \bibinfo{person}{Thomas~Ferreira de Lima}, \bibinfo{person}{Ellen Zhou}, {et~al\mbox{.}}} \bibinfo{year}{2017}\natexlab{}.
\newblock \showarticletitle{Neuromorphic photonic networks using silicon photonic weight banks}.
\newblock \bibinfo{journal}{\emph{Sci. Rep.}} (\bibinfo{year}{2017}).
\newblock


\bibitem[Tamura et~al\mbox{.}(2015)]%
        {tamura2015silica}
\bibfield{author}{\bibinfo{person}{Takuya Tamura}, \bibinfo{person}{Keisuke Kondo}, \bibinfo{person}{Yosuke Terada}, \bibinfo{person}{Yosuke Hinakura}, \bibinfo{person}{Norihiro Ishikura}, {and} \bibinfo{person}{Toshihiko Baba}.} \bibinfo{year}{2015}\natexlab{}.
\newblock \showarticletitle{Silica-clad silicon photonic crystal waveguides for wideband dispersion-free slow light}.
\newblock \bibinfo{journal}{\emph{Journal of Lightwave Technology}} \bibinfo{volume}{33}, \bibinfo{number}{14} (\bibinfo{year}{2015}), \bibinfo{pages}{3034--3040}.
\newblock


\bibitem[Tang et~al\mbox{.}(2026)]%
        {ref-SOA-103}
\bibfield{author}{\bibinfo{person}{Guobiao Tang}, \bibinfo{person}{Haibo Kuang}, \bibinfo{person}{Yao Fu}, \bibinfo{person}{Shiao Zhao}, \bibinfo{person}{Yu Zhang}, {and} \bibinfo{person}{Jian Wang}.} \bibinfo{year}{2026}\natexlab{}.
\newblock \showarticletitle{L-Band widely tunable laser and high gain semiconductor optical amplifier based on heterogeneous integration of III-V on silicon}.
\newblock \bibinfo{journal}{\emph{Opt. Express}} \bibinfo{volume}{34}, \bibinfo{number}{2} (\bibinfo{date}{Jan} \bibinfo{year}{2026}), \bibinfo{pages}{1826--1836}.
\newblock
\href{https://doi.org/10.1364/OE.586013}{doi:\nolinkurl{10.1364/OE.586013}}


\bibitem[Tang et~al\mbox{.}(2025)]%
        {tang2025waveguide}
\bibfield{author}{\bibinfo{person}{Rui Tang}, \bibinfo{person}{Makoto Okano}, \bibinfo{person}{Chao Zhang}, \bibinfo{person}{Kasidit Toprasertpong}, \bibinfo{person}{Shinichi Takagi}, {and} \bibinfo{person}{Mitsuru Takenaka}.} \bibinfo{year}{2025}\natexlab{}.
\newblock \showarticletitle{Waveguide-multiplexed photonic matrix--vector multiplication processor using multiport photodetectors}.
\newblock \bibinfo{journal}{\emph{Optica}} \bibinfo{volume}{12}, \bibinfo{number}{6} (\bibinfo{year}{2025}), \bibinfo{pages}{812--820}.
\newblock


\bibitem[Terada et~al\mbox{.}(2014)]%
        {Baba2014PhC}
\bibfield{author}{\bibinfo{person}{Yosuke Terada}, \bibinfo{person}{Hiroyuki Ito}, \bibinfo{person}{Hong~C. Nguyen}, {and} \bibinfo{person}{Toshihiko Baba}.} \bibinfo{year}{2014}\natexlab{}.
\newblock \showarticletitle{Theoretical and experimental investigation of low-volgage and low-loss 25-Gbps Si photonic crystal slow light Mach–Zehnder modulators with interleaved p/n junction}.
\newblock \bibinfo{journal}{\emph{Frontiers in Physics}}  \bibinfo{volume}{Volume 2 - 2014} (\bibinfo{year}{2014}).
\newblock
\showISSN{2296-424X}
\href{https://doi.org/10.3389/fphy.2014.00061}{doi:\nolinkurl{10.3389/fphy.2014.00061}}


\bibitem[Tian et~al\mbox{.}(2023)]%
        {tian2023designing}
\bibfield{author}{\bibinfo{person}{Keyu Tian}, \bibinfo{person}{Yi Jiang}, \bibinfo{person}{Qishuai Diao}, \bibinfo{person}{Chen Lin}, \bibinfo{person}{Liwei Wang}, {and} \bibinfo{person}{Zehuan Yuan}.} \bibinfo{year}{2023}\natexlab{}.
\newblock \showarticletitle{Designing BERT for Convolutional Networks: Sparse and Hierarchical Masked Modeling}.
\newblock \bibinfo{journal}{\emph{arXiv:2301.03580}} (\bibinfo{year}{2023}).
\newblock


\bibitem[Timurdogan et~al\mbox{.}(2018)]%
        {timurdogan2018aim}
\bibfield{author}{\bibinfo{person}{Erman Timurdogan}, \bibinfo{person}{Zhan Su}, \bibinfo{person}{Christopher~V Poulton}, \bibinfo{person}{Matthew~J Byrd}, \bibinfo{person}{Simon Xin}, \bibinfo{person}{Ren-Jye Shiue}, \bibinfo{person}{Benjamin~R Moss}, \bibinfo{person}{Ehsan~S Hosseini}, {and} \bibinfo{person}{Michael~R Watts}.} \bibinfo{year}{2018}\natexlab{}.
\newblock \showarticletitle{AIM process design kit (AIMPDKv2. 0): Silicon photonics passive and active component libraries on a 300mm wafer}. In \bibinfo{booktitle}{\emph{Optical Fiber Communication Conference}}. Optica Publishing Group, \bibinfo{pages}{M3F--1}.
\newblock


\bibitem[Totovic et~al\mbox{.}(2022)]%
        {Totovic2022WDMUniversalLinearOptics}
\bibfield{author}{\bibinfo{person}{Angelina Totovic}, \bibinfo{person}{Christos Pappas}, \bibinfo{person}{Manos Kirtas}, \bibinfo{person}{Apostolos Tsakyridis}, \bibinfo{person}{George Giamougiannis}, \bibinfo{person}{Nikolaos Passalis}, \bibinfo{person}{Miltiadis Moralis-Pegios}, \bibinfo{person}{Anastasios Tefas}, {and} \bibinfo{person}{Nikos Pleros}.} \bibinfo{year}{2022}\natexlab{}.
\newblock \showarticletitle{WDM equipped universal linear optics for programmable neuromorphic photonic processors}.
\newblock \bibinfo{journal}{\emph{Neuromorphic Computing and Engineering}} \bibinfo{volume}{2}, \bibinfo{number}{2} (\bibinfo{year}{2022}), \bibinfo{pages}{024010}.
\newblock
\href{https://doi.org/10.1088/2634-4386/ac724d}{doi:\nolinkurl{10.1088/2634-4386/ac724d}}


\bibitem[Wan et~al\mbox{.}(2026)]%
        {ref-SOA-104}
\bibfield{author}{\bibinfo{person}{Y. Wan}, \bibinfo{person}{W. He}, \bibinfo{person}{J. Jaussi}, \bibinfo{person}{L. Liao}, \bibinfo{person}{D. Pan}, \bibinfo{person}{J. Bowers}, {and} \bibinfo{person}{H. Rong}.} \bibinfo{year}{2026}\natexlab{}.
\newblock \showarticletitle{Integrating silicon photonics with complementary metal–oxide–semiconductor technologies.}
\newblock \bibinfo{journal}{\emph{Nat Rev Electr Eng 3, 15–31}} (\bibinfo{year}{2026}).
\newblock


\bibitem[Wang et~al\mbox{.}(2014)]%
        {wang_low-loss_2014}
\bibfield{author}{\bibinfo{person}{Jing Wang}, \bibinfo{person}{Zhen Sheng}, \bibinfo{person}{Le Li}, \bibinfo{person}{Albert Pang}, \bibinfo{person}{Aimin Wu}, \bibinfo{person}{Wei Li}, \bibinfo{person}{Xi Wang}, \bibinfo{person}{Shichang Zou}, \bibinfo{person}{Minghao Qi}, {and} \bibinfo{person}{Fuwan Gan}.} \bibinfo{year}{2014}\natexlab{}.
\newblock \showarticletitle{Low-loss and low-crosstalk 8 × 8 silicon nanowire {AWG} routers fabricated with {CMOS} technology}.
\newblock \bibinfo{journal}{\emph{Optics Express}} \bibinfo{volume}{22}, \bibinfo{number}{8} (\bibinfo{date}{April} \bibinfo{year}{2014}), \bibinfo{pages}{9395}.
\newblock
\showISSN{1094-4087}
\href{https://doi.org/10.1364/OE.22.009395}{doi:\nolinkurl{10.1364/OE.22.009395}}


\bibitem[Wang et~al\mbox{.}(2021)]%
        {9052469}
\bibfield{author}{\bibinfo{person}{Jingdong Wang}, \bibinfo{person}{Ke Sun}, \bibinfo{person}{Tianheng Cheng}, \bibinfo{person}{Borui Jiang}, \bibinfo{person}{Chaorui Deng}, \bibinfo{person}{Yang Zhao}, \bibinfo{person}{Dong Liu}, \bibinfo{person}{Yadong Mu}, \bibinfo{person}{Mingkui Tan}, \bibinfo{person}{Xinggang Wang}, \bibinfo{person}{Wenyu Liu}, {and} \bibinfo{person}{Bin Xiao}.} \bibinfo{year}{2021}\natexlab{}.
\newblock \showarticletitle{{ Deep High-Resolution Representation Learning for Visual Recognition }}.
\newblock \bibinfo{journal}{\emph{IEEE Transactions on Pattern Analysis \& Machine Intelligence}} \bibinfo{volume}{43}, \bibinfo{number}{10} (\bibinfo{date}{Oct.} \bibinfo{year}{2021}), \bibinfo{pages}{3349--3364}.
\newblock
\showISSN{1939-3539}
\href{https://doi.org/10.1109/TPAMI.2020.2983686}{doi:\nolinkurl{10.1109/TPAMI.2020.2983686}}


\bibitem[Więckowska et~al\mbox{.}(2020)]%
        {ref-VCSEL-406}
\bibfield{author}{\bibinfo{person}{M. Więckowska}, \bibinfo{person}{R.P. Sarzała}, \bibinfo{person}{R. Ledzion}, {and} \bibinfo{person}{M. Dems}.} \bibinfo{year}{2020}\natexlab{}.
\newblock \showarticletitle{Impact of an Antiresonant Oxide Island on the Lasing of Lateral Modes in VCSELs}.
\newblock \bibinfo{journal}{\emph{Materials, 13, 2195}} (\bibinfo{year}{2020}).
\newblock


\bibitem[Xia et~al\mbox{.}(2024)]%
        {Xia2024}
\bibfield{author}{\bibinfo{person}{J. Xia}, \bibinfo{person}{T. Wang}, \bibinfo{person}{Z. Wang}, \bibinfo{person}{J. Gong}, \bibinfo{person}{Y. Dong}, \bibinfo{person}{R. Yang}, {and} \bibinfo{person}{X. Miao}.} \bibinfo{year}{2024}\natexlab{}.
\newblock \showarticletitle{7 Bit Nonvolatile Electrically Programmable Photonics Based on Phase-Change Materials}.
\newblock \bibinfo{journal}{\emph{ACS Photonics}} (\bibinfo{date}{Jan.} \bibinfo{year}{2024}).
\newblock
\href{https://doi.org/10.1021/acsphotonics.3c01598}{doi:\nolinkurl{10.1021/acsphotonics.3c01598}}


\bibitem[Xiao et~al\mbox{.}(2024)]%
        {ref-VCSEL-407}
\bibfield{author}{\bibinfo{person}{Yao Xiao}, \bibinfo{person}{Pei Miao}, \bibinfo{person}{Jun Wang}, \bibinfo{person}{Heng Liu}, \bibinfo{person}{Yudan Gou}, \bibinfo{person}{Zhicheng Zhang}, \bibinfo{person}{Bangguo Wang}, \bibinfo{person}{Wuling Liu}, \bibinfo{person}{Qijie Wang}, \bibinfo{person}{Guoliang Deng}, {and} \bibinfo{person}{Shouhuan Zhou}.} \bibinfo{year}{2024}\natexlab{}.
\newblock \showarticletitle{Twenty-milliwatt, high-power, high-efficiency, single-mode, multi-junction vertical-cavity surface-emitting lasers using surface microstructures}.
\newblock \bibinfo{journal}{\emph{Photon. Res.}} \bibinfo{volume}{12}, \bibinfo{number}{9} (\bibinfo{date}{Sep} \bibinfo{year}{2024}), \bibinfo{pages}{1899--1906}.
\newblock
\urldef\tempurl%
\url{https://opg.optica.org/prj/abstract.cfm?URI=prj-12-9-1899}
\showURL{%
\tempurl}


\bibitem[Xue et~al\mbox{.}(2017)]%
        {Xue_kerr_2017}
\bibfield{author}{\bibinfo{person}{Xiaoxiao Xue}, \bibinfo{person}{Pei-Hsun Wang}, \bibinfo{person}{Yi Xuan}, \bibinfo{person}{Minghao Qi}, {and} \bibinfo{person}{Andrew~M. Weiner}.} \bibinfo{year}{2017}\natexlab{}.
\newblock \showarticletitle{Microresonator {Kerr} frequency combs with high conversion efficiency}.
\newblock \bibinfo{journal}{\emph{Laser \& Photonics Reviews}} \bibinfo{volume}{11}, \bibinfo{number}{1} (\bibinfo{year}{2017}), \bibinfo{pages}{1600276}.
\newblock
\href{https://doi.org/10.1002/lpor.201600276}{doi:\nolinkurl{10.1002/lpor.201600276}}


\bibitem[Yan et~al\mbox{.}(2017)]%
        {yan2017slow}
\bibfield{author}{\bibinfo{person}{Siqi Yan}, \bibinfo{person}{Xiaolong Zhu}, \bibinfo{person}{Lars~Hagedorn Frandsen}, \bibinfo{person}{Sanshui Xiao}, \bibinfo{person}{N~Asger Mortensen}, \bibinfo{person}{Jianji Dong}, {and} \bibinfo{person}{Yunhong Ding}.} \bibinfo{year}{2017}\natexlab{}.
\newblock \showarticletitle{Slow-light-enhanced energy efficiency for graphene microheaters on silicon photonic crystal waveguides}.
\newblock \bibinfo{journal}{\emph{Nature Communications}} \bibinfo{volume}{8}, \bibinfo{number}{1} (\bibinfo{year}{2017}), \bibinfo{pages}{14411}.
\newblock


\bibitem[Yang et~al\mbox{.}(2022)]%
        {Yang2022MultiDimMicrocombs}
\bibfield{author}{\bibinfo{person}{Ki~Youl Yang}, \bibinfo{person}{Chinmay Shirpurkar}, \bibinfo{person}{Alexander~D. White}, \bibinfo{person}{Jizhao Zang}, \bibinfo{person}{Lin Chang}, \bibinfo{person}{Farshid Ashtiani}, \bibinfo{person}{Melissa~A. Guidry}, \bibinfo{person}{Daniil~M. Lukin}, \bibinfo{person}{Srinivas~V. Pericherla}, \bibinfo{person}{Joshua Yang}, \bibinfo{person}{Hyounghan Kwon}, \bibinfo{person}{Jesse Lu}, \bibinfo{person}{Geun~Ho Ahn}, \bibinfo{person}{Kasper Van~Gasse}, \bibinfo{person}{Yan Jin}, \bibinfo{person}{Su-Peng Yu}, \bibinfo{person}{Travis~C. Briles}, \bibinfo{person}{Jordan~R. Stone}, \bibinfo{person}{David~R. Carlson}, \bibinfo{person}{Hao Song}, \bibinfo{person}{Kaiheng Zou}, \bibinfo{person}{Huibin Zhou}, \bibinfo{person}{Kai Pang}, \bibinfo{person}{Han Hao}, {et~al\mbox{.}}} \bibinfo{year}{2022}\natexlab{}.
\newblock \showarticletitle{Multi-dimensional data transmission using inverse-designed silicon photonics and microcombs}.
\newblock \bibinfo{journal}{\emph{Nature Communications}}  \bibinfo{volume}{13} (\bibinfo{year}{2022}), \bibinfo{pages}{7862}.
\newblock
\href{https://doi.org/10.1038/s41467-022-35641-5}{doi:\nolinkurl{10.1038/s41467-022-35641-5}}


\bibitem[Yang et~al\mbox{.}(2023)]%
        {yangHighperformanceGratingCouplers2023}
\bibfield{author}{\bibinfo{person}{Mingxiang Yang}, \bibinfo{person}{Yunjie Yan}, \bibinfo{person}{Zhenlin Wu}, \bibinfo{person}{YiYing Gu}, \bibinfo{person}{Shiyuan Zhao}, \bibinfo{person}{Geert Morthier}, {and} \bibinfo{person}{Mingshan Zhao}.} \bibinfo{year}{2023}\natexlab{}.
\newblock \showarticletitle{High-Performance Grating Couplers on 220-Nm Thick Silicon by Inverse Design for Perfectly Vertical Coupling}.
\newblock \bibinfo{journal}{\emph{Scientific Reports}} \bibinfo{volume}{13}, \bibinfo{number}{1} (\bibinfo{date}{Oct.} \bibinfo{year}{2023}), \bibinfo{pages}{18112}.
\newblock
\showISSN{2045-2322}
\href{https://doi.org/10.1038/s41598-023-45168-2}{doi:\nolinkurl{10.1038/s41598-023-45168-2}}


\bibitem[Yin et~al\mbox{.}(2024)]%
        {ONN_ICCAD2024_Gu}
\bibfield{author}{\bibinfo{person}{Ziang Yin}, \bibinfo{person}{Nicholas Gangi}, \bibinfo{person}{Meng Zhang}, \bibinfo{person}{Jeff Zhang}, \bibinfo{person}{Rena Huang}, {and} \bibinfo{person}{Jiaqi Gu}.} \bibinfo{year}{2024}\natexlab{}.
\newblock \showarticletitle{{SCATTER: Algorithm-Circuit Co-Sparse Photonic Accelerator with Thermal-Tolerant, Power-Efficient In-situ Light Redistribution}}. In \bibinfo{booktitle}{\emph{Proc.~ICCAD}}.
\newblock


\bibitem[Yin et~al\mbox{.}(2025)]%
        {ONN_DAC2025_Gu_SimPhony}
\bibfield{author}{\bibinfo{person}{Ziang Yin}, \bibinfo{person}{Meng Zhang}, \bibinfo{person}{Amir Begovic}, \bibinfo{person}{Rena Huang}, \bibinfo{person}{Jeff Zhang}, {and} \bibinfo{person}{Jiaqi Gu}.} \bibinfo{year}{2025}\natexlab{}.
\newblock \showarticletitle{{SimPhony: A Device-Circuit-Architecture Cross-Layer Modeling and Simulation Framework for Heterogeneous Electronic-Photonic AI System}}. In \bibinfo{booktitle}{\emph{Proc.~DAC}}.
\newblock


\bibitem[Yoon et~al\mbox{.}(2023)]%
        {yoonInverseDesignSibased2023}
\bibfield{author}{\bibinfo{person}{Jinhyeong Yoon}, \bibinfo{person}{Jae-Yong Kim}, \bibinfo{person}{Junhyeong Kim}, \bibinfo{person}{Hyeonho Yoon}, \bibinfo{person}{Berkay Ne{\c s}eli}, \bibinfo{person}{Hyo-Hoon Park}, {and} \bibinfo{person}{Hamza Kurt}.} \bibinfo{year}{2023}\natexlab{}.
\newblock \showarticletitle{Inverse Design of a {{Si-based}} High-Performance Vertical-Emitting Meta-Grating Coupler on a 220 Nm Silicon-on-Insulator Platform}.
\newblock \bibinfo{journal}{\emph{Photonics Research}} \bibinfo{volume}{11}, \bibinfo{number}{6} (\bibinfo{date}{June} \bibinfo{year}{2023}), \bibinfo{pages}{897}.
\newblock
\showISSN{2327-9125}
\href{https://doi.org/10.1364/PRJ.473978}{doi:\nolinkurl{10.1364/PRJ.473978}}


\bibitem[Yuan et~al\mbox{.}(2020)]%
        {10.1007/978-3-030-58539-6_11}
\bibfield{author}{\bibinfo{person}{Yuhui Yuan}, \bibinfo{person}{Xilin Chen}, {and} \bibinfo{person}{Jingdong Wang}.} \bibinfo{year}{2020}\natexlab{}.
\newblock \showarticletitle{Object-Contextual Representations for Semantic Segmentation}. In \bibinfo{booktitle}{\emph{Computer Vision – ECCV 2020: 16th European Conference, Glasgow, UK, August 23–28, 2020, Proceedings, Part VI}} (Glasgow, United Kingdom). \bibinfo{publisher}{Springer-Verlag}, \bibinfo{address}{Berlin, Heidelberg}, \bibinfo{pages}{173–190}.
\newblock
\showISBNx{978-3-030-58538-9}
\href{https://doi.org/10.1007/978-3-030-58539-6_11}{doi:\nolinkurl{10.1007/978-3-030-58539-6_11}}


\bibitem[Zhang et~al\mbox{.}(2024a)]%
        {Zhang:24}
\bibfield{author}{\bibinfo{person}{Meng Zhang}, \bibinfo{person}{Amir Begovi\'{c}}, \bibinfo{person}{Dennis Yin}, \bibinfo{person}{Nicholas Gangi}, \bibinfo{person}{Jiaqi Gu}, {and} \bibinfo{person}{Rena Huang}.} \bibinfo{year}{2024}\natexlab{a}.
\newblock \showarticletitle{Foundry Manufactured 6-bit Resolution, 150$\mu$m Long Slow-Light Electro-Optic Modulator for On-Chip Photonic Tensor Computing}, In \bibinfo{booktitle}{CLEO 2024}.
\newblock \bibinfo{journal}{\emph{CLEO 2024}}, \bibinfo{pages}{AM4J.1}.
\newblock
\href{https://doi.org/10.1364/CLEO_AT.2024.AM4J.1}{doi:\nolinkurl{10.1364/CLEO_AT.2024.AM4J.1}}


\bibitem[Zhang et~al\mbox{.}(2019a)]%
        {Zhang_ring_2019}
\bibfield{author}{\bibinfo{person}{Mian Zhang}, \bibinfo{person}{Brandon Buscaino}, \bibinfo{person}{Cheng Wang}, \bibinfo{person}{Amirhassan Shams-Ansari}, \bibinfo{person}{Christian Reimer}, \bibinfo{person}{Rongrong Zhu}, \bibinfo{person}{Joseph~M. Kahn}, {and} \bibinfo{person}{Marko~Lon\v car}.} \bibinfo{year}{2019}\natexlab{a}.
\newblock \showarticletitle{Broadband electro-optic frequency comb generation in a lithium niobate microring resonator}.
\newblock \bibinfo{journal}{\emph{Nature}} \bibinfo{volume}{568}, \bibinfo{number}{7752} (\bibinfo{date}{March} \bibinfo{year}{2019}), \bibinfo{pages}{373--377}.
\newblock
\href{https://doi.org/10.1038/s41586-019-1008-7}{doi:\nolinkurl{10.1038/s41586-019-1008-7}}


\bibitem[Zhang et~al\mbox{.}(2025a)]%
        {zhang2025thermally}
\bibfield{author}{\bibinfo{person}{Meng Zhang}, \bibinfo{person}{Nicholas Gangi}, \bibinfo{person}{Amir Begovi{\'c}}, \bibinfo{person}{Alexander Chen}, \bibinfo{person}{Bowen Liu}, {and} \bibinfo{person}{Z~Rena Huang}.} \bibinfo{year}{2025}\natexlab{a}.
\newblock \showarticletitle{Thermally robust compact single Bragg grating silicon modulator for photonic computing}.
\newblock \bibinfo{journal}{\emph{Applied Optics}} \bibinfo{volume}{64}, \bibinfo{number}{16} (\bibinfo{year}{2025}), \bibinfo{pages}{4646--4653}.
\newblock


\bibitem[Zhang et~al\mbox{.}(2024b)]%
        {zhang2024tempo}
\bibfield{author}{\bibinfo{person}{Meng Zhang}, \bibinfo{person}{Dennis Yin}, \bibinfo{person}{Nicholas Gangi}, \bibinfo{person}{Amir Begovi{\'c}}, \bibinfo{person}{Alexander Chen}, \bibinfo{person}{Zhaoran~Rena Huang}, {and} \bibinfo{person}{Jiaqi Gu}.} \bibinfo{year}{2024}\natexlab{b}.
\newblock \showarticletitle{Tempo: efficient time-multiplexed dynamic photonic tensor core for edge AI with compact slow-light electro-optic modulator}.
\newblock \bibinfo{journal}{\emph{Journal of Applied Physics}} \bibinfo{volume}{135}, \bibinfo{number}{22} (\bibinfo{year}{2024}).
\newblock


\bibitem[Zhang et~al\mbox{.}(2025b)]%
        {ref-VCSE-412}
\bibfield{author}{\bibinfo{person}{Xuhao Zhang}, \bibinfo{person}{Zhiting Tang}, \bibinfo{person}{Yihua Wang}, \bibinfo{person}{Feiyun Zhao}, \bibinfo{person}{Qiao He}, \bibinfo{person}{Aobo Ren}, \bibinfo{person}{Jing Zhang}, \bibinfo{person}{Wuyang Ren}, {and} \bibinfo{person}{Jiang Wu}.} \bibinfo{year}{2025}\natexlab{b}.
\newblock \showarticletitle{High-fundamental-mode output of 1064 nm vertical-cavity surface-emitting laser using double embedded antiresonant oxide islands}.
\newblock \bibinfo{journal}{\emph{Information \& Functional Materials}} \bibinfo{volume}{2}, \bibinfo{number}{2} (\bibinfo{year}{2025}), \bibinfo{pages}{118--129}.
\newblock
\href{https://doi.org/10.1002/ifm2.34}{doi:\nolinkurl{10.1002/ifm2.34}}


\bibitem[Zhang et~al\mbox{.}(2019b)]%
        {Zhang2019}
\bibfield{author}{\bibinfo{person}{Y. Zhang}, \bibinfo{person}{J.~B. Chou}, \bibinfo{person}{J. Li}, \bibinfo{person}{H. Li}, \bibinfo{person}{Q. Du}, \bibinfo{person}{A. Yadav}, \bibinfo{person}{S. Zhou}, \bibinfo{person}{M.~Y. Shalaginov}, \bibinfo{person}{Z. Fang}, \bibinfo{person}{H. Zhong}, \bibinfo{person}{C. Roberts}, \bibinfo{person}{P. Robinson}, \bibinfo{person}{B. Bohlin}, \bibinfo{person}{C. R{\'i}os}, \bibinfo{person}{H. Lin}, \bibinfo{person}{M. Kang}, \bibinfo{person}{T. Gu}, \bibinfo{person}{J. Warner}, \bibinfo{person}{V. Liberman}, \bibinfo{person}{K. Richardson}, {and} \bibinfo{person}{J. Hu}.} \bibinfo{year}{2019}\natexlab{b}.
\newblock \showarticletitle{Broadband Transparent Optical Phase Change Materials for High-Performance Nonvolatile Photonics}.
\newblock \bibinfo{journal}{\emph{Nature Communications}} \bibinfo{volume}{10}, \bibinfo{number}{1} (\bibinfo{year}{2019}), \bibinfo{pages}{4279}.
\newblock
\href{https://doi.org/10.1038/s41467-019-12196-4}{doi:\nolinkurl{10.1038/s41467-019-12196-4}}


\bibitem[Zhou and Mawst(2002)]%
        {ref-VCSEL-410}
\bibfield{author}{\bibinfo{person}{Delai Zhou} {and} \bibinfo{person}{L.J. Mawst}.} \bibinfo{year}{2002}\natexlab{}.
\newblock \showarticletitle{High-power single-mode antiresonant reflecting optical waveguide-type vertical-cavity surface-emitting lasers}.
\newblock \bibinfo{journal}{\emph{IEEE Journal of Quantum Electronics}} \bibinfo{volume}{38}, \bibinfo{number}{12} (\bibinfo{year}{2002}), \bibinfo{pages}{1599--1606}.
\newblock
\href{https://doi.org/10.1109/JQE.2002.805107}{doi:\nolinkurl{10.1109/JQE.2002.805107}}


\bibitem[Zhou et~al\mbox{.}(2015)]%
        {ref-VCSEL-401}
\bibfield{author}{\bibinfo{person}{Delai Zhou}, \bibinfo{person}{Jean-Francois Seurin}, \bibinfo{person}{Guoyang Xu}, \bibinfo{person}{Pu Zhao}, \bibinfo{person}{Bing Xu}, \bibinfo{person}{Tong Chen}, {et~al\mbox{.}}} \bibinfo{year}{2015}\natexlab{}.
\newblock \showarticletitle{Progress on high-power high-brightness VCSELs and applications}.
\newblock \bibinfo{journal}{\emph{Proceedings Volume 9381, Vertical-Cavity Surface-Emitting Lasers XIX; 93810B}} (\bibinfo{year}{2015}).
\newblock


\bibitem[Zhou et~al\mbox{.}(2023)]%
        {Zhou2023}
\bibfield{author}{\bibinfo{person}{Wen Zhou}, \bibinfo{person}{Bowei Dong}, \bibinfo{person}{Nikolaos Farmakidis}, \bibinfo{person}{Xuan Li}, \bibinfo{person}{Nathan Youngblood}, \bibinfo{person}{Kairan Huang}, \bibinfo{person}{Yuhan He}, \bibinfo{person}{C. David~Wright}, \bibinfo{person}{Wolfram H.~P. Pernice}, {and} \bibinfo{person}{Harish Bhaskaran}.} \bibinfo{year}{2023}\natexlab{}.
\newblock \showarticletitle{In-memory photonic dot-product engine with electrically programmable weight banks}.
\newblock \bibinfo{journal}{\emph{Nature Communications}} \bibinfo{volume}{14}, \bibinfo{number}{1} (\bibinfo{date}{20 May} \bibinfo{year}{2023}), \bibinfo{pages}{2887}.
\newblock
\showISSN{2041-1723}
\href{https://doi.org/10.1038/s41467-023-38473-x}{doi:\nolinkurl{10.1038/s41467-023-38473-x}}


\bibitem[Zhou et~al\mbox{.}(2022)]%
        {Zhou2022PCMPhotonicMemory}
\bibfield{author}{\bibinfo{person}{Wei Zhou}, \bibinfo{person}{Nikolaos Farmakidis}, \bibinfo{person}{Jorik Feldmann}, \bibinfo{person}{Xuan Li}, \bibinfo{person}{Jie Tan}, \bibinfo{person}{Yujie He}, \bibinfo{person}{C.~David Wright}, \bibinfo{person}{Wolfram H.~P. Pernice}, {and} \bibinfo{person}{Harish Bhaskaran}.} \bibinfo{year}{2022}\natexlab{}.
\newblock \showarticletitle{Phase-Change Materials for Energy-Efficient Photonic Memory and Computing}.
\newblock \bibinfo{journal}{\emph{MRS Bulletin}} \bibinfo{volume}{47}, \bibinfo{number}{5} (\bibinfo{year}{2022}), \bibinfo{pages}{502--510}.
\newblock
\href{https://doi.org/10.1557/s43577-022-00358-7}{doi:\nolinkurl{10.1557/s43577-022-00358-7}}


\bibitem[Zhu et~al\mbox{.}(2021)]%
        {zhu2021compact}
\bibfield{author}{\bibinfo{person}{Junbo Zhu}, \bibinfo{person}{Qiu Chao}, \bibinfo{person}{Haiyang Huang}, \bibinfo{person}{Yingxuan Zhao}, \bibinfo{person}{Yang Li}, \bibinfo{person}{Lue Tao}, \bibinfo{person}{Xiaojuan She}, \bibinfo{person}{Han Liao}, \bibinfo{person}{Rui Huang}, \bibinfo{person}{Zijian Zhu}, \bibinfo{person}{Xiang Liu}, \bibinfo{person}{Zhen Sheng}, {and} \bibinfo{person}{Fuwan Gan}.} \bibinfo{year}{2021}\natexlab{}.
\newblock \showarticletitle{Compact, broadband, and low-loss silicon photonic arbitrary ratio power splitter using adiabatic taper}.
\newblock \bibinfo{journal}{\emph{Applied optics}} \bibinfo{volume}{60}, \bibinfo{number}{2} (\bibinfo{year}{2021}), \bibinfo{pages}{413--416}.
\newblock


\end{thebibliography}

\end{document}